\documentclass[AMA,Times1COL]{WileyNJDv5} 

\articletype{Research Article}%

\received{Date Month Year}
\revised{Date Month Year}
\accepted{Date Month Year}
\journal{Statistics in Medicine}
\volume{00}
\copyyear{2023}
\startpage{1}

\usepackage{bm, bbm}
\usepackage{multirow}
\usepackage{lettersp}
\usepackage{enumitem}
\usepackage{graphicx}
\usepackage{svg}
\def\Pr{P}
\def\Sp{\mathrm{Sp}}
\def\Se{\mathrm{Se}}

\def\AUC{\mathrm{AUC}}

\def\indep{\!\perp\!\!\!\perp}

\begin{document}
\title{Cumulative/Dynamic Time-Dependent ROC Analysis for Left-Truncated and Right-Censored Data}

\author[1]{Kendrick Li*}
\author[2]{Mithun Kumar Acharjee*}
\author[3]{Yuyao Wang}
\author[1]{Subodh Selukar}
\author[1]{Sedigheh Mirzaei}
\author[4]{Yutaka Yasui}
\author[1]{Deo Kumar Srivastava}

\authormark{Li et al.}

\titlemark{Time-Dependent ROC for LTRC Data}
\address[1]{\orgdiv{Department of Biostatistics}, \orgname{St. Jude Children's Research Hospital}, \orgaddress{\state{Tennessee}, \country{USA}}}

\address[2]{\orgdiv{School of Health Professions}, \orgname{University of Southern Mississippi}, \orgaddress{\state{Mississippi}, \country{USA}}}

\address[3]{\orgdiv{Department of Biostatistics}, \orgname{Brown University}, \orgaddress{\state{Rhode Island}, \country{USA}}}

\address[4]{\orgdiv{Department of Epidemiology and Cancer Control}, \orgname{St. Jude Children's Research Hospital}, \orgaddress{\state{Tennessee}, \country{USA}}}

\corres{Corresponding author: Kendrick Li, 262 Danny Thomas Place
Memphis,  TN 38105. \email{kendrick.li@stjude.org}}

\abstract[Abstract]{Time-dependent Receiver Operating Characteristics (ROC) analysis is a standard method to evaluate the ability of a biomarker or predictive score to discriminate between cases and non-cases for time-to-event outcomes. Extensions of this useful method to left-truncated right-censored data have been understudied, with the exception of Li 2016 under limited settings. In this paper, we first developed several extended estimators based on Li 2016 for settings with independent or covariate-induced dependent left truncation and right censoring. We further introduced novel inverse probability weighting estimators of sensitivity, specificity, and area under the ROC curve (AUC), where the weights simultaneously account for left truncation and right censoring, in various scenarios. We evaluated the proposed AUC estimators in comprehensive simulation studies and performed the proposed time-dependent ROC analysis to evaluate the discriminatory performance of a mortality prediction model in elderly end-stage renal disease patients using the Organ Procurement and Transplantation Network data.}


\jnlcitation{\cname{%
\author{Li K.} and \author{Acharjee M.} }.
\ctitle{Cumulative/Dynamic Time-Dependent ROC Analysis for Left-Truncated and Right-Censored Data: Estimators and Comparison} \cjournal{\it Statistics in Medicine} \cvol{0000;00(00):1--18}.}

\maketitle

\renewcommand\thefootnote{}
\footnotetext{* These authors contributed equally to this work.}
\footnotetext{\textbf{Abbreviations:} ROC, receiver operating characteristics; AUC, area under the ROC curve; LTRC, left-truncated and right-censored; IPW: inverse probability weighting; CIPW: conditional inverse probability weighting; ESRD: end-stage renal disease; OPTN: Organ Procurement and Transplantation Network; MSE: mean squared errorl; CI: confidence interval.}

\renewcommand\thefootnote{\fnsymbol{footnote}}
\setcounter{footnote}{1}

\section{Introduction}\label{sec:intro}

The Receiver Operating Characteristics (ROC) analysis is a statistical method widely used to evaluate the ability of a continuous marker, often a biomarker or a score derived from a prediction model, to discriminate subjects for a binary outcome. Assuming a marker is used to predict the outcome of interest such that higher markers indicate greater likelihood that the outcome occurs, the ROC curve plots sensitivity (or true positive rate) against one minus specificity (or false positive rate) at all possible thresholds of the marker. The area under the ROC curve (AUC) is used to describe how well the marker discriminates the outcome status, with $\AUC=1$ indicating perfect discrimination and $\AUC=0.5$ indicating that the discriminative performance of the marker is no better than a random guess~\cite{metz1978basic, fawcett2006introduction}.

Cumulative/dynamic time-dependent ROC (hereafter referred to as time-dependent ROC for short) is a natural extension of the ROC analysis when the outcome is time-to-events~\cite{heagerty2000time}. In the time-dependent ROC analysis, the cases are defined as observations with events occurring before a pre-determined time point, and the non-cases are those with events occurring after the time point. A numeric marker from a prediction model or a biomarker  (referred to as ``the marker'' hereafter) is evaluated for their ability to discriminate between cases and non-cases. Several estimators of time-dependent sensitivity, specificity, and AUC have been proposed, which account for right censoring, including the Kaplan-Meier-based estimators by Heagerty et al.\cite{heagerty2000time} and the inverse probability weighting (IPW) estimators\cite{uno2007evaluating,hung2010optimal,blanche2013estimating}. Kamarudin et al. provided a comprehensive review of existing methods of time-dependent ROC analysis and its broad applications in biomedical research.~\cite{kamarudin2017time}

In some observational studies, the entry time of participants may be later than the intended time origin. This may naturally occur in prospective cohort studies or registry-based studies where participants are recruited sometime after the time origin, or may arise from the choice of a different time axis by the researchers. Such delayed entries result in left truncation, where only a subset of the underlying population who survived long enough to enter the study are observed~\cite {howards2007conditions,schisterman2013accuracy,betensky2015recognizing}. We refer to the data where left truncation and right censoring of the event times of interest are both present as left-truncated and right-censored (LTRC) data. The standard implementation of time-dependent ROC analysis, such as the R package timeROC~\cite{blanche2013estimating} or the inbuilt functions in the SAS PHREG procedure, does not account for left truncation and may give biased estimates of sensitivity, specificity, and AUC when applied to LTRC data.  Extensions of time-dependent ROC analysis for LTRC data have been understudied except Li 2016~\cite{li2017estimating}. However, Li's work has a few limitations: (1) Li's estimators of time-dependent sensitivity, specificity, and AUC were developed under the scenario where right censoring can only after study entry. Scenarios where censoring may occur before the left-truncation time are possible in some studies but were not considered.~\cite{qian2014assumptions} (2) Li developed regression-type estimators to account for marker-induced association between the left-truncation times and event times. If other factors besides the marker are associated with both the event times and the truncation times, Li's regression-type estimators may be biased. We provide a brief review of Li's estimators in Appendix~\ref{append:li-review}.

As an example, we consider the problem of evaluating of prediction model of mortality among elderly patients (60-75 years old at the initiation of dialysis to treat ESRD) with End Stage Renal Disease (ESRD) since the initiation of dialysis using the Organ Procurement and Transplantation Network (OPTN) data. Hartman et al. analyzed a similar problem to illustrate their proposed concordance index estimator with LTRC data~\citep{hartman2023concordance}. Time of dialysis initiation is the intended time origin as a proxy to time of ESRD diagnosis because patients often started dialysis immediately after their diagnosis. However, the OPTN data collection for the patients only started at their waitlist registration. The patients who died before they got a chance to register for the waitlist were unobserved. Therefore, the time from dialysis initiation to death is left truncated by time from dialysis initiation to waitlist registration.

In this paper, we developed novel estimators for time-dependent ROC analysis with LTRC data, including the Kaplan-Meier-based (KM-based) estimators, regression-based estimators, and two different versions inverse truncation-and-censoring probability weighting (IPW) estimators. The KM-based and regression-based estimators extend the nonparametric and semiparametric estimators in Li 2016, respectively, and are based on direct estimation of the (conditional) survival function of time-to-events. The semiparametric estimators in Li 2016 are limited to settings where right censoring occurs only after left truncation and does not accommodate general covariate-dependent truncation, which our regression-based approach is able to directly address.  The IPW approach weighted the observations by their inverse probabilities of being uncensored and included in the LTRC sample. In the first version of IPW estimators, only uncensored observations were included in the calculation of the specificity and AUC estimators, while in the second version, all observations at risk at the index time were included.  While inspired by the general weighting approach of Vakulenko-Lagun et al.~\citep{vakulenko2022nonparametric} and the IPW concordance index estimators of Hartman et al.~\citep{hartman2023concordance}, our proposed IPW estimators are the first to provide consistent estimation of time-dependent sensitivity, specificity, and AUC in general LTRC settings. Our methods thus address key limitations of prior work and provide a unified framework for ROC analysis in complex survival data.

The rest of the paper is organized as follows. We introduce the notation and parameters of interest in a time-dependent ROC analysis in Section~\ref{sec:concepts}, as well as different scenarios relating the censoring time to the left truncation times that we shall consider. In Section~\ref{sec:new-estimators}, we first propose the KM-based and regression-based estimators as extensions of those in Li 2016 in Sections~\ref{sec:km-based} and \ref{sec:reg-based}, and then introduce the two versions of novel IPW estimators of sensitivity, specificity, and AUC in Section~\ref{sec:ipw}. We perform comprehensive simulation studies to evaluate and compare the performance of the AUC estimators in Section~\ref{sec:simu}, and demonstrate the second version of the IPW estimators as the superior estimators. In Section~\ref{sec:optn}, we demonstrate the proposed ROC analyses by evaluating a mortality prediction model in elderly ESRD patients using the OPTN registry data. We conclude the paper with a brief discussion in Section~\ref{sec:discussion}. Proofs, additional results and discussions are deterred to the Appendices.

\section{Notation and Basic Concepts}\label{sec:concepts}

In the population of interest, we denote $T$ as time to event and $X$ as a real-valued marker for predicting the event occurrence. For an arbitrary cutoff $x_0$ of the marker and time point $t_0$, the cumulative sensitivity and dynamic specificity (using the terminology of Heagerty and Zheng 2005~\citep{heagerty2000time}), or simply sensitivity and specificity, are defined respectively as 
\begin{align}\text{Se}(x_0,t_0) &= P(X_i > x_0 \mid T_i \leq t_0),\label{eq:se}\\
\text{Sp}(x_0,t_0) &= P(X_i \leq x_0 \mid T_i > t_0).\label{eq:sp}\end{align}

The time-dependent ROC curve for time $t_0$, denoted as $\text{ROC}(t_0)$, plots $\Se(x_0,t_0)$ against $1-\Sp(x_0,t_0)$ for all possible values of $x_0$. The area under the ROC curve at time $t_0$, denoted as $\AUC(t_0)$, can be calculated as
\begin{align}\AUC(t_0)&=\int_{\infty}^{-\infty} \Se(x,t_0)\dfrac{\partial}{\partial x}[1 -\Sp(x,t_0)]\,\mathrm dx\label{eq:auc-def}\end{align}

The AUC satisfies that $\text{AUC}(t_0) = P(X_i > X_j \mid T_i \leq t_0, T_j > t_0)$, where $(X_i,T_i)$ and  $(X_j, T_j)$ are two independent marker-time pairs. With discrete $X$, it is common to define in practice that
    \begin{align*}\text{AUC}(t_0) &= P(X_i > X_j \mid T_i \leq t_0, T_j > t_0) +  0.5 * P(X_i = X_j \mid T_i \leq t_0, T_j > t_0)\end{align*}

We denote  $C$ as the right censoring time and $L$ as the left truncation time (study entry time). In the observed data, only those with $L<\min(T,C)$ were included. Due to the right censoring, only the minimum of $T$ and $C$ is observed for each observation. We denote $\widetilde T=\min(T,C)$ as the censored event time and $\Delta=\mathbbm 1(T\leq C)$ as the event indicator, where $\mathbbm 1(\cdot)$ is an indicator function. We say a subject is at risk at time $t_0$ if they have entered the study, haven't experienced an event, and haven't been censored before $t_0$. That is, they satisfy  $L<t_0$ and $\widetilde T_0\geq t_0$.

We let $Z$ be a vector of baseline covariates that may be associated with event times, censoring times, and/or truncation times. The marker $X$ can be written as a function of $Z$, that is, $Z$ determines $X$. For instance, $X$ can be a biomarker that predicts the outcome risks (and thus an element of $Z$) or a score from a prediction model based on the covariates $Z$. We shall write $X=\xi(Z)$. 

We assume that the underlying (untruncated, uncensored) population data consist of random samples drawn from the distribution of $(L, T, C, Z)$. The observed data contains a random sample of size $n$ drawn from the distribution of $(L, \widetilde T=\min(T,C), \Delta=\mathbbm 1(T<C),Z)$ conditioning on $L<\widetilde T$. We denote the data for observation $i$ as $(L_i, \widetilde T_i, \Delta_i,Z_i)$. 

We will use $S_T$ to denote the survival function (i.e. $S_T(t)=P(T>t)$) and $F_T$ the distribution function of $T$ (i.e. $F_T(t)=P(T\leq t)$), respectively. We will use $F_{T,X}$ to denote the joint distribution function of $(T,X)$ (i.e. $F_{T,X}(t,x)=P(T\leq t, X\leq x)$). We will use $S_{T\mid Z}$ and $F_{T\mid Z}$ for the conditional survival and distribution functions of $T$ conditioning on specific values of the covariates $Z$. With slight abuse of notation, we will use $\widehat S_T$, $\widehat F_T$, $\widehat F_{T,X}$, $\widehat S_{T\mid Z}$ and $\widehat F_{T\mid Z}$ to denote their estimators. Marginal or conditional survival, distribution, and hazard functions for other random variables in the paper will be denoted in the same manner.

\begin{figure}[!htbp]
    \centering
    \includegraphics[width=0.66\textwidth]{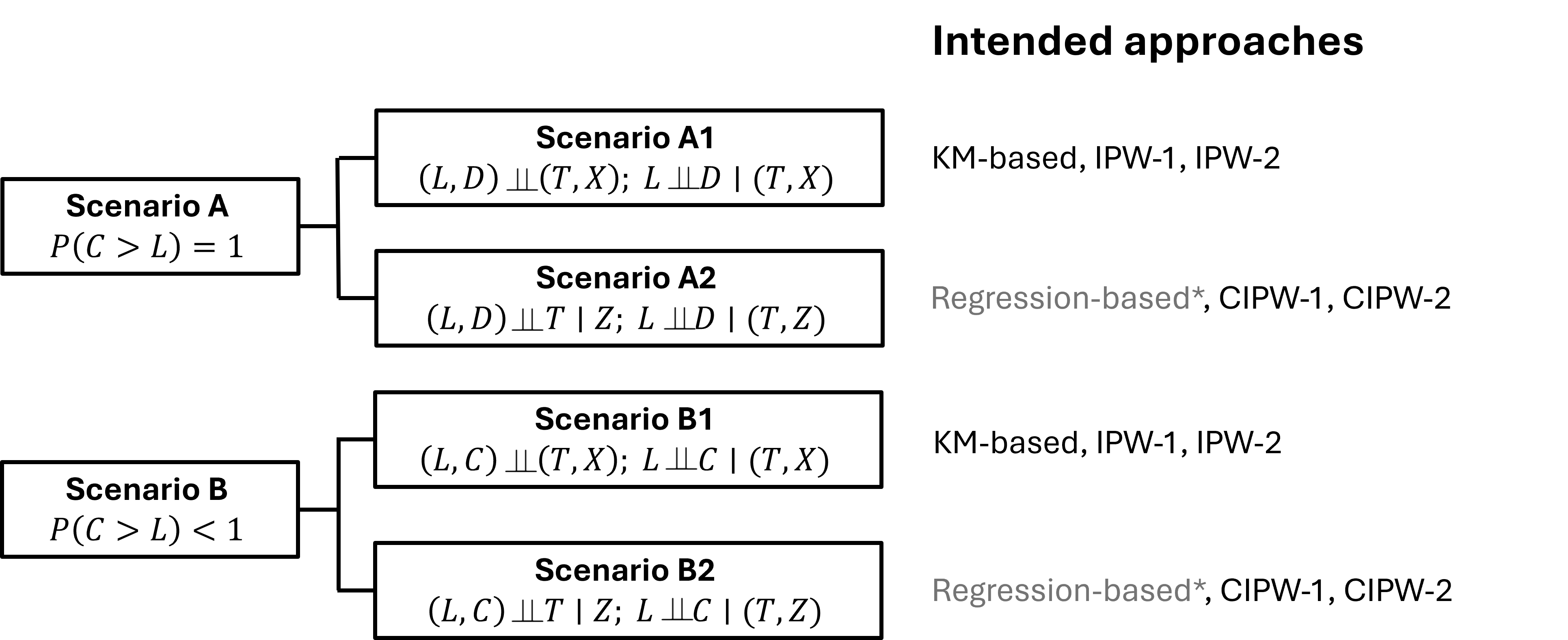}
    \caption{Four scenarios regarding the relations between event times, left truncation times, right censoring times, and covariates under consideration in this manuscript, as well as their intended approaches by our proposal. The regression-based approach was intended for Scenarios A2 and B2 but was found to be biased with moderate sample sizes in our simulation in Section~\ref{sec:simu}.}
    \label{fig:scenario}
\end{figure}

Figure~\ref{fig:scenario} illustrates various scenarios under consideration in later sections and the intended approaches under those scenarios. We consider two different scenarios for the relationship between right censoring and left truncation times, as originally
discussed in Qian and Betensky 2014\cite{qian2014assumptions}. We further divide each scenario into sub-scenarios based on whether the event times independent from tjeleft truncation and right censoring times.

\begin{itemize}
    \item[] \textbf{Scenario A:} $\Pr(C>L)=1$, that is, right censoring always occurs after study entry. This may be the case when the right censoring is due to loss of follow-up after study entry. In this case, we denote $D=C-L$ as the residual censoring time. 
    \begin{itemize}
        \item[] \textbf{Scenario A1} (marginal independent truncation/censoring): $(L,D)\indep (T,X)$ and $D\indep L\mid (T,X)$, that is, the left truncation time, residual censoring time, and the $(T,X)$-pair are mutually independent;
        \item[] \textbf{Scenario A2} (covariante-induced depent truncation/censoring): $(L,D)\indep T\mid Z$ and $D\indep L\mid T,Z$, that is, the left truncation time, residual censoring time, and event time are possibly associated, but the association can be entirely accounted for by conditioning on $Z$.
    \end{itemize}
\item[] \textbf{Scenario B}: $P(C>L)<1$, that is, right censoring or loss of contact may occur before study entry. This scenario sometimes occurs in a cross-sectional sample when censoring is caused by competing risks (e.g. death due to other reasons).~\citep{qian2014assumptions} In Section~\ref{sec:cprisk} and Appendix~\ref{append:cprisk}, we instead consider extensions of the ROC analysis with competing risks directly. Similarly, we consider two sets of independence assumptions below:
    \begin{itemize}
        \item[] \textbf{Scenario B1} (marginal independent truncation/censoring): $(L,C)\indep (T,X)$ and $L\indep C\mid (T,X)$;
        \item[] \textbf{Scenario B2} (covariante-induced depent truncation/censoring): $(L,C)\indep T\mid Z$ and $L\indep C\mid T,Z$.
    \end{itemize}
\end{itemize}

As underscored in Section~\ref{sec:intro}, Li 2016 only focused on Scenario A1 and a special case of Scenario A2 where $Z=X$. Therefore, to our knowledge, this paper is the first work to develop suitable time-dependent ROC analysis for LTRC data for each of the four scenarios above in a unified framework. Hereafter, we assume that one of the four scenarios can be determined before the analysis based on the study design. 

Denote $\tau_L= \liminf\{l:F_L(l)>0\} $, i.e. the lower limit of time where $L$ still has a positive density. Throughout the manuscript, we present results based on the implicit assumption that $\tau_L=0$, namely it is possible for subjects to enter the study immediately after the time origin.  When $\tau_L>0$, we only observe data with $\widetilde T>\tau_L$. The proposed ROC analysis should be understood as describing the discrimination of a biomarker or a risk score in the region $\widetilde T>\tau_L$.~\citep{woodroofe1985estimating}

\section{Novel Estimators for Time-dependent ROC analysis under LTRC data}\label{sec:new-estimators}

\subsection{KM-based estimators}\label{sec:km-based}

We first consider Scenarios A1 and B1. From Equations~\eqref{eq:se} and \eqref{eq:sp} we immediately have the following result for $\Se(x_0,t_0)$ and $\Sp(x_0,t_0)$ regardless the censoring and truncation distribution.
\begin{proposition}[KM-based identification]\label{res:km-based} The time-dependent cumulative sensitivity, $\Se(x_0,t_0)$, and dynamic specificity, $\Sp(x_0,t_0)$, satisfy: 
\begin{equation}\label{eq:km-based}
    \begin{aligned}
    \Se(x_0,t_0) &= 
   \dfrac{1 - S_T(t_0) - F_{T,X}(t_0,x_0)}{1-S_T(t_0)},\\
    \Sp(x_0,t_0) &= 
    \dfrac{F_X(x_0) - F_{T,X}(t_0,x_0)}{S_T(t_0)}.\\
\end{aligned}
\end{equation}
\end{proposition}

From Proposition~\ref{res:km-based}, under Scenarios A1 and B1 of independent left-truncation and right censoring, we propose the following plug-in KM-based estimators for $\Se(x_0,t_0)$ and $\Sp(x_0,t_0)$:
\begin{align}
    \widehat{\Se}_{\text{KM}}(x_0,t_0) &= 
    \dfrac{1 - \widehat S_T(t_0) - \widehat{F}_{T,X}(t_0,x_0)}{1-\widehat S_T(t_0)},\label{eq:se-reg-np-form}\\
    \widehat{\Sp}_{\text{KM}}(x_0,t_0) &= 
    \dfrac{\widehat F_X(x_0) - \widehat F_{T,X}(t_0,x_0)}{\widehat S_T(t_0)}.\label{eq:sp-reg-np-form}
\end{align}

Our proposed KM-based estimator for $\Se(x_0,t_0)$ is exactly the nonparametric estimator proposed in Li 2016, whereas $\widehat{\Sp}_{\text{KM}}(x_0,t_0)$ has not been considered before. Under Scenarios A1 and B1,  we may obtain an unbiased estimator of $S_T(t)$ using the Kaplan-Meier estimator  risk set adjusted for left truncation~\cite{kaplan1958nonparametric,lynden1971method}. Following Li 2016, we estimate $F_{T,X}(t_0,x_0)$ by
\begin{equation}\label{eq:F_TX}
    \widehat F_{T,X}(t_0,x_0)=\dfrac{1}{n}\sum_{i=1}^n\dfrac{\mathbbm 1(\widetilde T_i\leq t_0, X_i\leq x_0, \Delta_i=1)\widehat S_T(\widetilde T_i-)}{\widehat R(\widetilde T_i)},
\end{equation}
where \begin{equation}\widehat R(s)=\dfrac{1}{n}\sum_{i=1}^n\mathbbm 1(L_i< s\leq \widetilde T_i)\label{eq:Rhat}\end{equation}
is the estimated proportion of observations at risk at time $t$ and $T-$ indicates the time point right before $T$. Although Li derived $\widehat F_{T,X}$ under Scenario A1, in Appendix \ref{append:est-nuisance} we will show that $\widehat F_{T,X}$ is also  a consistent estimator under Scenario B1. In Equation~\eqref{eq:F_TX}, letting $t=\infty$, we obtain the estimator
\begin{align}
        \widehat F_{X}(x_0)=\dfrac{1}{n}\sum_{i=1}^n\dfrac{\mathbbm 1( X_i\leq x_0, \Delta_i=1)\widehat S_T(\widetilde T_i-)}{\widehat R(\widetilde T_i)}.
\end{align}

Therefore, plugging in the estimators of $\widehat S_T(t_0)$, $\widehat F_{T,X}(t_0,x_0)$, and $\widehat F_X(c)$ in Equations~\eqref{eq:se-reg-np-form} and \eqref{eq:sp-reg-np-form}, we have
\begin{align}
    \widehat{\Se}_{\text{KM}}(x_0,t_0) &= 
    1 -  \dfrac{1}{n}\sum_{i=1}^n\dfrac{\mathbbm 1(\widetilde T_i\leq t_0, X_i\leq x_0, \Delta_i=1)\widehat S_T(\widetilde T_i-)}{\widehat R(\widetilde T_i)\{1 - \widehat S_T(t_0)\}},\label{eq:se-km}\\
    \widehat{\Sp}_{\text{KM}}(x_0,t_0) &= 
    \dfrac{1}{n}\sum_{i=1}^n\dfrac{\mathbbm 1(\widetilde T_i> t_0, X_i\leq x_0, \Delta_i=1)\widehat S_T(\widetilde T_i-)}{\widehat R(\widetilde T_i)\widehat S_T(t_0)}.\label{eq:sp-km}
\end{align}

An estimator of $\AUC(t)$ may be obtained as

\begin{align}
    \widehat{\AUC}_{\text{KM}}(t_0)&=\int \widehat{\Se}_{\text{KM}}(x_0,t_0)\,\mathrm d\{1 - \widehat{\Sp}_{\text{KM}}(x_0,t_0)\}\nonumber\\
    &= 1 - \dfrac{1}{n^2}\sum_{i=1}^n \sum_{j=1}^n \dfrac{\Delta_i\Delta_j\mathbbm 1(X_i\leq X_j, \widetilde T_i\leq t_0, \widetilde T_j>t_0)\widehat S_T(\widetilde T_i-)\widehat S_T(\widetilde T_j-)}{\widehat R(\widetilde T_i)\widehat R(\widetilde T_j)}/\left[\widehat S_T(t_0)\{1 - \widehat S_T(t_0)\}\right].\label{eq:auc-km}
\end{align}

One well-known issue of estimating (conditional) survival functions under LTRC is the instability of the common estimators, such as the Kaplan-Meier estimator, when events occur at times with few at-risk observations. Instability due to early events further propagates to and causes bias at all later time points. This issue may affect not only the KM-based estimators, which explicitly use the Kaplan-Meier estimator $\widehat S_T$, but also the regression-based estimators below since the estimators of conditional survival functions of $T$ (and $C$ in Scenario B2) are needed. In Section~\ref{sec:ipw}, estimators of the (conditional) survival functions of $T$, $C$ and $L$ are used in the calculation of weights and may subject to the same instability if events, censoring, or study entry occur at time points with few at-risk observations. To mitigate such instability, one may use the approach of Lai and Ying in practice, which ignores the factors in the product-limit estimator for events at times with an overly small risk set, defined as number at risk less than $\epsilon n^\gamma$, where $n$ is the sample size, $0<\epsilon\leq 1$ and $0<\gamma<1$.~\citep{lai1991estimating} In our simulation in Section~\ref{sec:simu} and data analysis in Section~\ref{sec:optn}, we set $\epsilon=1$ and $\gamma=0.25$. 

\subsection{Regression-based estimators}\label{sec:reg-based}

We proceed to consider ovariate-induced dependent left truncation and right censoring. Due to the fact that $X=\chi(Z)$, it is straightforward to derive the following expressions of sensitivity, specificity, and AUC: 

\begin{proposition}[Regressioned-based identification]\label{res:reg-based} Under Scenario A2 or B2, we have

\begin{equation}\label{eq:reg-based}
    \begin{aligned}
   \Se(x_0,t_0) &= \dfrac{\int \mathbbm 1(\chi(z)>x_0) \{1 - S_{T\mid Z}(t\mid z)\}\,\mathrm d F_Z(z)}{\int \{1 - S_{T\mid Z}(t\mid z)\}\,\mathrm d F_Z(z)},\\
    \Sp(x_0,t_0) &= 
   \dfrac{\int \mathbbm 1(\chi(z)\leq x_0)S_{T\mid Z}(t\mid z)\,\mathrm d F_Z(z)}{\int  S_{T\mid Z}(t\mid z)\,\mathrm d F_Z(z)},\\
   \AUC(t_0) &=  \dfrac{\int\int \mathbbm 1(\chi(z_i)> \chi(z_j))\{1 - S_{T\mid Z}(t_0\mid z_i)\}S_{T\mid Z}(t_0\mid z_j)\,\mathrm d F_Z(z_i)\mathrm d F_Z(z_j)}{\int\int \{1 - S_{T\mid Z}(t_0\mid z_i)\}S_{T\mid Z}(t_0\mid z_j)\,\mathrm d F_Z(z_i)\mathrm d F_Z(z_j)}.
\end{aligned}
\end{equation}

\end{proposition}

The above ROC parameters can be estimated by replacing $S_{T\mid Z}$ and $dF_Z(z)$ with their corresponding estimators. The estimator $\widehat S_{T\mid Z}$ for the conditional survival function $S_{T\mid Z}$ may be derived a  regression model for time-to-event outcomes, such as Cox proportional hazards regression or the accelerated failure time model.~\cite{cox1972regression,wei1992accelerated} In Appendix \ref{append:est-nuisance}, under Scenarios A2 and B2,  we show that $dF_Z(z)$ -- the density function of $Z$ with respect to a dominating measure -- may be estimated as $$d\widehat F_Z(z) = \int_{l\in(0,\infty)} \dfrac{\{\widehat S_{T\mid Z}(l\mid z)\}^{-1}\widehat  F_{L,Z}(dl,dz\mid L<T)}{\sum_{i=1}^n \{\widehat S_{T\mid Z}(L_i\mid Z_i)\}^{-1}}$$
and

$$d\widehat F_Z(z) = \int_{l\in(0,\infty)} \dfrac{\{\widehat S_{T\mid Z}(l\mid z)\widehat S_{C\mid Z}(l\mid z)\}^{-1}\widehat  F_{L,Z}(dl,dz\mid L<\widetilde T)}{\sum_{i=1}^n \{\widehat S_{T\mid Z}(L_i\mid Z_i)\widehat S_{C\mid Z}(L_i\mid Z_i)\}^{-1}}$$
respectively, where $\widehat F_{L,Z}(\cdot,\cdot\mid L<\widetilde T)$ is the empirical joint distribution function of $(L,Z)$ in the LTRC data. Plugging in the estimators $\widehat S_{T\mid Z}$ and $d\widehat F_Z$ in Equation~\eqref{eq:reg-based},  we obtain the following regression-based estimators

\begin{align}
    \widehat{\Se}_{\text{REG}}(x_0,t_0) &= \dfrac{\sum_{i=1}^n \mathbbm 1(X_i> x_0)\{1 - \widehat S_{T\mid Z}(t_0\mid Z_i)\}/\widehat H(L_i,Z_i)}{\sum_{i=1}^n \{1 - \widehat S_{T\mid Z}(t_0\mid Z_i)\}/\widehat H(L_i,Z_i)},\label{eq:se-reg}\\
    \widehat{\Sp}_{\text{REG}}(x_0,t_0) &= \dfrac{\sum_{i=1}^n \mathbbm 1(X_i\leq x_0)\widehat S_{T\mid Z}(t_0\mid Z_i)/\widehat H(L_i,Z_i)}{\sum_{i=1}^n \widehat S_{T\mid Z}(t_0\mid Z_i)/\widehat H(L_i,Z_i)},\label{eq:sp-reg}\\
    \widehat{\AUC}_{\text{REG}}(t_0) &= \dfrac{\sum_{i=1}^n \mathbbm 1(X_i\geq X_j)\{1 - \widehat S_{T\mid Z}(t_0\mid Z_i)\}\widehat S_{T\mid Z}(t_0\mid Z_j)/\left\{\widehat H(L_i,Z_i)\widehat H(L_j,Z_j)\right\}}{\sum_{i=1}^n \{1 - \widehat S_{T\mid Z}(t_0\mid Z_i)\}\widehat S_{T\mid Z}(t_0\mid Z_j)/\left\{\widehat H(L_i,Z_i)\widehat H(L_j,Z_j)\right\}},\label{eq:auc-reg}
\end{align}
where we set $\widehat H(u,z) = \widehat S_{T\mid Z}(u\mid z)$ under Scenario A2, and $\widehat H(u,z) = \widehat S_{T\mid Z}(u\mid z)\widehat S_{C\mid Z}(u\mid z)$ under Scenario B2. Note that if $Z_i$ contains only the marker $X_i$, these estimators would be identical to the semiparametric estimators in Li 2016 under Scenario A2.

The conditional survival function estimators may be subject to the same bias due to early failures as the Kaplan-Meier estimators. To mitigate this bias, a similar approach to Lai and Ying's proposal may be adopted by ignoring the events that occur at times with an overly small risk set in the Breslow estimator of the baseline cumulative hazard function. 

Proof of consistency and asymptotic normality of the above estimators directly follows Theorem 1 in Li 2016~\cite{li2017estimating}. However, these appealing properties rely on the consistency and asymptotic normality of $\widehat S_{T\mid Z}$ and, under Scenario B2, both $\widehat S_{T\mid Z}$ and  $\widehat S_{C\mid Z}$.  When an incorrect model for the conditional distribution of $T\mid Z$ is assumed,  the regression-based estimators may be biased. In addition, our simulation in Section~\ref{sec:reg-based} revealed unfavorable operating characteristics in both the semiparametric estimators of Li 2016 and the regression-based estimators, including notable final-sample bias and uncalibrated bootstrap confidence intervals.

\subsection{Inverse probability weighting (IPW) estimators}\label{sec:ipw}

In this section, we develop IPW estimators for time-dependent sensitivity, specificity, and AUC that don't rely on estimating $S_{T\mid Z}$. 

\subsubsection{Estimation under independent left truncation and right censoring}\label{sec:ind-ipw}

We first consider estimation of the numerator and denominator of $\Se(x_0,t_0)$ in Equation~\eqref{eq:se}. For a fixed time $t$, the true time to event $T$ would be known to be no larger than $t$ if only an uncensored event occurs at or before $t$. Therefore, for the estimation, we can only include the uncensored observations. We then adjust for the selection of untruncated and uncensored observations via inverse probability weighting. More formally, we have the following result:

\begin{proposition}[IPW Identification I]\label{res:ipw} 
Under Scenarios A1, we have
\begin{equation}\label{eq:ipw-ident-1}
    \begin{aligned}
   \Se(x_0,t_0) &= \dfrac{E\left\{\dfrac{\Delta}{K(\widetilde T)}\mathbbm 1(X> x_0, \widetilde T \leq t_0)\mid L < \widetilde T\right\}}{E\left\{\dfrac{\Delta}{K(\widetilde T)}\mathbbm 1(\widetilde T \leq t_0)\mid L < \widetilde T\right\}},\\
    \Sp(x_0,t_0) &= 
   \dfrac{E\left\{\dfrac{\Delta}{ K(\widetilde T)}\mathbbm 1(X\leq x_0, \widetilde T > t_0)\mid L < \widetilde T\right\}}{E\left\{\dfrac{\Delta}{K(\widetilde T)}\mathbbm 1(\widetilde T > t_0)\mid L < \widetilde T\right\}},\\
    {\AUC}(t_0)&=\dfrac{ E\left\{\dfrac{\Delta_i\Delta_j}{ K(\widetilde T_i)K(\widetilde T_j)}\mathbbm 1(X_i> X_j, \widetilde T_i\leq t_0, \widetilde T_j>t_0)\mid L_i < \widetilde T_i, L_j < \widetilde T_j \right\}}{ E\left\{ \dfrac{\Delta_i\Delta_j}{K(\widetilde T_i) K(\widetilde T_j)}\mathbbm 1(\widetilde T_i\leq t_0, \widetilde T_j>t_0) \mid L_i < \widetilde T_i, L_j < \widetilde T_j\right\}}
\end{aligned}
\end{equation}
where the weight function $K(u) = \Pr(L < u, C > u)$ and the subscripts $i,j$ indicate two independent observations. 
\end{proposition}

Proposition~\ref{res:ipw} motivates the following IPW estimators:
\begin{align}
    \widehat{\Se}_{\text{IPW}}(x_0,t_0) &= \dfrac{\sum_{i=1}^n\dfrac{\Delta_i}{\widehat K( \widetilde T_i)}\mathbbm 1(X_i> x_0, \widetilde T_i\leq t_0)}{\sum_{i=1}^n\dfrac{\Delta_i}{\widehat K(\widetilde T_i)}\mathbbm 1(\widetilde T_i<t_0)},\label{eq:se-ipw-1}\\
    \widehat{\Sp}_{\text{IPW-1}}(x_0,t_0) &= \dfrac{\sum_{i=1}^n \dfrac{\Delta_i}{\widehat K(\widetilde T_i)} \mathbbm 1(X_i\leq x_0, \widetilde T_i>t) }{\sum_{i=1}^n \dfrac{\Delta_i}{\widehat K(\widetilde T_i)} \mathbbm 1(\widetilde T_i>t)},\label{eq:sp-ipw-1}\\
    \widehat {\AUC}_{\text{IPW-1}}(t_0)&=\dfrac{ \sum_{i=1}^n\sum_{j=1}^n \dfrac{\Delta_i\Delta_j}{\widehat K(\widetilde T_i)\widehat K(\widetilde T_j)}\mathbbm 1(X_i> X_j, \widetilde T_i\leq t_0, \widetilde T_j>t_0) }{ \sum_{i=1}^n\sum_{j=1}^n \dfrac{\Delta_i\Delta_j}{\widehat K(\widetilde T_i)\widehat K( \widetilde T_j)}\mathbbm 1(\widetilde T_i\leq t_0, \widetilde T_j>t_0) }.\label{eq:auc-ipw-1}
\end{align}
where $\widehat K(u)$ is an estimator of $K(u)=\Pr(L < u, C > u)$. Under Scenario A1, we have
$$\Pr(L<u, C>u)=\Pr(L<u, D>u-L)=\int_0^uS_D(u-s)dF_L(s),$$
which suggests the estimator 
\begin{equation*}
    \widehat K(u)=\int_0^u\widehat S_D(u-s)d\hat F_L(s),
\end{equation*}
where $\widehat S_D$ is the Kaplan-Meier estimator for the survival function of the residual censoring time. To estimate $F_L(s)$, under Scenario A1, we use the IPW estimator
$$\widehat F_L(s) =\left(\sum_{i=1}^n \dfrac{\mathbbm 1(L_i\leq s)}{\widehat S_T(L_i)}\right) /\left(\sum_{i=1}^n \dfrac{1}{\widehat S_T(L_i)}\right)$$
proposed by Wang 1991.~\cite{wang1991nonparametric}

Under Scenario B1, we may simply use $\widehat K(u)=\widehat S_C(u)\widehat F_L(v)$, where $\widehat S_C(u)$ is the Kaplan-Meier estimator of $S_C(u)$ with the risk set adjusted for left truncation. In. Appendix \ref{append:est-nuisance}, we show that an  estimator for $F_L(s)$ is
$$\widehat F_L(s) =\left(\sum_{i=1}^n \dfrac{\mathbbm 1(L_i\leq s)}{\widehat S_T(L_i)\widehat S_C(L_i)}\right) /\left(\sum_{i=1}^n \dfrac{1}{\widehat S_T(L_i)\widehat S_C(L_i)}\right).$$

In Equation~\eqref{eq:sp-ipw-1}, the calculation of $\Sp(x_0, t_0)$ only uses uncensored observations of which the event time is greater than $t_0$. An alternative approach identifies $\Sp(x_0, t_0)$ by only using the observations that are at risk at $t_0$, including those which are censored after time $t_0$. If the number of censored observations far exceeds the number at risk at $t_0$, estimation based on the following identification result may produce more precise estimate of $\Sp(x_0, t_0)$ and $\AUC(t_0)$.

\begin{proposition}[IPW Identification II]\label{res:ipw2} 
Under Scenarios A1 and B1, we have
\begin{equation}\label{eq:ipw-ident-2}
    \begin{aligned}
    \Sp(x_0,t_0) &= 
   \dfrac{E\left\{\mathbbm 1(X\leq x_0, L < t_0<\widetilde T)\mid L < \widetilde T\right\}}{E\left\{\mathbbm 1(L< t_0<\widetilde T)\mid L < \widetilde T\right\}}.
\end{aligned}
\end{equation}
\end{proposition}

Proposition~\ref{res:ipw2} motivates alternative estimators $\Sp(x_0,t_0)$ and $\AUC(t_0)$: 
\begin{align}
\widehat{\Sp}_{\text{IPW-2}}(x_0,t_0) &= \dfrac{\sum_{i=1}^n \mathbbm 1(L_i<t_0, X_i\leq x_0, \tilde T_i>t_0) }{\sum_{i=1}^n  \mathbbm 1(L_i<t_0< \tilde T_i)}\label{eq:sp-ipw-2},\\
\widehat {\AUC}_{\text{IPW-2}}(t_0)&=\dfrac{ \sum_{i=1}^n\sum_{j=1}^n \dfrac{\Delta_i}{\widehat K(\widetilde T_i)}\mathbbm 1(X_i> X_j, \widetilde T_i\leq t_0, L_j<t_0<\widetilde T_j) }{ \sum_{i=1}^n\sum_{j=1}^n \dfrac{\Delta_i}{\widehat K(\widetilde T_i)}\mathbbm 1( \widetilde T_i\leq t_0, L_j<t_0<\widetilde T_j) }.\label{eq:auc-ipw-2}
\end{align}

The specificity estimator $\widehat{\Sp}_{\text{IPW-2}}(x_0,t_0)$ is identical to the nonparametric estimator of specificity in Li 2016 (See Appendix~\ref{append:li-review}). In fact, Li's proposed nonparametric estimator of AUC can be written as

\begin{equation}
    \widehat{\AUC}_{\text{LI-np}}(t)=\int \widehat{\Se}_{\text{KM}}(x_0,t_0)\,\mathrm d\{1 - \widehat{\Sp}_{\text{IPW-2}}(x_0,t_0)\}.\label{eq:auc-li-np-alt}
\end{equation}

Proof of consistency and asymptotic normality of the proposed sensitivity, specificity, and AUC estimators directly follows the same arguments from Web Appendix D of Hartman et al. 2023.~\citep{hartman2023concordance} In practice, we propose to obtain the standard error or pointwise confidence intervals of these estimators by nonparametric bootstrap~\cite{efron1994introduction}. We propose to use the same adjustment by Lai and Ying (1991)~\citep{lai1991estimating} as in Section~\ref{sec:km-based} to mitigate the bias of  $\widehat S_T$ and $\widehat S_D$ or $\widehat S_C$ in the inverse probability weights due to early failures. 

\subsubsection{Estimation under covariate-induced dependent left truncation and right censoring}\label{sec:dep-ipw}

Results in Propositions \ref{res:reg-based}-\ref{res:ipw2} can be easily extended to settings with covariate-induced dependent left truncation and right censoring as formalized in Scenarios A2 and B2. We introduce the results in Proposition~\ref{res:cipw-ident} below. 

\begin{proposition}[Conditional IPW identification]\label{res:cipw-ident} Under Scenario A2 or B2, we have
\begin{equation}\label{eq:cipw-ident}
\begin{aligned}
     \Se(x_0,t_0) &= \dfrac{E\left\{\dfrac{\Delta}{K_C( \widetilde T, Z)}\mathbbm 1(X> x_0, \widetilde T\leq t_0)\mid L < \widetilde T\right\}}{E\left\{\dfrac{\Delta}{K_C(\widetilde T, Z)}\mathbbm 1(\widetilde T \leq t_0)\mid L < \widetilde T\right\}},\\
    \Sp(x_0,t_0) &= 
   \dfrac{E\left\{\dfrac{\Delta}{ K_C( \widetilde T,Z)}\mathbbm 1(X\leq x_0, \widetilde T > t_0)\mid L < \widetilde T\right\}}{E\left\{\dfrac{\Delta}{K_C( \widetilde T, Z)}\mathbbm 1(\widetilde T > t_0)\mid L < \widetilde T\right\}}\\
   &=  \dfrac{E\left\{\dfrac{1}{K_C( t_0, Z)}\mathbbm 1(X\leq x_0, L < t_0<\widetilde T)\mid L < \widetilde T\right\}}{E\left\{\dfrac{1}{K_C( t_0, Z)}\mathbbm 1(L< t_0<\widetilde T)\mid L < \widetilde T\right\}},\\
    {\AUC}(t_0)&=\dfrac{ E\left\{\dfrac{\Delta_i\Delta_j}{ K_C( \widetilde T_i, Z_i)K(\widetilde T_j, Z_j)}\mathbbm 1(X_i> X_j, \widetilde T_i\leq t_0, \widetilde T_j>t_0)\mid L_i < \widetilde T_i, L_j < \widetilde T_j \right\}}{E\left\{ \dfrac{\Delta_i\Delta_j}{K_C(\widetilde T_i, Z_i) K_C(\widetilde T_j, Z_j)}\mathbbm 1(\widetilde T_i\leq t_0, \widetilde T_j>t_0) \mid L_i < \widetilde T_i, L_j < \widetilde T_j\right\}}\\
    &= \dfrac{ E\left\{\dfrac{\Delta_i}{ K_C(\widetilde T_i,Z_i)K_C(t_0, Z_j)}\mathbbm 1(X_i> X_j, \widetilde T_i\leq t_0, L_j < t_0 < \widetilde T_j )\mid L_i < \widetilde T_i, L_j < \widetilde T_j \right\}}{ E\left\{ \dfrac{\Delta_i}{K_C(\widetilde T_i, Z_i)K_C(t_0, Z_j) }\mathbbm 1(\widetilde T_i\leq t_0, L_j<t_0< \widetilde T_j \mid L_i < \widetilde T_i, L_j < \widetilde T_j\right\}},
\end{aligned}
\end{equation}
where $$K_C(u,z)=\Pr(L<u, C>u\mid Z=z)=\begin{cases}\int_0^u S_{D\mid Z}(u-s\mid z)dF_{L\mid Z }(s\mid z) & \text{(under Scenario A2)}\\
S_{C\mid Z}(u\mid z)F_{L\mid Z }(u\mid z) & \text{(under Scenario B2)}\end{cases}.$$
\end{proposition}

Following Proposition~\ref{res:cipw-ident}, the conditional inverse probability weighting (CIPW) estimators corresponding to $\widehat{\Se}_{\text{IPW}}(x_0,t_0)$, $\widehat{\Sp}_{\text{IPW-1}}(x_0,t_0)$, $\widehat{\AUC}_{\text{IPW-1}}(t_0)$, $\widehat{\Sp}_{\text{IPW-2}}(x_0,t_0)$, and $\widehat{\AUC}_{\text{IPW-2}}(t_0)$ are:
\begin{align}
    \widehat{\Se}_{\text{CIPW}}(x_0,t_0)&= \dfrac{\sum_{i=1}^n\dfrac{\Delta_i}{\widehat K_{C}(\widetilde T_i,Z_i)}\mathbbm 1(X_i> x_0, \widetilde T_i\leq t_0)}{\sum_{i=1}^n\dfrac{\Delta_i}{\widehat K_{C}(\widetilde T_i,Z_i)}\mathbbm 1(\widetilde T_i<t_0)},\label{eq:se-cipw}\\
     \widehat{\Sp}_{\text{CIPW-1}}(x_0,t_0)&= \dfrac{\sum_{i=1}^n \dfrac{\Delta_i}{\widehat K_{C}( \widetilde T_i, Z_i)} \mathbbm 1(X_i\leq x_0, \widetilde T_i>t_0) }{\sum_{i=1}^n \dfrac{\Delta_i}{\widehat K_C(\widetilde T_i,Z_i)} \mathbbm 1(\widetilde T_i>t_0)}\label{eq:sp-cipw-1}\\
     \widehat {\AUC}_{\text{CIPW-1}}(t_0)&=\dfrac{ \sum_{i=1}^n\sum_{j=1}^n \dfrac{\Delta_i\Delta_j}{\widehat K_{C}(T_i,Z_i)\widehat K_{C}(T_j,Z_j)}\mathbbm 1(X_i> X_j, T_i\leq t_0, T_j>t_0) }{ \sum_{i=1}^n\sum_{j=1}^n \dfrac{\Delta_i\Delta_j}{\widehat K_{C}(T_i,Z_i)\widehat K_{C}(T_j,Z_j)}\mathbbm 1( T_i\leq t_0, T_j>t_0) }\label{eq:auc-cipw-1}\\
     \widehat{\Sp}_{\text{CIPW-2}}(x_0,t_0) &= \dfrac{\sum_{i=1}^n \dfrac{1}{\widehat K_{C}( t_0,Z_i)} \mathbbm 1(L_i<t_0<\widetilde T_i, X_i\leq x_0) }{\sum_{i=1}^n \dfrac{1}{\widehat K_{C}( t_0,Z_i)} \mathbbm 1(L_i<t_0< \tilde T_i)},\label{eq:sp-cipw-2}\\
     \widehat {\AUC}_{\text{CIPW-2}}(t_0)&=\dfrac{ \sum_{i=1}^n\sum_{j=1}^n \dfrac{\Delta_i}{\widehat K_{C}(\widetilde T_i,Z_i)\widehat K_{C}( t_0,Z_j)}\mathbbm 1(X_i> X_j, T_i\leq t_0, L_j<t_0<T_j) }{ \sum_{i=1}^n\sum_{j=1}^n \dfrac{\Delta_i}{\widehat K_{C}(\widetilde T_i,Z_i)\widehat K_{C}(t_0,Z_j)}\mathbbm 1( T_i\leq t_0, L_j<t_0<T_j) }.\label{eq:auc-cipw-2}
\end{align}
where $\widehat K_{C}(u,z)$ is an estimator of $K_C(u,z)$. We propose to use the estimator
\begin{align*}
\widehat K_{C}(u,z)&=\begin{cases}
    \int_0^u\widehat S_{D\mid Z}(u-s\mid z)d\widehat F_{L\mid Z}(s\mid z) &\text{(under Scenario A2)}\\
    S_{C\mid Z}(u\mid z)F_{L\mid Z}(u\mid z) &\text{(under Scenario B2)}
\end{cases},
\end{align*}
where $\widehat S_{D\mid Z}$ (or $\widehat S_{C\mid Z}$) may come from a regression model for $S_{D\mid Z}$ (or $S_{C\mid Z}$). As proposed by Vakulenko-Lagun et al., the estimator $\widehat F_{L\mid Z}(\cdot\mid z)$ may come from a regression model for the survival function of the reverse entry times $\tau-L$ for a large number $\tau$ with the risk set adjusted for the left truncation by the reverse event times $\tau-\widetilde T$. Under Scenario A2, the regression model for $\widehat F_{L\mid Z}(\cdot\mid z)$ uses only uncensored observations, weighted by the inverse probability of censoring $1/\hat S_{D\mid Z}(\widetilde T_i - L_i\mid Z_i)$. In contrast, under Scenario B2, the regression model for $\widehat F_{L\mid Z}(\cdot\mid z)$ uses all observations and no inverse censoring weights are needed since the right censoring is assumed to be the conditionally independent.\cite{vakulenko2022nonparametric}

\section{Time-Dependent ROC with competing risks}\label{sec:cprisk}

The competing risks setting frequently occurs in clinical and public health research when evaluating discrimination of a prognostic model or a biomarker. In the presence of competing risks, the cases at time $t_0$ are the subjects who experienced by the event of interest by time $t_0$, which defines the cumulative sensitivity in a similar manner as before, whereas non-cases may either refer to event-free subjects, i.e. those who didn't experience any events, or anyone who is not a case (the literal ``non-cases''), i.e. the subjects who didn't experience the event of interest by time $t_0$, leading to two different definitions of specificity and the corresponding AUC in the previous literature.~\cite{blanche2013estimating,saha2010time,zheng2012evaluating} With the former definition, the ROC analysis aims to evaluate the marker's discriminatory performance between subjects who experienced event of interest and event-free subjects; with the latter, the evaluation is between subjects who experienced event of interest and all the other subjects. Our proposed KM-based, regression-based, and IPW estimators for time-dependent ROC analysis in Section~\ref{sec:new-estimators} can be extended to the competing risk setting, with both sets of definitions for sensitivity, specificity, and AUC, under each of the four scenarios described in Section~\ref{sec:concepts}. We leave these extensions to Appendix~\ref{append:cprisk}.

\section{Simulation for the  AUC estimators}\label{sec:simu}

In this section, we perform a simulation study to evaluate the AUC estimators in Sections~\ref{sec:km-based} and \ref{sec:ipw}, with LTRC data under various data-generating mechanisms. 

For all scenarios, independent samples were generated until the LTRC sample size reaches $N=300$ or $600$. We consider a vector of two iid baseline covariates $Z=(Z_1,Z_2)$ where $Z_j\sim \text{Unif}(-1,1)$, $j=1,2$. We consider two scenarios of event time distributions:
\begin{enumerate}[label = {(T\arabic*)}]
    \item $T\mid Z\sim \text{Weibull}(2, 2\exp(-\{Z_1+Z_2/2\}/2)$;
    \item $T\mid Z\sim \text{Weibull}(2, 2\exp(-\{2(Z_1-0.33)^++(Z_1+0.33)^++Z_2/5\}/2)$;
\end{enumerate}
Here $\text{Weibull}(a,b)$ denotes the Weibull distribution with the shape parameter $a$ and scale parameter $b$, and $(a)^+=\max(a,0)$. Note that the proportional hazard assumption about $Z_1$ and $Z_2$ holds in the T1 but not T2. We consider the following scenarios for the distribution of left truncation times:
\begin{enumerate}[label = {(L\arabic*)}]
    \item $L\mid Z\sim \text{Unif}(0,5)$;
    \item $$S_{L\mid Z}(u\mid Z)= 1 - \left(\dfrac{5-u}{5}\right)^{\exp(2Z_1/5+Z_2/10)}$$ if  $u\leq 5$ and $S_{L\mid Z}(u\mid Z)=0$ if $u>5$;
    \item $$S_{L\mid Z}(u\mid Z)= 1 - \left(\dfrac{5-u}{5}\right)^{\exp(2\text{sign}(|Z_1|-0.33)/5+Z_2/5)}$$
    if $u\leq 5$ and $S_{L\mid Z}(u\mid Z)=0$ if $u>5$, where $\text{sign}(a)$ is the sign of $a$. 
\end{enumerate}
Scenario L1 encodes the independent left truncation. The proportional hazard assumption about $Z_1$ and $Z_2$ holds for the reverse entry time in L2 but not L3. The baseline hazard function for both L2 and L3 is that of a $\text{Unif}(0,5)$ distribution. 
Finally, we consider two scenarios of independent right censoring. We do not consider covariate-induced dependent right censoring, as this paper primarily focuses on bias of time-dependent ROC estimators due to left truncation.
\begin{enumerate}[label = {(C\arabic*)}]
    \item $C=L+D$ where $D\sim \text{Unif}(0.5,5)$ so that right censoring always occurs after study entry, corresponding to Scenario A;
    \item $C\sim\text{Weibull}(0.5,10)$ so that the independent right censoring may occur before study entry, corresponding to Scenario B.
\end{enumerate}

We compare the bias, the square root of mean squared error ($\sqrt{\text{MSE}}$), and the coverage rate of the 95\% bootstrap confidence intervals (with 1,000 resamples) over 1,000 repeated simulations for the following 8 AUC estimators:
\begin{enumerate}
    \item AUC estimators developed for right-censored time-to-event data but not for LTRC data, including $\widehat{\AUC}_{\text{RC-IPW}}$ and $\widehat{\AUC}_{\text{RC-CIPW}}$, the IPW estimator by Uno et al. 2011, and the conditional IPW estimator by Blanche et al. 2013 (modified for general covariate-induced dependent censoring);
    \item The KM-based estimator $\widehat{\AUC}_{\text{KM}}$ and regression-based estimator $\widehat{\AUC}_{\text{REG}}$ in Sections~\ref{sec:km-based} and \ref{sec:reg-based};
    \item The proposed two versions of IPW estimators $\widehat {\AUC}_{\text{IPW-1}}$ and $\widehat {\AUC}_{\text{IPW-2}}$, as well as their conditional IPW extensions $\widehat {\AUC}_{\text{CIPW-1}}$ and $\widehat {\AUC}_{\text{CIPW-2}}$, introduced in Section~\ref{sec:ipw}.
\end{enumerate}

In each scenario, Cox proportional hazard regression models adjusting for $Z_1$ and $Z_2$ are used for the estimation of all conditional survival functions. Time-dependent AUC was evaluated at $t=1.0$, $1.6$, and $2.3$, which are roughly the quartiles of the marginal distribution of $T$.

\begin{sidewaystable}[!htbp]
    \centering   
    \caption{\label{tab:sim_common_a1a2_1}Bias, square root of MSE, and coverage rate of 95\% bootstrap intervals of $\widehat \AUC_{\text{RC-IPW}}$, $\widehat \AUC_{\text{RC-CIPW}}$, $\widehat \AUC_{\text{KM}}$, and $\widehat \AUC_{\text{REG}}$ in simulation studies with 1,000 replications under different scenarios of event time and left truncation time distribution within Scenarios A1 or A2 as described in Section~\ref{sec:simu}. The censoring time distribution follows scenario C1. }
\begin{tabular}{|rr|rr|rrr|rrr|rrr|rrr|}
  \hline
\multicolumn{2}{|c|}{Scenario} & \multirow{2}{*}{$t$} & \multirow{2}{*}{$N$}  & \multicolumn{3}{c|}{$\widehat \AUC_{\text{RC-IPW}}$} & \multicolumn{3}{c|}{$\widehat \AUC_{\text{RC-CIPW}}$} & \multicolumn{3}{c|}{$\widehat \AUC_{\text{KM}}$} &\multicolumn{3}{c|}{$\widehat \AUC_{\text{REG}}$} \\ 
\cline{1-2}\cline{5-16}
$T$ & $L$ & &  & Bias &$\sqrt{\text{MSE}}$ & Coverage & Bias &$\sqrt{\text{MSE}}$ & Coverage & Bias &$\sqrt{\text{MSE}}$ & Coverage & Bias &$\sqrt{\text{MSE}}$ & Coverage \\
  \hline
T1 & L1 & 1.0 & 300 & 0.046 & 0.071 & 81.4 & 0.045 & 0.071 & 81.5 & 0.004 & 0.071 & 93.1 & -0.012 & 0.031 & 91.2 \\ 
  T1 & L1 & 1.0 & 600 & 0.044 & 0.058 & 76.5 & 0.044 & 0.057 & 76.5 & 0.002 & 0.044 & 94.0 & -0.012 & 0.023 & 88.7 \\ 
  T1 & L1 & 1.6 & 300 & 0.025 & 0.042 & 86.9 & 0.025 & 0.042 & 87.4 & 0.010 & 0.047 & 94.8 & -0.017 & 0.037 & 90.4 \\ 
  T1 & L1 & 1.6 & 600 & 0.024 & 0.035 & 80.1 & 0.023 & 0.034 & 81.2 & 0.007 & 0.031 & 94.5 & -0.017 & 0.028 & 85.9 \\ 
  T1 & L1 & 2.3 & 300 & 0.001 & 0.030 & 95.6 & 0.000 & 0.030 & 95.6 & 0.016 & 0.046 & 94.8 & -0.024 & 0.044 & 87.6 \\ 
  T1 & L1 & 2.3 & 600 & 0.001 & 0.022 & 92.9 & -0.001 & 0.022 & 92.8 & 0.011 & 0.031 & 94.1 & -0.023 & 0.034 & 82.5 \\ 
  \hline
  T1 & L2 & 1.0 & 300 & -0.032 & 0.071 & 93.8 & -0.032 & 0.071 & 93.7 & -0.075 & 0.105 & 79.8 & -0.019 & 0.034 & 87.8 \\ 
  T1 & L2 & 1.0 & 600 & -0.032 & 0.055 & 89.6 & -0.032 & 0.055 & 89.6 & -0.076 & 0.091 & 64.6 & -0.019 & 0.028 & 82.2 \\ 
  T1 & L2 & 1.6 & 300 & -0.040 & 0.055 & 84.5 & -0.039 & 0.054 & 85.1 & -0.064 & 0.079 & 66.7 & -0.025 & 0.042 & 85.8 \\ 
  T1 & L2 & 1.6 & 600 & -0.040 & 0.049 & 68.6 & -0.039 & 0.048 & 69.6 & -0.066 & 0.074 & 38.6 & -0.025 & 0.035 & 79.2 \\ 
  T1 & L2 & 2.3 & 300 & -0.056 & 0.066 & 58.2 & -0.055 & 0.064 & 59.4 & -0.056 & 0.072 & 67.4 & -0.032 & 0.050 & 84.6 \\ 
  T1 & L2 & 2.3 & 600 & -0.057 & 0.062 & 28.7 & -0.056 & 0.061 & 29.7 & -0.062 & 0.069 & 31.6 & -0.032 & 0.042 & 76.3 \\ 
  \hline
  T1 & L3 & 1.0 & 300 & 0.017 & 0.053 & 90.0 & 0.017 & 0.052 & 90.0 & -0.039 & 0.070 & 87.5 & -0.022 & 0.039 & 84.9 \\ 
  T1 & L3 & 1.0 & 600 & 0.017 & 0.038 & 91.7 & 0.016 & 0.037 & 92.1 & -0.041 & 0.056 & 79.8 & -0.022 & 0.032 & 79.4 \\ 
  T1 & L3 & 1.6 & 300 & 0.011 & 0.034 & 92.7 & 0.008 & 0.034 & 93.5 & -0.022 & 0.047 & 89.1 & -0.028 & 0.047 & 83.7 \\ 
  T1 & L3 & 1.6 & 600 & 0.012 & 0.027 & 90.7 & 0.009 & 0.025 & 92.4 & -0.024 & 0.036 & 82.5 & -0.028 & 0.039 & 76.1 \\ 
  T1 & L3 & 2.3 & 300 & 0.007 & 0.031 & 92.8 & 0.001 & 0.030 & 93.2 & 0.006 & 0.039 & 95.5 & -0.035 & 0.055 & 82.9 \\ 
  T1 & L3 & 2.3 & 600 & 0.007 & 0.023 & 92.1 & 0.001 & 0.021 & 94.3 & 0.002 & 0.027 & 94.5 & -0.035 & 0.046 & 77.0 \\ 
  \hline
  T2 & L1 & 1.0 & 300 & 0.030 & 0.063 & 91.6 & 0.030 & 0.063 & 91.6 & 0.006 & 0.071 & 94.3 & -0.043 & 0.048 & 47.9 \\ 
  T2 & L1 & 1.0 & 600 & 0.029 & 0.048 & 86.7 & 0.029 & 0.048 & 86.7 & 0.003 & 0.045 & 95.0 & -0.042 & 0.045 & 21.0 \\ 
  T2 & L1 & 1.6 & 300 & 0.010 & 0.037 & 94.3 & 0.009 & 0.037 & 94.0 & 0.012 & 0.048 & 94.0 & -0.036 & 0.045 & 71.4 \\ 
  T2 & L1 & 1.6 & 600 & 0.009 & 0.027 & 92.5 & 0.008 & 0.026 & 93.0 & 0.005 & 0.030 & 95.2 & -0.035 & 0.039 & 50.2 \\ 
  T2 & L1 & 2.3 & 300 & -0.009 & 0.034 & 94.1 & -0.010 & 0.034 & 93.9 & 0.022 & 0.050 & 93.5 & -0.021 & 0.040 & 89.9 \\ 
  T2 & L1 & 2.3 & 600 & -0.009 & 0.024 & 92.0 & -0.010 & 0.025 & 91.3 & 0.011 & 0.029 & 94.6 & -0.020 & 0.030 & 86.3 \\ 
  \hline
  T2 & L2 & 1.0 & 300 & -0.063 & 0.089 & 82.2 & -0.063 & 0.088 & 82.3 & -0.093 & 0.121 & 68.8 & -0.053 & 0.058 & 35.6 \\ 
  T2 & L2 & 1.0 & 600 & -0.067 & 0.080 & 66.0 & -0.066 & 0.079 & 66.1 & -0.098 & 0.109 & 45.9 & -0.053 & 0.055 & 8.6 \\ 
  T2 & L2 & 1.6 & 300 & -0.064 & 0.074 & 62.5 & -0.063 & 0.073 & 63.5 & -0.077 & 0.090 & 53.4 & -0.049 & 0.056 & 56.6 \\ 
  T2 & L2 & 1.6 & 600 & -0.064 & 0.069 & 32.7 & -0.063 & 0.068 & 33.7 & -0.082 & 0.087 & 19.1 & -0.049 & 0.052 & 29.6 \\ 
  T2 & L2 & 2.3 & 300 & -0.070 & 0.077 & 45.6 & -0.068 & 0.075 & 47.3 & -0.060 & 0.075 & 63.3 & -0.037 & 0.051 & 78.9 \\ 
  T2 & L2 & 2.3 & 600 & -0.072 & 0.076 & 13.4 & -0.070 & 0.074 & 15.5 & -0.072 & 0.078 & 19.6 & -0.037 & 0.043 & 69.5 \\ 
  \hline
  T2 & L3 & 1.0 & 300 & -0.017 & 0.056 & 93.4 & -0.018 & 0.056 & 93.3 & -0.047 & 0.075 & 84.7 & -0.057 & 0.062 & 34.6 \\ 
  T2 & L3 & 1.0 & 600 & -0.019 & 0.040 & 91.6 & -0.019 & 0.041 & 91.3 & -0.051 & 0.064 & 72.2 & -0.057 & 0.059 & 9.6 \\ 
  T2 & L3 & 1.6 & 300 & -0.020 & 0.039 & 92.5 & -0.022 & 0.040 & 91.9 & -0.030 & 0.049 & 86.6 & -0.054 & 0.061 & 52.6 \\ 
  T2 & L3 & 1.6 & 600 & -0.019 & 0.031 & 88.0 & -0.021 & 0.033 & 85.2 & -0.035 & 0.045 & 70.4 & -0.054 & 0.057 & 25.6 \\ 
  T2 & L3 & 2.3 & 300 & -0.018 & 0.036 & 91.4 & -0.023 & 0.039 & 89.6 & -0.002 & 0.039 & 96.2 & -0.044 & 0.057 & 78.2 \\ 
  T2 & L3 & 2.3 & 600 & -0.019 & 0.031 & 86.8 & -0.025 & 0.034 & 81.7 & -0.014 & 0.030 & 88.5 & -0.044 & 0.051 & 59.4 \\ 
   \hline
\end{tabular}
\end{sidewaystable}

\begin{sidewaystable}[!htbp]
    \centering   
    \caption{\label{tab:sim_common_a1a2_2}Bias, square root of MSE, and coverage rate of 95\% bootstrap intervals of $\widehat \AUC_{\text{IPW-1}}$, $\widehat \AUC_{\text{CIPW-1}}$, $\widehat \AUC_{\text{IPW-2}}$, and $\widehat \AUC_{\text{CIPW-2}}$ in simulation studies with 1,000 replications under different scenarios of event time and left truncation time distribution within Scenarios A1 or A2 as described in Section~\ref{sec:simu}. The censoring time distribution follows scenario C1. }
\begin{tabular}{|rr|rr|rrr|rrr|rrr|rrr|}
  \hline
\multicolumn{2}{|c|}{Scenario} & \multirow{2}{*}{$t$} & \multirow{2}{*}{$N$}  & \multicolumn{3}{c|}{$\widehat \AUC_{\text{IPW-1}}$} & \multicolumn{3}{c|}{$\widehat \AUC_{\text{CIPW-1}}$} & \multicolumn{3}{c|}{$\widehat \AUC_{\text{IPW-2}}$} &\multicolumn{3}{c|}{$\widehat \AUC_{\text{CIPW-2}}$} \\ 
\cline{1-2}\cline{5-16}
$T$ & $L$ & &  & Bias &$\sqrt{\text{MSE}}$ & Coverage & Bias &$\sqrt{\text{MSE}}$ & Coverage & Bias &$\sqrt{\text{MSE}}$ & Coverage & Bias &$\sqrt{\text{MSE}}$ & Coverage \\
  \hline
T1 & L1 & 1.0 & 300 & -0.002 & 0.065 & 93.1 & -0.013 & 0.074 & 94.4 & -0.002 & 0.066 & 93.9 & -0.004 & 0.072 & 94.7 \\ 
  T1 & L1 & 1.0 & 600 & -0.001 & 0.043 & 94.1 & -0.011 & 0.050 & 93.6 & -0.001 & 0.046 & 93.9 & -0.004 & 0.049 & 93.8 \\ 
  T1 & L1 & 1.6 & 300 & 0.001 & 0.041 & 95.0 & -0.009 & 0.050 & 94.9 & 0.001 & 0.043 & 94.7 & -0.002 & 0.049 & 95.6 \\ 
  T1 & L1 & 1.6 & 600 & 0.001 & 0.029 & 94.2 & -0.008 & 0.035 & 93.7 & 0.001 & 0.030 & 94.6 & -0.002 & 0.034 & 94.8 \\ 
  T1 & L1 & 2.3 & 300 & 0.000 & 0.036 & 94.4 & -0.008 & 0.044 & 94.6 & 0.001 & 0.036 & 94.9 & -0.002 & 0.043 & 95.1 \\ 
  T1 & L1 & 2.3 & 600 & 0.002 & 0.025 & 93.8 & -0.007 & 0.032 & 93.5 & 0.002 & 0.025 & 94.4 & -0.002 & 0.030 & 93.9 \\ 
  \hline
  T1 & L2 & 1.0 & 300 & -0.082 & 0.110 & 78.5 & -0.037 & 0.084 & 92.0 & -0.042 & 0.083 & 92.8 & -0.026 & 0.081 & 94.4 \\ 
  T1 & L2 & 1.0 & 600 & -0.080 & 0.094 & 63.0 & -0.023 & 0.058 & 91.0 & -0.041 & 0.065 & 87.9 & -0.014 & 0.055 & 93.5 \\ 
  T1 & L2 & 1.6 & 300 & -0.074 & 0.086 & 62.1 & -0.019 & 0.054 & 93.9 & -0.049 & 0.066 & 81.1 & -0.011 & 0.052 & 95.9 \\ 
  T1 & L2 & 1.6 & 600 & -0.072 & 0.079 & 33.4 & -0.013 & 0.039 & 93.0 & -0.049 & 0.059 & 64.8 & -0.007 & 0.037 & 94.4 \\ 
  T1 & L2 & 2.3 & 300 & -0.073 & 0.082 & 49.3 & -0.012 & 0.047 & 95.1 & -0.057 & 0.068 & 67.0 & -0.004 & 0.046 & 95.8 \\ 
  T1 & L2 & 2.3 & 600 & -0.072 & 0.077 & 20.1 & -0.009 & 0.033 & 93.9 & -0.057 & 0.063 & 38.8 & -0.003 & 0.032 & 95.3 \\ 
  \hline
  T1 & L3 & 1.0 & 300 & -0.044 & 0.072 & 87.1 & -0.055 & 0.085 & 87.7 & -0.061 & 0.090 & 83.6 & -0.056 & 0.091 & 89.8 \\ 
  T1 & L3 & 1.0 & 600 & -0.044 & 0.058 & 78.8 & -0.056 & 0.071 & 76.0 & -0.061 & 0.076 & 71.3 & -0.059 & 0.077 & 79.3 \\ 
  T1 & L3 & 1.6 & 300 & -0.030 & 0.050 & 87.6 & -0.048 & 0.065 & 82.9 & -0.050 & 0.066 & 78.8 & -0.053 & 0.071 & 82.7 \\ 
  T1 & L3 & 1.6 & 600 & -0.028 & 0.039 & 81.5 & -0.046 & 0.055 & 65.8 & -0.048 & 0.057 & 62.3 & -0.052 & 0.061 & 65.0 \\ 
  T1 & L3 & 2.3 & 300 & -0.008 & 0.035 & 95.1 & -0.031 & 0.050 & 87.3 & -0.025 & 0.044 & 89.8 & -0.036 & 0.054 & 87.3 \\ 
  T1 & L3 & 2.3 & 600 & -0.006 & 0.025 & 93.4 & -0.029 & 0.040 & 80.8 & -0.024 & 0.035 & 84.1 & -0.035 & 0.045 & 76.1 \\ 
  \hline
  T2 & L1 & 1.0 & 300 & -0.001 & 0.065 & 93.9 & -0.005 & 0.072 & 95.0 & -0.001 & 0.066 & 93.5 & -0.002 & 0.073 & 95.2 \\ 
  T2 & L1 & 1.0 & 600 & -0.001 & 0.045 & 94.9 & -0.006 & 0.052 & 94.4 & -0.001 & 0.047 & 93.9 & -0.003 & 0.052 & 93.9 \\ 
  T2 & L1 & 1.6 & 300 & 0.000 & 0.042 & 94.6 & -0.003 & 0.050 & 94.9 & 0.000 & 0.043 & 94.6 & -0.001 & 0.051 & 95.9 \\ 
  T2 & L1 & 1.6 & 600 & 0.000 & 0.029 & 95.1 & -0.005 & 0.035 & 95.5 & 0.000 & 0.030 & 94.8 & -0.002 & 0.036 & 95.0 \\ 
  T2 & L1 & 2.3 & 300 & -0.001 & 0.037 & 94.1 & -0.002 & 0.044 & 95.7 & 0.000 & 0.038 & 93.6 & 0.000 & 0.045 & 94.6 \\ 
  T2 & L1 & 2.3 & 600 & 0.000 & 0.026 & 94.4 & -0.004 & 0.032 & 94.4 & 0.000 & 0.026 & 94.4 & -0.002 & 0.032 & 94.6 \\ 
  \hline
  T2 & L2 & 1.0 & 300 & -0.101 & 0.123 & 66.3 & -0.027 & 0.087 & 91.9 & -0.062 & 0.093 & 85.6 & -0.021 & 0.085 & 93.1 \\ 
  T2 & L2 & 1.0 & 600 & -0.103 & 0.113 & 44.1 & -0.018 & 0.060 & 93.0 & -0.064 & 0.080 & 73.2 & -0.014 & 0.059 & 93.7 \\ 
  T2 & L2 & 1.6 & 300 & -0.089 & 0.099 & 44.5 & -0.013 & 0.057 & 94.6 & -0.065 & 0.078 & 68.4 & -0.009 & 0.057 & 94.5 \\ 
  T2 & L2 & 1.6 & 600 & -0.088 & 0.093 & 15.3 & -0.008 & 0.040 & 93.7 & -0.065 & 0.072 & 43.3 & -0.006 & 0.040 & 94.5 \\ 
  T2 & L2 & 2.3 & 300 & -0.082 & 0.091 & 40.7 & -0.007 & 0.050 & 94.8 & -0.067 & 0.077 & 58.4 & -0.005 & 0.050 & 95.8 \\ 
  T2 & L2 & 2.3 & 600 & -0.084 & 0.088 & 10.9 & -0.007 & 0.036 & 95.3 & -0.069 & 0.074 & 27.3 & -0.004 & 0.035 & 95.6 \\ 
  \hline
  T2 & L3 & 1.0 & 300 & -0.054 & 0.079 & 82.5 & -0.069 & 0.093 & 81.9 & -0.075 & 0.100 & 78.3 & -0.079 & 0.106 & 82.8 \\ 
  T2 & L3 & 1.0 & 600 & -0.055 & 0.067 & 69.9 & -0.071 & 0.083 & 63.1 & -0.077 & 0.089 & 60.7 & -0.083 & 0.096 & 61.0 \\ 
  T2 & L3 & 1.6 & 300 & -0.041 & 0.056 & 83.0 & -0.060 & 0.073 & 73.4 & -0.063 & 0.076 & 67.8 & -0.071 & 0.085 & 68.3 \\ 
  T2 & L3 & 1.6 & 600 & -0.040 & 0.048 & 67.0 & -0.059 & 0.067 & 48.3 & -0.062 & 0.069 & 44.4 & -0.072 & 0.079 & 39.7 \\ 
  T2 & L3 & 2.3 & 300 & -0.022 & 0.042 & 91.3 & -0.044 & 0.058 & 83.3 & -0.042 & 0.056 & 82.5 & -0.054 & 0.067 & 78.7 \\ 
  T2 & L3 & 2.3 & 600 & -0.023 & 0.035 & 83.9 & -0.046 & 0.054 & 61.6 & -0.043 & 0.051 & 66.1 & -0.056 & 0.064 & 53.7 \\  
   \hline
\end{tabular}
\end{sidewaystable}

We present the detailed results under Scenarios A1 and A2, where the censoring distribution follows (C1), in Tables~\ref{tab:sim_common_a1a2_1} and \ref{tab:sim_common_a1a2_2}. Detailed simulation setting and additional results are presented in Appendix~\ref{append:sim} due to space limit. Results under scenarios (C1) and (C2) are similar. In all scenarios considered, the estimators overlooking left truncation are severely biased, with a negative bias up to -0.07 or a positive bias up to 0.05 in some cases, highlighting the need to explicitly account for left truncation (Tables~\ref{tab:sim_common_a1a2_1}, \ref{tab:sim_common_a1a2_2}, \ref{tab:sim_common_b1b2_1}, and \ref{tab:sim_common_b1b2_2}).

Under independent truncation (L1), the regression-type nonparametric estimator ($\widehat{\AUC}_{\text{KM}}$) and the proposed IPW estimators ($\widehat \AUC_{\text{IPW-1}}$, $\widehat \AUC_{\text{CIPW-1}}$, $\widehat \AUC_{\text{IPW-2}}$, and $\widehat \AUC_{\text{CIPW-2}}$) are all unbiased with calibrated 95\% bootstrap CIs and indistinguishable  MSEs. Under scenario (L2) with covariate-induced dependent left truncation, the estimators $\widehat \AUC_{\text{IPW-1}}$ and $\widehat \AUC_{\text{IPW-2}}$ are biased while the conditional IPW estimators $\widehat \AUC_{\text{CIPW-1}}$ and $\widehat \AUC_{\text{CIPW-2}}$ still perform well. Finally, under scenario (L3) where $F_{L\mid Z}$ is misspecified, both $\widehat \AUC_{\text{CIPW-1}}$ and $\widehat \AUC_{\text{CIPW-2}}$ are biased with under-covered bootstrap CIs, but the bias of $\widehat \AUC_{\text{CIPW-1}}$ is smaller.

To our surprise, the regression-type semiparametric estimator $\widehat{\AUC}_{\text{REG}}$ is negatively biased in all scenarios with the 95\% bootstrap CI subject to moderate under-coverage, even under scenario (T1) with correctly specific event time models. Under scenario (T2), where the model for $S_{T\mid Z}$ is misspecified, severe negative bias and under-coverage of 95\% bootstrap CI ensue. We also evaluated $\widehat{\AUC}_{\text{REG}}$ under the setting with only one covariate in the models of event time and left-truncation time distributions and the covariate is identical to the marker, in which it is identical to the semiparametric estimator $\widehat{\AUC}_{\text{LI-SP}}$ in Li 2016. Under these scenarios, the semiparametric estimator is unbiased with calibrated 95\% confidence intervals if the model for $S_{T\mid X}$ is correctly specified model but is biased otherwise, more aligned with our expectation~(Appendix~\ref{append:sim}, Tables~\ref{tab:sim_common_a1a2_regonly_v2_0}, \ref{tab:sim_common_a1a2_regonly_v2_1}). Under the simulation settings with two covariates as above, however, the estimator $\widehat{AUC}_{\text{LI-SP}}$ is biased as expected~(Table \ref{tab:sim_common_a1a2_regonly_v1}).

Finally, we compared the estimators under settings with less frequent outcomes, where the event time distributions follow a mixture cure model~(Appendix \ref{append:sim-rare}). Under these settings, most estimators under perform with the exceptions of $\widehat\AUC_{\text{IPW-2}}$ and $\widehat\AUC_{\text{CIPW-2}}$ (Tables~\ref{tab:sim_rare_a1a2_1}, \ref{tab:sim_rare_a1a2_2}, \ref{tab:sim_rare_b1b2_1}, \ref{tab:sim_rare_b1b2_2}): under independent left truncation, both $\widehat\AUC_{\text{IPW-2}}$ and $\widehat\AUC_{\text{CIPW-2}}$ are unbiased and their 95\% confidence intervals have nominal coverage; under covariate-induced dependent left truncation with correctly specific left truncation time model, only $\widehat\AUC_{\text{CIPW-2}}$ remains unbiased. These results highlighted the robustness of $\widehat\AUC_{\text{IPW-2}}$ and $\widehat\AUC_{\text{CIPW-2}}$ with less frequent outcomes.

\section{Data Application: Evaluation of a mortality prediction model in elderly ESRD patients}\label{sec:optn}

We demonstrate the proposed ROC analyses by evaluating a mortality prediction model in elderly ESRD patients using the OPTN data, as described in Section~\ref{sec:intro}. We used the data downloaded on June 29, 2026. The analysis included ESRD patients who (1) were age 60-75 years old at dialysis initiation, (2) joined the wait-list between September 1, 2010 and August 31, 2012, (3) started dialysis within 2 years before waitlist registration, and (4) had complete covariates (n=7,838). To avoid the influence on mortality due to the global COVID-19 pandemic, we employed the administrative censoring on December 1, 2019. We randomly split the study sample into 70\% of the training data (n=5,487) and 30\% of the test data (n=2,351). We developed the mortality prediction model in the training data using Cox proportional hazards regression model with the same predictors selected in Hartman et al. 2023, including age at dialysis, sex, race and ethnicity, serum albumin level, body mass index (BMI), and diabetes. Race and ethnicity was categorized into Non-Hispanic White, Non-Hispanic Black, Hispanic/Latino, and Others or Unknown. Body mass index was categorized into underweight (BMI$<$18.5), healthy weight (18.5$\leq$BMI$<$25),  overweight (25$\leq$BMI$<$30) and obese (BMI$\geq$30). Serum albumin level was categorized into low ($<$3.5 g/dL), normal ($\geq$3.5 g/dL and $\leq$5.5 g/dL), and high ($>$5.5 g/dL). We report the sample characteristics of the training and test data in Table~\ref{tab:optn_summary} and results of the mortality prediction model in Table~\ref{tab:optn_reg}.

We used the linear predictors from the motality prediction model as the marker and evaluated its discriminatory performance of death in the test data at 1-9 years after dialysis initiation. Since we employed administrative censoring, we did not adjust for covariates for the distribution of $C$. In the model for $F_{L\mid Z}$, we adjusted for the baseline demographic covariates including age at dialysis, sex, and race and ethnicity.

\begin{figure}[!htbp]
    \centering
    \includegraphics[width=0.66\linewidth]{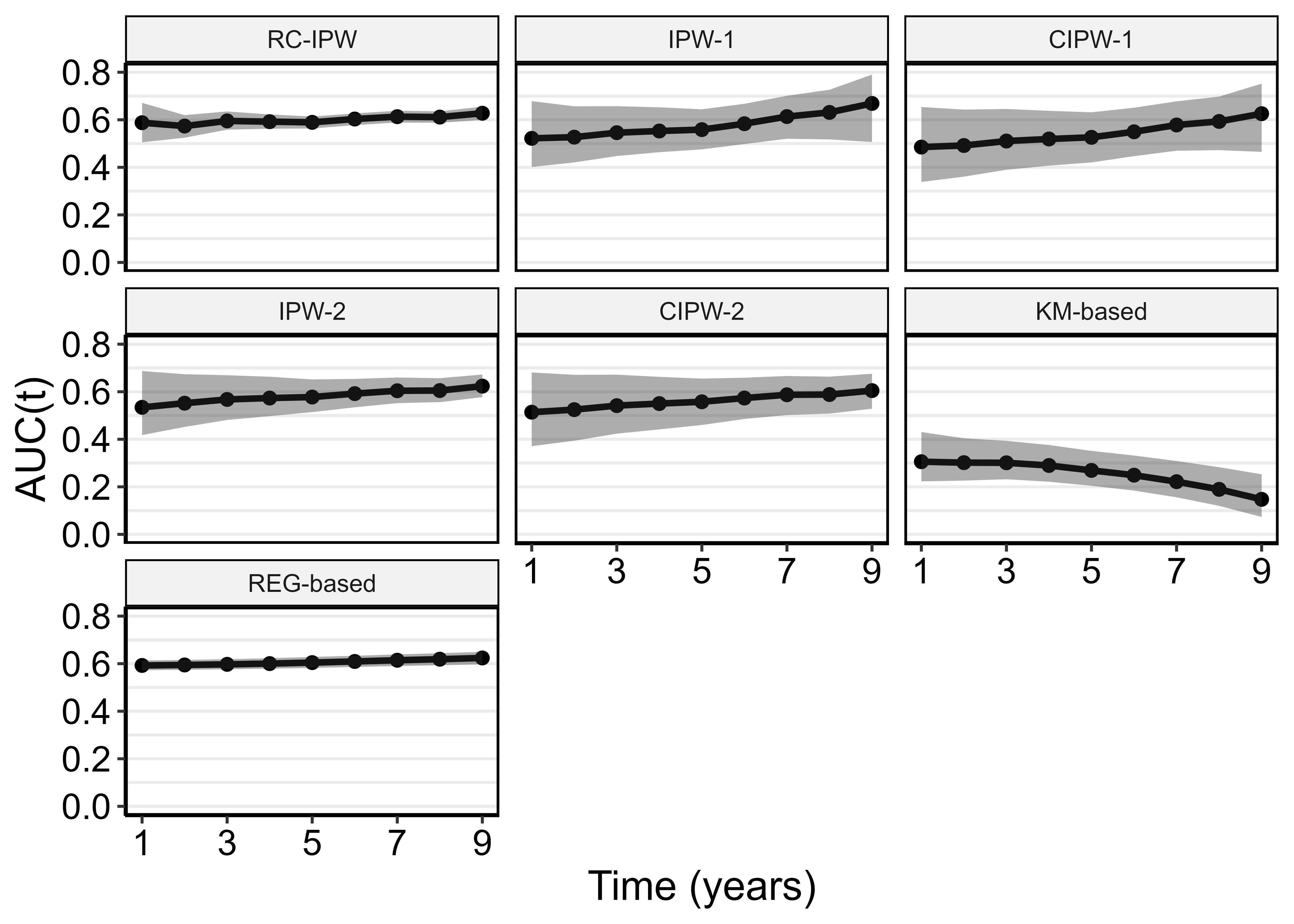}
    \caption{Time-dependent AUC estimators for the descriminatory performances of the mortality prediction model since dialysis initiation in elderly ESRD patients using the OPTN data. RC-IPW: the IPW estimator adjusting for only right censoring by Uno et al.~\citep{uno2007evaluating}; IPW-1: the IPW estimator in Equation \eqref{eq:auc-ipw-1}; CIPW-1: the conditional IPW estimator in Equation \eqref{eq:auc-cipw-1}; IPW-2: the IPW estimator in Equation \eqref{eq:auc-ipw-2}; CIPW-2: the conditional IPW estimator in Equation \eqref{eq:auc-cipw-2}; KM-based: the KM-based estimator in Equation~\eqref{eq:auc-km}; REG-based: the regression-based estimator in Equation~\eqref{eq:auc-reg}.}
    \label{fig:optn_auc}
\end{figure}

Figure~\ref{fig:optn_auc} and Figures~\ref{fig:roc-optn-1}-\ref{fig:roc-optn-9} show the time-dependent AUC and ROC curves, respectively. The KM-based AUC estimates are considerably lower than the rest, suggesting that the left-truncation times and event times are not independent. All other AUC estimators across 1-9 years after dialysis initiation are modest and range from 0.5 to 0.7. These results suggest that baseline covariates cannot adequately predict long-term mortality in elderly ESRD patients.

\section{Discussion}\label{sec:discussion}

In this article, we developed and evaluated cumulative/dynamic time-dependent ROC analysis using a regression-based or IPW-based approach for the LTRC data. The KM-based and regression-based estimators are closely connected to Li 2016\cite{li2017estimating}, although we have extend their methods in several directions: (1) we considered both the setting where right censoring can only occur after study entry and the setting where right censoring may occur before study entry; (2) the regression-based estimators can account for association between the left truncation times and event times induced by covariates other than the marker of interest. Our proposed ROC analysis methods apply to a wide range of settings where we are interested in evaluating a risk prediction model or a biomarker using time-to-event data, but delayed study entry in the validation data may present an issue. Among several proposed estimators, we have also identified a particular version of IPW estimators (IPW-2 and CIPW-2) as more robust ones under less frequent outcomes.

Our simulation studies revealed ubiquitous biases of the regression-based AUC estimators in settings with more than one covariate. This does not conflict with the results in Li 2016, as their regression-based semiparametric AUC estimator is proposed for settings with one covariate. We hypothesize that the biases may result from the difficulty of estimating a multivariate distribution function for the covariates compared with a univariate one.

From our data analysis of the OPTN data, we concluded that the predictive performances of baseline covariates on long-term mortality are at best modest. Our analysis has two limitations. First, time-varying covariates including serum albumin levels and BMI were measured at waitlist registration, instead of the time origin, which may lead to overestimation of the predictive performances.  Second, the OPTN data only included patients who had registered for organ transplantation, but not all ESRD patients were suitable for kidney transplant, which limits the generalizability of our conclusions. 

The proposed conditional IPW estimators are general. In this paper, to estimate the nuisance functions including $F_{L\mid Z}$ and $S_{D\mid Z}$ or $S_{C\mid Z}$,  we have only used Cox proportional hazards regression. However, more flexible regression models for the time-to-event outcomes may be considered. With moderate or large samples, the kernel-based Kaplan-Meier type estimators or flexible modeling methods such as general additive models provide nonparametric alternatives.~\cite{akritas1994nearest} In practice, some preliminary model comparison may be employed to choose the suitable regression models and covariates. As a future direction, doubly robust estimation of the time-dependent ROC parameters may be developed based on the recent work of doubly robust estimation with LTRC data, which allows flexible modeling of the nuisance functions while retaining the asymptotic normality of the estimators.\cite{morenz2024debiased,wang2024doubly,wang2024learning}

\bibliography{wileyNJD-AMA}

\bmsection*{Acknowledgment}

We would like to express our gratitude for Prof. Ronghui (Lily) Xu at University of California, San Diego for their helpful discussion and advice to improve the manuscript. The analysis of OPTN registry data in Section~\ref{sec:optn} was supported in part by Health Resources and Services Administration contract HHSH250-2019-00001C. The content is the responsibility of the authors alone and does not necessarily reflect the views or policies of the Department of Health and Human Services, nor does mention of trade names, commercial products, or organizations imply endorsement by the U.S. Government.

\bmsection*{Supporting information}

Additional supporting information may be found in the
online version of the article at the publisher’s website.

\clearpage

\appendix

\renewcommand{\theproposition}{A\arabic{proposition}}
\setcounter{proposition}{0}

\renewcommand{\thetable}{A\arabic{table}}
\setcounter{table}{0}

\renewcommand{\thefigure}{A\arabic{figure}}
\setcounter{figure}{0}

\renewcommand{\theequation}{A\arabic{equation}}
\setcounter{equation}{0}

\bmsection{Proofs of Results}
\label{append:proof}
\bmsubsection{Proof of Proposition~\ref{res:km-based}}

By definition, we have

\begin{align*}
    \Se(x_0, t_0) &= P(X_i>x_0\mid T_i\leq t_0)\\
    &= \dfrac{P(X_i>x_0, T_i\leq t_0)}{P(T_i\leq t_0)}\\
    &= \dfrac{P(T_i\leq t_0) - P(X_i\leq x_0, T_i\leq t_0)}{P(T_i\leq t_0)}\\
    &= \dfrac{1 - S_T(t_0)-F_{T,X}(t_0, x_0)}{1 - S_T(t_0)}
\end{align*}
and
\begin{align*}
    \Sp(x_0, t_0) &= P(X_i \leq x_0 \mid T_i > t_0)\\
    &= \dfrac{P(X_i\leq x_0, T_i > t_0)}{P(T_i > t_0)}\\
    &= \dfrac{P(X_i\leq x_0) - P(X_i\leq x_0, T_i \leq t_0)}{P(T_i > t_0)}\\
    &= \dfrac{F_X(x_0)-F_{T,X}(t_0, x_0)}{S_T(t_0)}
\end{align*}

\subsection{Proof of Proposition~\ref{res:reg-based}}
\begin{align*}
    \Se(x_0, t_0) &= P(X_i\geq x_0\mid T_i\leq t_0)\\
    &=\dfrac{P(X_i\geq x_0, T_i\leq t_0)}{P(T_i\leq t_0)}\\
    &= \dfrac{P(\chi(Z_i)\geq x_0, T_i\leq t_0)}{P(\chi(Z_i)\in\mathbb R, T_i\leq t_0)}\\
    &= \dfrac{E\left\{P(T_i\leq t_0\mid Z_i)\mathbbm 1(\chi(Z_i)\geq x_0)\right\}}{E\left\{P(T_i\leq t_0\mid Z_i)\right\}}\\
    &= \dfrac{\int\mathbbm 1(\chi(z)\geq x_0)P(T_i\leq t_0\mid Z_i=z)dF_Z(z)}{\int\mathbbm 1(\chi(z)\in\mathbb R)P(T_i\leq t_0\mid Z_i=z)dF_Z(z)}\\
    &= \dfrac{\int\mathbbm 1(\chi(z)\geq x_0)\{1-S_{T\mid Z}(t_0\mid z)\}dF_Z(z)}{\int\mathbbm \{1-S_{T\mid Z}(t_0\mid z)\}dF_Z(z)}
\end{align*}
and
\begin{align*}
    \Sp(x_0, t_0) &=  P(X_i\leq x_0\mid T_i>t_0)\\
    &=\dfrac{P(X_i\leq x_0, T_i> t_0)}{P(T_i> t_0)}\\
    &= \dfrac{P(\chi(Z_i)\leq x_0, T_i> t_0)}{P(\chi(Z_i)\in\mathbb R, T_i > t_0)}\\
    &= \dfrac{E\left\{P(T_i > t_0\mid Z_i)\mathbbm 1(\chi(Z_i)\geq x_0)\right\}}{E\left\{P(T_i > t_0\mid Z_i)\right\}}\\
    &= \dfrac{\int\mathbbm 1(\chi(z)\leq x_0)P(T_i> t_0\mid Z_i=z)dF_Z(z)}{\int\mathbbm 1(\chi(z)\in\mathbb R)P(T_i> t_0\mid Z_i=z)dF_Z(z)}\\
    &= \dfrac{\int\mathbbm 1(\chi(z)\leq x_0)S_{T\mid Z}(t_0\mid z)dF_Z(z)}{\int  S_{T\mid Z}(t_0\mid z)dF_Z(z)}
\end{align*}

The expression for $\AUC(t_0)$ immediately follows.

\subsection{Proof of Proposition~\ref{res:ipw}}

First, in Scenario A1, because of the relation $C = D + L$ and the assumption $(L, D)\indep (T,X)$, we also have $(L, C)\indep (T, X)$.
\begin{align*}
&E\left\{\dfrac{\Delta}{K(\widetilde T)}\mathbbm 1(X> x_0, \widetilde T\leq t_0)\mid L < \widetilde T\right\} \\&= E\left\{\dfrac{\Delta}{K(\widetilde T)}\mathbbm 1(X> x_0, \widetilde T\leq t_0)\dfrac{\mathbbm 1(L < \widetilde T)}{P(L < \widetilde T)} \right\}\\
&= \dfrac{1}{P(L < \widetilde T)}E\left\{\dfrac{\Delta}{K(\widetilde T)}\mathbbm 1(X> x_0, L < \widetilde T \leq t_0)\right\}\\
&= \dfrac{1}{P(L < \widetilde T)}E\left\{\dfrac{\Delta}{K(T)}\mathbbm 1(X> x_0, L < T\leq t_0)\right\}\\
&=  \dfrac{1}{P(L < \widetilde T)}E\left[E\left\{\dfrac{\Delta}{K(T)}\mathbbm 1(X> x_0, L < T\leq t_0)\mid T\right\}\right]\\
&= \dfrac{1}{P(L < \widetilde T)}E\left[\dfrac{1}{K(T)}E\left\{\mathbbm 1(C \geq T, L < T)\mid T\right\}\mathbbm 1(X > x_0, T \leq  t_0)\right]&&\text{(Since $(L, C)\indep (T, X)$)}\\
&= \dfrac{1}{P(L < \widetilde T)}E\left\{\mathbbm 1(X > x_0, T \leq  t_0)\right\}\\
&= \dfrac{P(X>x_0, T\leq t_0)}{P(L<\widetilde T)}.
\end{align*}

Similarly, we have
\begin{align*}
    &E\left\{\dfrac{\Delta}{K(\widetilde T)}\mathbbm 1(\widetilde T\leq t_0)\mid L < \widetilde T\right\} = \dfrac{P(T\leq t_0)}{P(L<\widetilde T)},\\
    &E\left\{\dfrac{\Delta}{K(\widetilde T)}\mathbbm 1(X\leq x_0, \widetilde T>t_0)\mid L < \widetilde T\right\} = \dfrac{P(X\leq x_0, T >  t_0)}{P(L<\widetilde T)},\\
    &E\left\{\dfrac{\Delta}{K(\widetilde T)}\mathbbm 1(\widetilde T > t_0)\mid L < \widetilde T\right\} = \dfrac{P(T > t_0)}{P(L<\widetilde T)}.
\end{align*}

The results follow.

\subsection{Proof of Proposition~\ref{res:ipw2}}

The results follow from the observation that
\begin{align*}
    & E\left\{\mathbbm 1(X\leq x_0, L < t_0 < \widetilde T)\mid L < \widetilde T\right\}\\
    = & \dfrac{1}{P(L<\widetilde T)}E\left\{\mathbbm 1(X\leq x_0, L < t_0 < \widetilde T)\right\}\\
    =& \dfrac{1}{P(L<\widetilde T)}E\left\{\mathbbm 1(X\leq x_0, L < t_0 , T > t_0, C > t_0)\right\}\\
    =& \dfrac{1}{P(L<\widetilde T)}E\left\{\mathbbm 1(X\leq x_0, T > t_0)\right\}E\left\{\mathbbm 1( L < t_0 , C > t_0)\right\}&&\text{(Because $(C,L)\indep (T,X)$)}\\
    =& \dfrac{K(t_0, t_0)}{P(L < \widetilde T)}P(X \leq x_0, T > t_0)
\end{align*}
and similarly,
\begin{align*}
    E\left\{\mathbbm 1(L < t_0 < \widetilde T)\mid L < \widetilde T\right\} = \dfrac{K(t_0, t_0)}{P(L < \widetilde T)}P(T > t_0).
\end{align*}

\subsection{Proof of Proposition~\ref{res:cipw-ident}}

Proofs of these results are similar to before. Because $C = L + D$, under Scenarios A2 and B2, we both have $(L,C)\indep T\mid Z$. It follows that
\begin{align*}
&E\left\{\dfrac{\Delta}{K_C(\widetilde T,Z)}\mathbbm 1(X> x_0, \widetilde T\leq t_0)\mid L < \widetilde T\right\} \\&= E\left\{\dfrac{\Delta}{K_C(\widetilde T,Z)}\mathbbm 1(X> x_0, \widetilde T\leq t_0)\dfrac{\mathbbm 1(L < \widetilde T)}{P(L < \widetilde T)} \right\}\\
&= \dfrac{1}{P(L < \widetilde T)}E\left\{\dfrac{\Delta}{K_C( T,Z)}\mathbbm 1(X> x_0, L < \widetilde T \leq t_0)\right\}\\
&= \dfrac{1}{P(L < \widetilde T)}E\left\{\dfrac{\Delta}{K_C( T,Z)}\mathbbm 1(X> x_0, L < T\leq t_0)\right\}\\
&=  \dfrac{1}{P(L < \widetilde T)}E\left[E\left\{\dfrac{\Delta}{K_C( T,Z)}\mathbbm 1(X> x_0, L < T\leq t_0)\mid T, Z\right\}\right]\\
&= \dfrac{1}{P(L < \widetilde T)}E\left[\dfrac{1}{K_C( T,Z)}E\left\{\mathbbm 1(C \geq T, L < T)\mid T, Z\right\}\mathbbm 1(X > x_0, T \leq  t_0)\right]&&\text{(Because $(L, C)\indep T\mid Z$ and $X=\chi(Z)$)}\\
&= \dfrac{1}{P(L < \widetilde T)}E\left\{\mathbbm 1(X > x_0, T \leq  t_0)\right\}\\
&= \dfrac{P(X>x_0, T\leq t_0)}{P(L<\widetilde T)}.
\end{align*}

Similarly, we have
\begin{align*}
    &E\left\{\dfrac{\Delta}{K_C(\widetilde T,Z)}\mathbbm 1(\widetilde T\leq t_0)\mid L < \widetilde T\right\} = \dfrac{P(T\leq t_0)}{P(L<\widetilde T)},\\
    &E\left\{\dfrac{\Delta}{K_C(\widetilde T,Z)}\mathbbm 1(X\leq x_0, \widetilde T>t_0)\mid L < \widetilde T\right\} = \dfrac{P(X\leq x_0, T >  t_0)}{P(L<\widetilde T)},\\
    &E\left\{\dfrac{\Delta}{K_C(\widetilde T,Z)}\mathbbm 1(\widetilde T > t_0)\mid L < \widetilde T\right\} = \dfrac{P(T > t_0)}{P(L<\widetilde T)}.
\end{align*}

\newpage
\bmsection{Review of estimators in Li 2016}\label{append:li-review}

For time-dependent ROC analysis with LTRC data, Li developed two sets of regression-type estimators of sensitivity, specificity and AUC under the scenario $\Pr(C>L)=1$: the nonparametric estimators which assume that censoring and truncation times are independent from the event times, and the semiparametric estimators which assume that censoring and truncation times are conditionally independent from the event times given the marker. The nonparametric estimators are 

\begin{align}
    \widehat{\Se}_{\text{LI-np}}(x_0,t_0) &= \dfrac{1 - \widehat S_T(t_0)-\widehat F_{T,X}(t_0,x_0)}{1 - \widehat S_T(t_0)},\label{eq:se-li-np}\\ 
    \widehat{\Sp}_{\text{LI-np}}(x_0,t_0) &= \dfrac{\sum_{i=1}^n\mathbbm 1(X_i>x_0,L_i< t_0<\tilde T_i)}{\sum_{i=1}^n \mathbbm 1(L_i< t_0<\widetilde T_i)},\label{eq:sp-li-np}\\
\widehat{\AUC}_{\text{LI-np}}(x_0,t_0) &= 1-\dfrac{\sum_{i=1}^n\sum_{j=1}^n \Delta_i\mathbbm 1(\widetilde T_i\leq t_0, X_i<X_j,L_j< t_0 <\widetilde T_j)\hat S_T(\widetilde T_i-)\widehat R^{-1}(\widetilde T_i)}{n^2 \widehat R(t_0)\{1 - \widehat S_T(t_0)\}}.\label{eq:auc-li-np}
\end{align}

Here $\widehat S_T(t_0)$ is the Kaplan-Meier estimator of $S_T(t_0)$ with risk set adjusted for left truncation~\cite{kaplan1958nonparametric,lynden1971method}, $\widehat F_{T,X}(t_0,x_0)$ is an estimator of $\Pr(T\leq t_0, X\leq x_0)$ given by
\begin{equation*}
    \widehat F_{T,X}(t_0,x_0)=\dfrac{1}{n}\sum_{i=1}^n\dfrac{\mathbbm 1(\widetilde T_i\leq t_0, X_i\leq x_0, \Delta_i=1)\widehat S(\widetilde T_i-)}{\widehat R(\widetilde T_i)},
\end{equation*}
and $$\widehat R(t_0)=\dfrac{1}{n}\sum_{i=1}^n\mathbbm 1(L_i< t_0\leq \widetilde T_i)$$
is the proportion of observations at risk at time $t_0$. We refer the interested readers to Li 2016 for the motivation and derivation of these estimators. 

The semiparametric estimators, which account for marker-induced dependent left truncation, can be written as

\begin{align*}
    \widehat{\Se}_{\text{LI-sp}}(x_0,t_0) &= \dfrac{\sum_{i=1}^n \mathbbm 1(X_i> x_0)\{1 - \widehat S_{T\mid X}(t_0\mid X_i)\}/\widehat S_{T\mid X}(L_i\mid X_i)}{\sum_{i=1}^n \{1 - \widehat S_{T\mid X}(t_0\mid X_i)\}/\widehat S_{T\mid X}(L_i\mid X_i)},\\
    \widehat{\Sp}_{\text{LI-sp}}(x_0,t_0) &= \dfrac{\sum_{i=1}^n \mathbbm 1(X_i\leq  x_0)\widehat S_{T\mid X}(t_0\mid X_i)/\widehat S_{T\mid X}(L_i\mid X_i)}{\sum_{i=1}^n \widehat S_{T\mid X}(t_0\mid X_i)/\widehat S_{T\mid X}(L_i\mid X_i)},\\
    \widehat{\AUC}_{\text{LI-sp}}(x_0,t_0) &= \dfrac{\sum_{i=1}^n \mathbbm 1(X_i> X_j)\{1 - \widehat S_{T\mid X}(t_0\mid X_i)\}\widehat S_{T\mid X}(t_0\mid X_j)/\left\{\widehat S_{T\mid X}(L_i\mid X_i)\widehat S_{T\mid X}(L_j\mid X_j)\right\}}{\sum_{i=1}^n \{1 - \widehat S_{T\mid X}(t_0\mid X_i)\}\widehat S_{T\mid X}(t_0\mid X_j)/\left\{\widehat S_{T\mid X}(L_i\mid X_i)\widehat S_{T\mid X}(L_j\mid X_j)\right\}}.
\end{align*}

 Li 2016 proposed using the proportional odds model to obtain  $\widehat S_{T\mid X}$, although the Cox proportional hazards model or other regression methods for event times may also be used, with risk set adjusted to account for left truncation~\cite{cox1972regression}. In the original semiparametric estimators, it is assumed that the dependence between the left truncation times, right censoring times, and event times is only induced by the marker $X$, which is often unrealistic in practice. Furthermore, when it is of interest to compare the predictive performances of several markers, the above semiparametric estimators would make different conditional independence assumptions when evaluating each marker. Finally, estimators under the scenario $\Pr(C>L)<1$ were not discussed.

\newpage
\bmsection{Estimators of $F_{T,X}$, $F_L$, and $F_Z$}\label{append:est-nuisance}

In this section, we will consider the estimation of $F_{T,X}(t,c)$  and $F_L(u)$ under scenarios A1 and B1, and $dF_Z(z)$ under scenarios A2 and B2.

We first consider $F_{T,X}(t,x) = \Pr(T\leq t, X\leq x)$. Under Scenario A1, the derivation follows Li 2016. We have
\begin{align*}
    \Pr(\widetilde T\leq t, \Delta=1,X\leq x \mid L < T) &= \Pr(T\leq t, L + D\geq T, X\leq x\mid L < T)\\
    &= \Pr(L<T\leq t, L+D\geq T, X\leq x) / \Pr(L< T)\\
    &= \int_{-\infty}^x\int_0^t \Pr(L<u\leq L+D)F_{T,X}(du, dv)/P(L<T).
\end{align*}

Noticing that the probability of being at risk at time $u$ is 
$$R(u) = \Pr(L<u\leq \widetilde T\mid L < T) = P(L<u\leq L + D, T\geq u)/\Pr(L<T)=\Pr(L<u\leq L+D)S_T(u-)/\Pr(L<T),$$
so we have
$$ \Pr(\widetilde T\leq t, \Delta=1,X\leq x \mid L < T)=\int_{-\infty}^x\int_0^t \dfrac{R(u)}{S_T(u-)}F_{T,X}(du, dv).$$
Taking derivatives with respect to $(t,c)$ and evaluating at $(u,v)$, we have
$$F_{T,X}(du, dv) =\dfrac{S_T(u-)}{R(u)} d \Pr(\widetilde T\leq u, \Delta=1,X\leq v \mid L < T)$$
and therefore,
$$F_{T,X}(t, x) =\int_{u\leq t, v\leq x}\dfrac{S_T(u-)}{R(u)} d \Pr(\widetilde T\leq u, \Delta=1,X\leq v \mid L < T).$$

Replacing $S_T(u-)$ with $\widehat S_T(u-)$, $R(u)$ with $\widehat R(u)$ in Equation~\eqref{eq:Rhat}, and $\Pr(\widetilde T\leq u, \Delta=1,X\leq v \mid L < T)$ with its empirical estimator
$$\widehat \Pr(\widetilde T\leq u, \Delta=1,X\leq v \mid L < T)=\dfrac{1}{n}\sum_{i=1}^n \mathbbm 1(\widetilde T_i\leq u, \Delta_i=1, X_i\leq v),$$
we obtain the estimator $\widehat F_{T,X}(t,c)$ in Equation~\eqref{eq:F_TX}.

Under Scenario B1, we have
\begin{align*}
    \Pr(\widetilde T\leq t, \Delta = 1, X\leq x \mid L < \widetilde T) &= \Pr(L<T\leq t, C\geq T, X\leq x_0)/\Pr(L<\widetilde T)\\
    &= \int_{-\infty}^x\int_0^t \Pr(L<u, C\geq u)F_{T,X}(du, dv)/\Pr(L<\widetilde T).
\end{align*}

The probability of being at risk at time $u$ is
\begin{align*}
    R(u) &=\Pr(L<u\leq \widetilde T\mid L<\widetilde T)=P(L<u, C\geq u)S_T(u-)/\Pr(L<\widetilde T).
\end{align*}
Again, we have

$$\Pr(\widetilde T\leq t, \Delta = 1, X\leq x\mid L < \widetilde T)=\int_{-\infty}^x\int_0^t \dfrac{R(u)}{S_T(u-)}F_{T,X}(du, dv).$$

Similar derivation as before gives the same estimator $\widehat F_{T,X}(t,c)$.

Next, we consider the estimator for $F_L(u)$. Under Scenario A1, we have
\begin{align*}
    \Pr(L \leq u\mid L<T) &\propto \Pr(L\leq u, L<T)\\
    &= \int_0^u \Pr(T>s\mid L=s)dF_L(s)\\
    &= \int_0^u S_T(s)dF_L(s)&&\text{(Since $T\indep L$)}.
\end{align*}

Taking the derivative with respect to $u$ on both sides and rearranging the terms, we have
$$dF_L(u) \propto S_T(u)^{-1} d\Pr(L\leq u\mid L<T).$$
Since $\int_0^\infty dF_L(u)=1$, we have
$$dF_L(u) = \dfrac{S_T(u)^{-1} d\Pr(L\leq u\mid L<T)}{\int S_T(s)^{-1} d\Pr(L\leq s\mid L<T)}.$$

So 
$$F_L(u) = \int_0^u dF_L(s) = \dfrac{\int_0^u  S_T(s)^{-1} d\Pr(L\leq s\mid L<T)}{\int  S_T(s)^{-1} d\Pr(L\leq s\mid L<T)}. $$

Replacing $S_T(s)$ with $\widehat S_T(s)$ and $\Pr(L\leq s\mid L<T)$ with its empirical distribution function estimator, we obtain 
$$\widehat F_L(u) =\left(\sum_{i=1}^n \dfrac{\mathbbm 1(L_i\leq u)}{\widehat S_T(L_i)}\right) /\left(\sum_{i=1}^n \dfrac{1}{\widehat S_T(L_i)}\right).$$

 Under Scenario B2, we have
\begin{align*}
    \Pr(L \leq u\mid L<\widetilde T) &\propto \Pr(L\leq u, L<\widetilde T)\\
    &= \int_0^u \Pr(T>s,C>s\mid L=s)dF_L(s)\\
    &= \int_0^u S_T(s)S_C(s)dF_L(s)&&\text{(Since $T\indep L$)}.
\end{align*}

With the same derivation, we can show that
$$F_L(u) = \int_0^u dF_L(s) = \dfrac{\int_0^u  \{S_T(s)S_C(s)\}^{-1} d\Pr(L\leq s\mid L<\widetilde T)}{\int  \{S_T(s)S_C(s)\}^{-1} d\Pr(L\leq s\mid L<\widetilde T)}. $$

We obtain the estimator
$$\widehat F_L(u) =\left(\sum_{i=1}^n \dfrac{\mathbbm 1(L_i\leq u)}{\widehat S_T(L_i)\widehat S_C(L_i)}\right) /\left(\sum_{i=1}^n \dfrac{1}{\widehat S_T(L_i)\widehat S_C(L_i)}\right).$$

Finally, we consider the estimator for $dF_Z(z)$. Under Scenario A2, for any measurable set $\zeta$, we have
\begin{align*}
    \Pr(L\leq u, Z\in \zeta\mid L<T)&\propto \int_{z\in\zeta}\int_0^u \Pr(T>l\mid Z=z, L=l)F_{L,Z}(dl,dz)\\
    &= \int_{z\in\zeta}\int_0^u S_{T\mid Z}(l\mid z)F_{L,Z}(dl,dz) &&\text{(Since $T\indep L\mid Z$)}.\\
\end{align*}
Taking the derivative on both sides, we have
$$F_{L,Z}(dl,dz\mid L<T)\propto S_{T\mid Z}(l\mid z)F_{L,Z}(dl, dz) $$
and so
$$ F_{L,Z}(dl, dz) \propto\{S_{T\mid Z}(l\mid z)\}^{-1} F_{L,Z}(dl,dz\mid L<T).$$

Since $\int\int F_{L,Z}(dl,dz)=1$, we have
 $$F_{L,Z}(dl, dz)=\dfrac{\{S_{T\mid Z}(l\mid z)\}^{-1} F_{L,Z}(ds,dz\mid L<T)}{\int\int\{S_{T\mid Z}(l\mid z)\}^{-1} F_{L,Z}(ds,dz\mid L<T)}.$$

 Integrating over $l$, we have
 $$F_{Z}( dz)=\dfrac{\int_{l\in(0,\infty)}\{S_{T\mid Z}(l\mid z)\}^{-1} F_{L,Z}(dl,dz\mid L<T)}{\int\int\{S_{T\mid Z}(s\mid z)\}^{-1} F_{L,Z}(ds,dz\mid L<T)}.$$

Replacing $S_{T\mid Z}(\cdot\mid z)$ with its estimator $\widehat S_{T\mid Z}(\cdot\mid z)$ and $F_{L,Z}(\cdot, \cdot\mid L<T)$ with the empirical distribution function, we obtain the estimator
$$d\widehat F_Z(z) = \int_{l\in(0,\infty)} \dfrac{\{\widehat S_{T\mid Z}(l\mid z)\}^{-1}\widehat  F_{L,Z}(dl,dz\mid L<T)}{\sum_{i=1}^n \{\widehat S_{T\mid Z}(L_i\mid Z_i)\}^{-1}} $$

Under Scenario B2, for any measurable set $\zeta$, we have
\begin{align*}
    \Pr(L\leq u, Z\in \zeta\mid L<\widetilde T)&\propto \int_{z\in\zeta}\int_0^u \Pr(T>l, C>l\mid Z=z, L=l)F_{L,Z}(dl,dz)\\
    &= \int_{z\in\zeta}\int_0^u S_{T\mid Z}(l\mid z)S_{C\mid Z}(l\mid z)F_{L,Z}(dl,dz)
\end{align*}
The second equivalence is because $T$,$C$, and $L$ are conditionally mutually independent given $Z$. Taking the derivative on both sides, we have
$$F_{L,Z}(dl,dz\mid L<\widetilde T)\propto S_{T\mid Z}(l\mid z)S_{C\mid Z}(l\mid z)F_{L,Z}(dl, dz)$$
and therefore
$$F_{L,Z}(dl, dz) \propto \{S_{T\mid Z}(l\mid z)S_{C\mid Z}(l\mid z)\}^{-1}F_{L,Z}(dl,dz\mid L<\widetilde T).$$

With the same derivation as before, we can show that
 $$F_{Z}( dz)=\dfrac{\int_{l\in(0,\infty)}\{S_{T\mid Z}(l\mid z)S_{C\mid Z}(l\mid z)\}^{-1} F_{L,Z}(dl,dz\mid L<T)}{\int\int\{S_{T\mid Z}(s\mid z)S_{C\mid Z}(s\mid z)\}^{-1} F_{L,Z}(ds,dz\mid L<T)}.$$
 
We obtain the estimator

$$d\widehat F_Z(z) = \int_{l\in(0,\infty)} \dfrac{\{\widehat S_{T\mid Z}(l\mid z)\widehat S_{C\mid Z}(l\mid z)\}^{-1}\widehat  F_{L,Z}(dl,dz\mid L<T)}{\sum_{i=1}^n \{\widehat S_{T\mid Z}(L_i\mid Z_i)\widehat S_{C\mid Z}(L_i\mid Z_i)\}^{-1}} $$

\newpage
\bmsection{Time-dependent ROC for LTRC data with competing risks}\label{append:cprisk}
Similar as before, we denote $T$ as time to an event, $X$ as a real-valued marker, $Z$ as a vector of baseline covariates, $L$ as the left truncation time, and $C$ as the censoring time. In the presence of competing risks, we denote $\eta$ as the event type at time $T$, where $\eta=1$ indicates the event of interest, and $\eta=2$ indicates the competing risk. We denote $\widetilde T=\min(T,C)$ as the observed event time and $\widetilde \eta=\eta\mathbbm 1(T\leq C)$ as the observed event type. The observed data are realizations of the random vector $$(\widetilde T, \widetilde \eta,L, X,Z)$$
conditioning on the selection process $L<\widetilde T$.

Similar to the previous literature~\cite{blanche2013estimating,saha2010time,zheng2012evaluating}, we define the cause-specific sensitivity as
$$\Se(x_0,t_0)=P(X_i>x_0\mid T_i\leq t_0,\eta_i=1).$$

If we consider the non-cases as subjects who didn't experience any events, the specificity is defined as

$$\Sp^1(x_0, t_0)=P(X_i\leq x_0\mid T_i> t_0).$$

Otherwise, if we define the non-cases as all subjects who were not a case, the specificity is defined as 
$$\Sp^2(x_0,t_0) = P(X_i\leq x_0\mid T_i>t_0 \text{ or }\{T_i\leq t_0,\eta_i\neq 1\}).$$

We use the superscript 1 or 2 to distinguish between the two definitions of specificity. Corresponding to the two definitions of specificity, the time-dependent AUC is defined as

$$\AUC^1(t_0)=P(X_i>0, X_j\leq x_0\mid T_i\leq t_0, \eta_i=1, T_j>t_0)$$
or
$$\AUC^2(t_0)=P(X_i>0, X_j\leq x_0\mid T_i\leq t_0, \eta_i=1, T_j>t_0\text{ or }\{T_j\leq t_0,\eta_j\neq 1\}).$$

Since the identification and estimation of $\Sp^1(x_0, t_0)$ is entirely the same as the specificity without competing risk, below we will focus on $\Se(x_0, t_0)$, $\Sp^2(x_0, t_0)$, and $\AUC^2(t_0)$.

We consider the following four scenarios with respect to different relations between right censoring time, left truncation time, and event times, corresponding to the four scenarios introduced in Section~\ref{sec:concepts}.

\begin{itemize}
    \item[] \textbf{Scenario A:} $\Pr(C>L)=1$. We denote $D=C-L$ as the residual censoring time. 
    \begin{itemize}
        \item[] \textbf{Scenario A1} (marginal mutual independence): $(L,D)\indep (T, \eta,X)$ and $D\indep L\mid (T,, \eta, X)$;
        \item[] \textbf{Scenario A2} (conditional mutual independence given $Z$): $(L,D)\indep (T,\eta)\mid Z$ and $D\indep L\mid T,\eta,Z$.
    \end{itemize}
    \item[] \textbf{Scenario B}: $P(C>L)<1$.
    \begin{itemize}
        \item[] \textbf{Scenario B1} (marginal mutual independence): $(L,C)\indep (T,\eta,X)$ and $L\indep C\mid (T,\eta, X)$;
        \item[] \textbf{Scenario B2} (conditional mutual independence): $(L,C)\indep (T, \eta)\mid Z$ and $L\indep C\mid T,\eta, Z$.
    \end{itemize}
\end{itemize}

Similar to Propositions~\ref{res:km-based} to \ref{res:cipw-ident}, below we introduce the identification and estimation of the time-dependent ROC estimators under different scenarios. We omit the proofs since they mimic the ones of Propositions~\ref{res:km-based} to \ref{res:cipw-ident} in Appendix~\ref{append:proof}.

\begin{proposition}[KM-based identification]\label{prop:km-cprisk}
    \begin{align*}
        &\Se(x_0, t_0)=\dfrac{F_{1T}(t_0) - F_{1T,X}(t_0, x_0)}{F_{1T}(t_0)},\\
        &\Sp^2(x_0, t_0)=\dfrac{F_{X}(x_0) - F_{1T,X}(t_0, x_0)}{1 - F_{1T}(t_0)}
    \end{align*}
    where $F_{1T}(t_0)=P(T\leq t_0, \eta=1)$ is the cause-specific cumulative incidence function of the primary event and
    $$F_{1T,X}(t_0, x_0) = P(T\leq t_0, \eta=1, X\leq x_0).$$
\end{proposition}

Under Scenarios A1 and B1, the estimation of $\Sp^1(x_0, t_0)$ is entirely the same as before. We can estimate $F_{1T}(t_0)$ using the Aalen-Johansen estimator of cumulative incidence functions, with the risk set adjusted to account for left truncation. Similar to before, we can estimate $\widehat F_{1T,X}(t_0, x_0)$ with
\begin{equation*}
    \widehat F_{1T,X}(t_0,x_0)=\dfrac{1}{n}\sum_{i=1}^n\dfrac{\mathbbm 1(\widetilde T_i\leq t_0, X_i\leq x_0, \widetilde \eta_i=1)\widehat S_T(\widetilde T_i-)}{\widehat R(\widetilde T_i)}
\end{equation*}
and estimate $F_X(x_0)$ with $$\widehat F_X(x_0)=\dfrac{1}{n}\sum_{i=1}^n\dfrac{\mathbbm 1(X_i\leq x_0, \widetilde \Delta_i = 1)\widehat S_T(\widetilde T_i-)}{\widehat R(\widetilde T_i)}$$
where $$\widehat R(t_0)=\dfrac{1}{n}\sum_{i=1}^n\mathbbm 1(L_i< t_0\leq \widetilde T_i).$$

\begin{proposition}[Regressioned-based identification]\label{res:reg-based-cprisk} Under Scenario A2 or B2, we have

\begin{equation}\label{eq:reg-based-cprisk}
    \begin{aligned}
   \Se(x_0,t_0) &= \dfrac{\int \mathbbm 1(\chi(z)>x_0) F_{1T\mid Z}(t\mid z)\,\mathrm d F_Z(z)}{\int F_{1T\mid Z}(t\mid z)\,\mathrm d F_Z(z)},\\
    \Sp^2(x_0,t_0) &= 
   \dfrac{\int \mathbbm 1(\chi(z)\leq x_0)\{1 - F_{1T\mid Z}(t\mid z)\}\,\mathrm d F_Z(z)}{\int  \{1 - F_{1T\mid Z}(t\mid z)\}\,\mathrm d F_Z(z)},\\
   \AUC(t_0) &=  \dfrac{\int\int \mathbbm 1(\chi(z_i)> \chi(z_j))F_{1T\mid Z}(t\mid z_j)\{1 - F_{1T\mid Z}(t\mid z_j)\}\,\mathrm d F_Z(z_i)\mathrm d F_Z(z_j)}{\int\int F_{1T\mid Z}(t\mid z)\{1 - F_{1T\mid Z}(t\mid z_j)\}\,\mathrm d F_Z(z_i)\mathrm d F_Z(z_j)}.
\end{aligned}
\end{equation}
\end{proposition}

 Proposition~\ref{res:reg-based-cprisk} motivates the following regression-based estimators

\begin{align*}
    \widehat{\Se}_{\text{REG}}(x_0,t_0) &= \dfrac{\sum_{i=1}^n \mathbbm 1(X_i> x_0)\widehat F_{1T\mid Z}(t_0\mid Z_i)/\widehat H(L_i,Z_i)}{\sum_{i=1}^n \widehat F_{1T\mid Z}(t_0\mid Z_i)/\widehat H(L_i,Z_i)},\\
    \widehat{\Sp}_{\text{REG}}^2(x_0,t_0) &= \dfrac{\sum_{i=1}^n \mathbbm 1(X_i\leq x_0)\{1 - \widehat F_{1T\mid Z}(t_0\mid Z_i)\}/\widehat H(L_i,Z_i)}{\sum_{i=1}^n \{\widehat 1 - F_{1T\mid Z}(t_0\mid Z_i)\}/\widehat H(L_i,Z_i)},\\
    \widehat{\AUC}_{\text{REG}}^2(t_0) &= \dfrac{\sum_{i=1}^n \mathbbm 1(X_i> X_j)\widehat F_{1T\mid Z}(t_0\mid Z_i)\{1 - \widehat F_{1T\mid Z}(t_0\mid Z_j)\}/\left\{\widehat H(L_i,Z_i)\widehat H(L_j,Z_j)\right\}}{\sum_{i=1}^n \widehat F_{1T\mid Z}(t_0\mid Z_i)\{\widehat 1 - F_{1T\mid Z}(t_0\mid Z_j)\}/\left\{\widehat H(L_i,Z_i)\widehat H(L_j,Z_j)\right\}},
\end{align*}
where we set $\widehat H(u,z) = \widehat S_{T\mid Z}(u\mid z)$ under Scenario A2, and $\widehat H(u,z) = \widehat S_{T\mid Z}(u\mid z)\widehat S_{C\mid Z}(u\mid z)$ under Scenario B2. The conditional cumulative incidence function $F_{1T\mid Z}(t\mid z)$ may be estimated from a regression model, for example the Fine-Gray regression~\cite{fine1999proportional}. We suspect that the regression-based estimators may suffer from the same final-sample bias as their counterparts without competing risks, and thus caution their use in practice.

\begin{proposition}[IPW Identification I]\label{res:ipw-cprisk} 
Under Scenarios A1, we have
\begin{equation}\label{eq:ipw-ident-1-cprisk}
    \begin{aligned}
   \Se(x_0,t_0) &= \dfrac{E\left\{\dfrac{\Delta}{K(\widetilde T)}\mathbbm 1(X> x_0, \widetilde T \leq t_0, \widetilde\eta = 1)\mid L < \widetilde T\right\}}{E\left\{\dfrac{\Delta}{K(\widetilde T)}\mathbbm 1(\widetilde T \leq t_0,\widetilde\eta = 1)\mid L < \widetilde T\right\}},\\
    \Sp^2(x_0,t_0) &= 
   \dfrac{E\left\{\dfrac{\Delta}{ K(\widetilde T)}\mathbbm 1(X\leq x_0, \widetilde T > t_0\text{ or }\{\widetilde T \leq t_0,\widetilde\eta\neq 1\})\mid L < \widetilde T\right\}}{E\left\{\dfrac{\Delta}{K(\widetilde T)}\mathbbm 1(\widetilde T > t_0\text{ or }\{\widetilde T \leq t_0,\widetilde\eta\neq 1\}))\mid L < \widetilde T\right\}},\\
    {\AUC}^2(t_0)&=\dfrac{ E\left\{\dfrac{\Delta_i\Delta_j}{ K(\widetilde T_i)K(\widetilde T_j)}\mathbbm 1(X_i> X_j, \widetilde T_i\leq t_0, \widetilde\eta_i=1, \widetilde T_j>t_0\text{ or }\{\widetilde T_j \leq t_0,\widetilde\eta_j\neq 1\}))\mid L_i < \widetilde T_i, L_j < \widetilde T_j \right\}}{ E\left\{ \dfrac{\Delta_i\Delta_j}{K(\widetilde T_i) K(\widetilde T_j)}\mathbbm 1(\widetilde T_i\leq t_0,\widetilde\eta_i=1, \widetilde T_j>t_0\text{ or }\{\widetilde T_j \leq t_0,\widetilde\eta_j\neq 1\})) \mid L_i < \widetilde T_i, L_j < \widetilde T_j\right\}}
\end{aligned}
\end{equation}
where the weight function $K(u) = \Pr(L < u, C > u)$ and the subscripts $i,j$ indicate two independent observations. 
\end{proposition}

Proposition~\ref{res:ipw-cprisk} motivates the following IPW estimators:
\begin{align*}
    \widehat{\Se}_{\text{IPW}}(x_0,t_0) &= \dfrac{\sum_{i=1}^n\dfrac{\Delta_i}{\widehat K( \widetilde T_i)}\mathbbm 1(X_i> x_0, \widetilde T_i\leq t_0,\widetilde\eta_i=1)}{\sum_{i=1}^n\dfrac{\Delta_i}{\widehat K(\widetilde T_i)}\mathbbm 1(\widetilde T_i<t_0,\widetilde\eta_i=1)},\\
    \widehat{\Sp}^2_{\text{IPW-1}}(x_0,t_0) &= \dfrac{\sum_{i=1}^n \dfrac{\Delta_i}{\widehat K(\widetilde T_i)} \mathbbm 1(X_i\leq x_0, \widetilde T_i>t\text{ or }\{\widetilde T_i \leq t_0,\widetilde\eta_i\neq 1\}) }{\sum_{i=1}^n \dfrac{\Delta_i}{\widehat K(\widetilde T_i)} \mathbbm 1(\widetilde T_i>t\text{ or }\{\widetilde T_i \leq t_0,\widetilde\eta_i\neq 1\})},\\
    \widehat {\AUC}^2_{\text{IPW-1}}(t_0)&=\dfrac{ \sum_{i=1}^n\sum_{j=1}^n \dfrac{\Delta_i\Delta_j}{\widehat K(\widetilde T_i)\widehat K(\widetilde T_j)}\mathbbm 1(X_i> X_j, \widetilde T_i\leq t_0,\widetilde\eta_i=1, \widetilde T_j>t_0\text{ or }\{\widetilde T_j \leq t_0,\widetilde\eta_j\neq 1\}) }{ \sum_{i=1}^n\sum_{j=1}^n \dfrac{\Delta_i\Delta_j}{\widehat K(\widetilde T_i)\widehat K( \widetilde T_j)}\mathbbm 1(\widetilde T_i\leq t_0, \widetilde\eta_i=1, \widetilde T_j>t_0\text{ or }\{\widetilde T_j \leq t_0,\widetilde\eta_j\neq 1\}) }.
\end{align*}
where $\widehat K(u)$ is the same estimator of $K(u)=\Pr(L < u, C > u)$ as before.

\begin{proposition}[IPW Identification II]\label{res:ipw2-cprisk} 
Under Scenarios A1 and B1, we have
\begin{equation}\label{eq:ipw-ident-2-cprisk}
    \begin{aligned}
    \Sp^2(x_0,t_0) &= 
   \dfrac{E\left\{\mathbbm 1(X\leq x_0, L < t_0, \widetilde T > t_0 \text{ or }\{\widetilde T \leq t_0,\widetilde\eta\neq 1\})\mid L < \widetilde T\right\}}{E\left\{\mathbbm 1(L< t_0,\widetilde T > t_0 \text{ or }\{\widetilde T \leq t_0,\widetilde\eta\neq 1\})\mid L < \widetilde T\right\}}.
\end{aligned}
\end{equation}
\end{proposition}

Proposition~\ref{res:ipw2} motivates alternative estimators $\Sp(x_0,t_0)$ and $\AUC(t_0)$: 
\begin{align*}
\widehat{\Sp}^2_{\text{IPW-2}}(x_0,t_0) &= \dfrac{\sum_{i=1}^n \mathbbm 1(L_i<t_0, X_i\leq x_0, \tilde T_i>t_0\text{ or }\{\widetilde T_i \leq t_0,\widetilde\eta_i\neq 1\}) }{\sum_{i=1}^n  \mathbbm 1(L_i<t_0, \tilde T_i>t_0\text{ or }\{\widetilde T_i \leq t_0,\widetilde\eta_i\neq 1\})},\\
\widehat {\AUC}^2_{\text{IPW-2}}(t_0)&=\dfrac{ \sum_{i=1}^n\sum_{j=1}^n \dfrac{\Delta_i}{\widehat K(\widetilde T_i)}\mathbbm 1(X_i> X_j, \widetilde T_i\leq t_0, \widetilde\eta_i=1, L_j<t_0,\widetilde T_j>t_0\text{ or }\{\widetilde T_j \leq t_0,\widetilde\eta_j\neq 1\}) }{ \sum_{i=1}^n\sum_{j=1}^n \dfrac{\Delta_i}{\widehat K(\widetilde T_i)}\mathbbm 1( \widetilde T_i\leq t_0, \widetilde\eta_i=1, L_j<t_0,\widetilde T_j > t_0 \text{ or }\{\widetilde T_j \leq t_0,\widetilde\eta_j\neq 1\}) }.
\end{align*}

Finally, we consider estimation under covariate-induced dependent truncation and censoring.
\begin{proposition}[Conditional IPW identification]\label{res:cipw-ident-cprisk} Under Scenario A2 or B2, we have
\begin{align*}
     \Se(x_0,t_0) &= \dfrac{E\left\{\dfrac{\Delta}{K_C( \widetilde T, Z)}\mathbbm 1(X> x_0, \widetilde T\leq t_0, \widetilde\eta=1)\mid L < \widetilde T\right\}}{E\left\{\dfrac{\Delta}{K_C(\widetilde T, Z)}\mathbbm 1(\widetilde T \leq t_0,\widetilde\eta=1)\mid L < \widetilde T\right\}},\\
    \Sp^2(x_0,t_0) &= 
   \dfrac{E\left\{\dfrac{\Delta}{ K_C( \widetilde T,Z)}\mathbbm 1(X\leq x_0, \widetilde T > t_0\text{ or }\{\widetilde T \leq t_0,\widetilde\eta\neq 1\})\mid L < \widetilde T\right\}}{E\left\{\dfrac{\Delta}{K_C( \widetilde T, Z)}\mathbbm 1(\widetilde T > t_0\text{ or }\{\widetilde T \leq t_0,\widetilde\eta\neq 1\})\mid L < \widetilde T\right\}}\\
   &=  \dfrac{E\left\{\dfrac{1}{K_C( t_0, Z)}\mathbbm 1(X\leq x_0, L < t_0, \widetilde T > t_0\text{ or }\{\widetilde T \leq t_0,\widetilde\eta\neq 1\})\mid L < \widetilde T\right\}}{E\left\{\dfrac{1}{K_C( t_0, Z)}\mathbbm 1(L< t_0,\widetilde T>t_0 \text{ or }\{\widetilde T \leq t_0,\widetilde\eta\neq 1\})\mid L < \widetilde T\right\}},\\
    {\AUC}^2(t_0)&=\dfrac{ E\left\{\dfrac{\Delta_i\Delta_j}{ K_C( \widetilde T_i, Z_i)K(\widetilde T_j, Z_j)}\mathbbm 1(X_i> X_j, \widetilde T_i\leq t_0, \widetilde\eta_i=1, \widetilde T_j>t_0\text{ or }\{\widetilde T_j \leq t_0,\widetilde\eta_j\neq 1\})\mid L_i < \widetilde T_i, L_j < \widetilde T_j \right\}}{E\left\{ \dfrac{\Delta_i\Delta_j}{K_C(\widetilde T_i, Z_i) K_C(\widetilde T_j, Z_j)}\mathbbm 1(\widetilde T_i\leq t_0, \widetilde\eta_i=1,\widetilde T_j>t_0\text{ or }\{\widetilde T_j \leq t_0,\widetilde\eta_j\neq 1\}) \mid L_i < \widetilde T_i, L_j < \widetilde T_j\right\}}\\
    &= \dfrac{ E\left\{\dfrac{\Delta_i}{ K_C(\widetilde T_i,Z_i)K_C(t_0, Z_j)}\mathbbm 1(X_i> X_j, \widetilde T_i\leq t_0, \widetilde\eta_i=1,L_j < t_0, \widetilde T_j>t_0\text{ or }\{\widetilde T_j \leq t_0,\widetilde\eta_j\neq 1\} )\mid L_i < \widetilde T_i, L_j < \widetilde T_j \right\}}{ E\left\{ \dfrac{\Delta_i}{K_C(\widetilde T_i, Z_i)K_C(t_0, Z_j) }\mathbbm 1(\widetilde T_i\leq t_0, \widetilde\eta_i=1,L_j<t_0, \widetilde T_j >t_0 \text{ or }\{\widetilde T_j \leq t_0,\widetilde\eta_j\neq 1\} \mid L_i < \widetilde T_i, L_j < \widetilde T_j\right\}},
\end{align*}
where $$K_C(u,z)=\Pr(L<u, C>u\mid Z=z)=\begin{cases}\int_0^u S_{D\mid Z}(u-s\mid z)dF_{L\mid Z }(s\mid z) & \text{(under Scenario A2)}\\
S_{C\mid Z}(u\mid z)F_{L\mid Z }(u\mid z) & \text{(under Scenario B2)}\end{cases}.$$
\end{proposition}

This motivates the following conditional inverse probability weighting estimators:
\begin{align}
    \widehat{\Se}_{\text{CIPW}}(x_0,t_0)&= \dfrac{\sum_{i=1}^n\dfrac{\Delta_i}{\widehat K_{C}(\widetilde T_i,Z_i)}\mathbbm 1(X_i> x_0, \widetilde T_i\leq t_0, \widetilde\eta_i=1)}{\sum_{i=1}^n\dfrac{\Delta_i}{\widehat K_{C}(\widetilde T_i,Z_i)}\mathbbm 1(\widetilde T_i<t_0, \widetilde\eta_i=1)},\label{eq:se-cipw-cprisk}\\
     \widehat{\Sp}_{\text{CIPW-1}}(x_0,t_0)&= \dfrac{\sum_{i=1}^n \dfrac{\Delta_i}{\widehat K_{C}( \widetilde T_i, Z_i)} \mathbbm 1(X_i\leq x_0, \widetilde T_i>t_0) }{\sum_{i=1}^n \dfrac{\Delta_i}{\widehat K_C(\widetilde T_i,Z_i)} \mathbbm 1(\widetilde T_i>t_0)}\label{eq:sp-cipw-1-cprisk}\\
     \widehat {\AUC}_{\text{CIPW-1}}(t_0)&=\dfrac{ \sum_{i=1}^n\sum_{j=1}^n \dfrac{\Delta_i\Delta_j}{\widehat K_{C}(T_i,Z_i)\widehat K_{C}(T_j,Z_j)}\mathbbm 1(X_i> X_j, T_i\leq t_0, \widetilde\eta_i=1, T_j>t_0) }{ \sum_{i=1}^n\sum_{j=1}^n \dfrac{\Delta_i\Delta_j}{\widehat K_{C}(T_i,Z_i)\widehat K_{C}(T_j,Z_j)}\mathbbm 1( T_i\leq t_0, \widetilde\eta_i=1, T_j>t_0) }\label{eq:auc-cipw-1-cprisk}\\
     \widehat{\Sp}_{\text{CIPW-2}}(x_0,t_0) &= \dfrac{\sum_{i=1}^n \dfrac{1}{\widehat K_{C}( t_0,Z_i)} \mathbbm 1(L_i<t_0<\widetilde T_i, X_i\leq x_0) }{\sum_{i=1}^n \dfrac{1}{\widehat K_{C}( t_0,Z_i)} \mathbbm 1(L_i<t_0< \tilde T_i)},\label{eq:sp-cipw-2-cprisk}\\
     \widehat {\AUC}_{\text{CIPW-2}}(t_0)&=\dfrac{ \sum_{i=1}^n\sum_{j=1}^n \dfrac{\Delta_i}{\widehat K_{C}(\widetilde T_i,Z_i)\widehat K_{C}( t_0,Z_j)}\mathbbm 1(X_i> X_j, T_i\leq t_0, \widetilde\eta_i=1, L_j<t_0<T_j) }{ \sum_{i=1}^n\sum_{j=1}^n \dfrac{\Delta_i}{\widehat K_{C}(\widetilde T_i,Z_i)\widehat K_{C}(t_0,Z_j)}\mathbbm 1( T_i\leq t_0, \widetilde\eta_i=1, L_j<t_0<T_j) }.\label{eq:auc-cipw-2-cprisk}
\end{align}
where $\widehat K_{C}(u,z)$ is the same estimator of $K_C(u,z)$ as before.

\newpage
\bmsection{Results of the simulation studies in Section~5}\label{append:sim}

Due to the space limitation, we present the results of simulation studies in separate tables:
\begin{itemize}
    \item Table~\ref{tab:sim_common_a1a2_0}: Simulation setting summary under scenarios T1-T2, L1-L3, and C1;
    \item Tables~\ref{tab:sim_common_b1b2_0}-\ref{tab:sim_common_b1b2_2}: Simulation setting summary and results of different AUC estimators under scenarios T1-T2, L1-L3, and C2.

\end{itemize}

\clearpage

\begin{table}[!htbp]
    \centering   
    \caption{\label{tab:sim_common_a1a2_0} Summary of different simulation scenarios of event time and left truncation time distribution within Scenarios A1 or A2 as described in Section~\ref{sec:simu}. In each scenario and each time point, we report the true value of the time-dependent AUC ($\AUC(t)$) the average number and IQR of total events ($\overline{\#\text{Total events}}$), number of subjects that have entered the study ($\overline{\#\text{Entered by $t$}}$), number at risk  ($\overline{\#\text{At risk at $t$}}$), and number of events cumulative by $t$ ($\overline{\#\text{Events by $t$}}$)}
\begin{tabular}{rr|rr|r|cccc}
  \hline
\multicolumn{2}{c|}{Scenario} & \multirow{2}{*}{$t$} & \multirow{2}{*}{$N$} & \multirow{2}{*}{$AUC(t)$} & \multirow{2}{*}{$\#\text{ Total events}$} & \multirow{2}{*}{$\#\text{ Entered by $t$}$} & \multirow{2}{*}{$\#\text{ At risk at }t$} & \multirow{2}{*}{$\#\text{ Events by $t$}$} \\ 
\cline{1-2}
$T$ & $L$ & & & & & & \\
  \hline
  T1 & L1 & 1.0 & 300 & 0.679 & 242 (238, 247) & 147 (142, 153) & 118 (112, 124) & 25 (22, 29) \\ 
  T1 & L1 & 1.0 & 600 & 0.679 & 484 (477, 490) & 295 (286, 303) & 237 (228, 244) & 51 (46, 56) \\ 
  T1 & L1 & 1.6 & 300 & 0.707 & 242 (238, 247) & 208 (203, 214) & 121 (115, 126) & 73 (69, 79) \\ 
  T1 & L1 & 1.6 & 600 & 0.707 & 484 (477, 490) & 417 (409, 425) & 241 (233, 250) & 147 (140, 154) \\ 
  T1 & L1 & 2.3 & 300 & 0.745 & 242 (238, 247) & 254 (249, 258) & 91 (86, 96) & 134 (128, 139) \\ 
  T1 & L1 & 2.3 & 600 & 0.745 & 484 (477, 490) & 506 (500, 513) & 182 (175, 189) & 268 (260, 276) \\ 
  \hline
  T1 & L2 & 1.0 & 300 & 0.679 & 234 (229, 239) & 154 (149, 160) & 129 (123, 135) & 21 (18, 24) \\ 
  T1 & L2 & 1.0 & 600 & 0.679 & 467 (461, 474) & 310 (301, 319) & 258 (250, 267) & 43 (39, 47) \\ 
  T1 & L2 & 1.6 & 300 & 0.707 & 234 (229, 239) & 215 (210, 220) & 133 (127, 138) & 66 (61, 71) \\ 
  T1 & L2 & 1.6 & 600 & 0.707 & 467 (461, 474) & 430 (423, 438) & 265 (256, 274) & 132 (125, 139) \\ 
  T1 & L2 & 2.3 & 300 & 0.745 & 234 (229, 239) & 259 (255, 262) & 99 (95, 105) & 126 (120, 132) \\ 
  T1 & L2 & 2.3 & 600 & 0.745 & 467 (461, 474) & 517 (511, 522) & 198 (191, 205) & 252 (244, 260) \\ 
  \hline
  T1 & L3 & 1.0 & 300 & 0.679 & 241 (237, 246) & 143 (137, 149) & 116 (110, 121) & 23 (20, 26) \\ 
  T1 & L3 & 1.0 & 600 & 0.679 & 483 (475, 489) & 286 (277, 294) & 230 (222, 239) & 47 (42, 51) \\ 
  T1 & L3 & 1.6 & 300 & 0.707 & 241 (237, 246) & 201 (195, 206) & 117 (112, 123) & 68 (63, 73) \\ 
  T1 & L3 & 1.6 & 600 & 0.707 & 483 (475, 489) & 402 (393, 410) & 234 (226, 243) & 137 (130, 144) \\ 
  T1 & L3 & 2.3 & 300 & 0.745 & 241 (237, 246) & 246 (241, 250) & 89 (84, 95) & 127 (121, 133) \\ 
  T1 & L3 & 2.3 & 600 & 0.745 & 483 (475, 489) & 491 (485, 497) & 178 (171, 185) & 256 (248, 264) \\ 
  \hline
  T2 & L1 & 1.0 & 300 & 0.658 & 252 (247, 256) & 158 (151, 163) & 125 (119, 131) & 29 (26, 32) \\ 
  T2 & L1 & 1.0 & 600 & 0.658 & 504 (497, 510) & 315 (306, 323) & 250 (241, 257) & 57 (53, 62) \\ 
  T2 & L1 & 1.6 & 300 & 0.674 & 252 (247, 256) & 222 (216, 226) & 123 (118, 129) & 83 (78, 88) \\ 
  T2 & L1 & 1.6 & 600 & 0.674 & 504 (497, 510) & 442 (435, 449) & 246 (237, 254) & 166 (158, 174) \\ 
  T2 & L1 & 2.3 & 300 & 0.697 & 252 (247, 256) & 265 (262, 269) & 85 (80, 90) & 152 (146, 157) \\ 
  T2 & L1 & 2.3 & 600 & 0.697 & 504 (497, 510) & 530 (525, 535) & 170 (162, 177) & 303 (295, 312) \\ 
  \hline
  T2 & L2 & 1.0 & 300 & 0.658 & 246 (241, 251) & 164 (158, 169) & 134 (128, 139) & 25 (22, 28) \\ 
  T2 & L2 & 1.0 & 600 & 0.658 & 492 (485, 497) & 328 (319, 336) & 268 (259, 276) & 51 (46, 55) \\ 
  T2 & L2 & 1.6 & 300 & 0.674 & 246 (241, 251) & 226 (221, 230) & 132 (126, 138) & 77 (72, 82) \\ 
  T2 & L2 & 1.6 & 600 & 0.674 & 492 (485, 497) & 452 (445, 459) & 263 (255, 272) & 154 (147, 161) \\ 
  T2 & L2 & 2.3 & 300 & 0.697 & 246 (241, 251) & 269 (265, 272) & 91 (85, 96) & 145 (139, 151) \\ 
  T2 & L2 & 2.3 & 600 & 0.697 & 492 (485, 497) & 537 (532, 541) & 180 (172, 187) & 291 (284, 300) \\ 
  \hline
  T2 & L3 & 1.0 & 300 & 0.658 & 251 (246, 255) & 154 (149, 161) & 124 (118, 130) & 26 (23, 29) \\ 
  T2 & L3 & 1.0 & 600 & 0.658 & 500 (494, 506) & 309 (301, 317) & 247 (239, 255) & 52 (48, 57) \\ 
  T2 & L3 & 1.6 & 300 & 0.674 & 251 (246, 255) & 216 (209, 220) & 122 (116, 127) & 77 (72, 82) \\ 
  T2 & L3 & 1.6 & 600 & 0.674 & 500 (494, 506) & 431 (423, 439) & 243 (235, 252) & 155 (148, 162) \\ 
  T2 & L3 & 2.3 & 300 & 0.697 & 251 (246, 255) & 260 (256, 264) & 85 (80, 91) & 145 (139, 151) \\ 
  T2 & L3 & 2.3 & 600 & 0.697 & 500 (494, 506) & 520 (515, 526) & 170 (162, 177) & 291 (282, 299) \\
   \hline
\end{tabular}
\end{table}

\clearpage

\begin{table}[!htbp]
    \centering   
    \caption{\label{tab:sim_common_b1b2_0} Summary of different simulation scenarios of event time and left truncation time distribution within Scenarios B1 or B2 as described in Section~\ref{sec:simu}. In each scenario and each time point, we report the true value of the time-dependent AUC ($\AUC(t)$) the average number and IQR of total events ($\overline{\#\text{Total events}}$), number of subjects that have entered the study ($\overline{\#\text{Entered by $t$}}$), number at risk  ($\overline{\#\text{At risk at $t$}}$), and number of events cumulative by $t$ ($\overline{\#\text{Events by $t$}}$)}
\begin{tabular}{rr|rr|r|cccc}
  \hline
\multicolumn{2}{c|}{Scenario} & \multirow{2}{*}{$t$} & \multirow{2}{*}{$N$} & \multirow{2}{*}{$AUC(t)$} & \multirow{2}{*}{$\#\text{ Total events}$} & \multirow{2}{*}{$\#\text{ Entered by $t$}$} & \multirow{2}{*}{$\#\text{ At risk at }t$} & \multirow{2}{*}{$\#\text{ Events by $t$}$} \\ 
\cline{1-2}
$T$ & $L$ & & & & & & \\
  \hline
 T1 & L1 & 1.0 & 300 & 0.680 & 257 (253, 261) & 160 (154, 166) & 126 (120, 133) & 28 (24, 31) \\ 
  T1 & L1 & 1.0 & 600 & 0.680 & 514 (508, 520) & 320 (311, 328) & 253 (244, 260) & 55 (50, 60) \\ 
  T1 & L1 & 1.6 & 300 & 0.712 & 257 (253, 261) & 222 (216, 227) & 128 (122, 134) & 79 (73, 84) \\ 
  T1 & L1 & 1.6 & 600 & 0.712 & 514 (508, 520) & 442 (435, 450) & 255 (247, 263) & 157 (150, 165) \\ 
  T1 & L1 & 2.3 & 300 & 0.756 & 257 (253, 261) & 263 (260, 267) & 96 (91, 101) & 142 (137, 148) \\ 
  T1 & L1 & 2.3 & 600 & 0.756 & 514 (508, 520) & 525 (520, 531) & 191 (184, 199) & 285 (277, 293) \\ 
  \hline
  T1 & L2 & 1.0 & 300 & 0.680 & 253 (248, 257) & 166 (161, 172) & 137 (131, 142) & 23 (20, 26) \\ 
  T1 & L2 & 1.0 & 600 & 0.680 & 506 (500, 511) & 334 (325, 342) & 275 (267, 283) & 46 (42, 50) \\ 
  T1 & L2 & 1.6 & 300 & 0.712 & 253 (248, 257) & 227 (222, 232) & 141 (135, 147) & 70 (65, 75) \\ 
  T1 & L2 & 1.6 & 600 & 0.712 & 506 (500, 511) & 454 (447, 461) & 282 (273, 290) & 140 (133, 147) \\ 
  T1 & L2 & 2.3 & 300 & 0.756 & 253 (248, 257) & 267 (263, 271) & 107 (101, 112) & 133 (127, 139) \\ 
  T1 & L2 & 2.3 & 600 & 0.756 & 506 (500, 511) & 534 (529, 539) & 212 (205, 220) & 267 (259, 275) \\ 
  \hline
  T1 & L3 & 1.0 & 300 & 0.680 & 257 (252, 261) & 156 (150, 162) & 125 (119, 130) & 25 (22, 28) \\ 
  T1 & L3 & 1.0 & 600 & 0.680 & 513 (507, 519) & 313 (304, 320) & 250 (242, 258) & 50 (46, 55) \\ 
  T1 & L3 & 1.6 & 300 & 0.712 & 257 (252, 261) & 215 (209, 219) & 126 (120, 131) & 74 (69, 78) \\ 
  T1 & L3 & 1.6 & 600 & 0.712 & 513 (507, 519) & 429 (422, 437) & 252 (243, 260) & 148 (141, 154) \\ 
  T1 & L3 & 2.3 & 300 & 0.756 & 257 (252, 261) & 257 (253, 261) & 95 (90, 100) & 137 (131, 143) \\ 
  T1 & L3 & 2.3 & 600 & 0.756 & 513 (507, 519) & 514 (509, 519) & 190 (183, 197) & 275 (267, 283) \\ 
  \hline
  T2 & L1 & 1.0 & 300 & 0.659 & 263 (259, 267) & 168 (163, 174) & 131 (126, 138) & 31 (27, 34) \\ 
  T2 & L1 & 1.0 & 600 & 0.659 & 526 (521, 532) & 337 (329, 345) & 263 (255, 271) & 61 (56, 66) \\ 
  T2 & L1 & 1.6 & 300 & 0.675 & 263 (259, 267) & 232 (227, 237) & 129 (123, 134) & 87 (82, 93) \\ 
  T2 & L1 & 1.6 & 600 & 0.675 & 526 (521, 532) & 463 (456, 470) & 257 (248, 266) & 175 (168, 183) \\ 
  T2 & L1 & 2.3 & 300 & 0.697 & 263 (259, 267) & 273 (269, 276) & 89 (83, 93) & 159 (153, 164) \\ 
  T2 & L1 & 2.3 & 600 & 0.697 & 526 (521, 532) & 544 (539, 549) & 176 (168, 184) & 318 (310, 327) \\ 
  \hline
  T2 & L2 & 1.0 & 300 & 0.659 & 260 (256, 265) & 174 (169, 180) & 141 (135, 146) & 27 (23, 30) \\ 
  T2 & L2 & 1.0 & 600 & 0.659 & 520 (514, 526) & 349 (341, 357) & 282 (274, 290) & 54 (49, 58) \\ 
  T2 & L2 & 1.6 & 300 & 0.675 & 260 (256, 265) & 236 (231, 240) & 138 (133, 144) & 81 (76, 86) \\ 
  T2 & L2 & 1.6 & 600 & 0.675 & 520 (514, 526) & 472 (465, 478) & 277 (269, 286) & 161 (154, 168) \\ 
  T2 & L2 & 2.3 & 300 & 0.697 & 260 (256, 265) & 275 (272, 278) & 95 (91, 101) & 152 (146, 158) \\ 
  T2 & L2 & 2.3 & 600 & 0.697 & 520 (514, 526) & 549 (545, 553) & 190 (182, 197) & 305 (297, 314) \\ 
  \hline
  T2 & L3 & 1.0 & 300 & 0.659 & 262 (258, 266) & 166 (160, 172) & 132 (126, 137) & 28 (25, 31) \\ 
  T2 & L3 & 1.0 & 600 & 0.659 & 525 (519, 530) & 332 (324, 340) & 264 (255, 272) & 56 (51, 61) \\ 
  T2 & L3 & 1.6 & 300 & 0.675 & 262 (258, 266) & 227 (221, 232) & 129 (123, 134) & 82 (77, 87) \\ 
  T2 & L3 & 1.6 & 600 & 0.675 & 525 (519, 530) & 454 (446, 461) & 258 (249, 266) & 164 (158, 172) \\ 
  T2 & L3 & 2.3 & 300 & 0.697 & 262 (258, 266) & 268 (265, 272) & 90 (84, 95) & 153 (147, 160) \\ 
  T2 & L3 & 2.3 & 600 & 0.697 & 525 (519, 530) & 536 (532, 542) & 178 (171, 186) & 308 (300, 316) \\ 
   \hline
\end{tabular}
\end{table}

\begin{sidewaystable}[!htbp]
    \centering   
    \caption{\label{tab:sim_common_b1b2_1}Bias, square root of MSE, and coverage rate of 95\% bootstrap intervals of $\widehat \AUC_{\text{RC-IPW}}$, $\widehat \AUC_{\text{RC-CIPW}}$, $\widehat \AUC_{\text{KM}}$, and $\widehat \AUC_{\text{REG}}$ in simulation studies with 1,000 replications under different scenarios of event time and left truncation time distribution within Scenarios B1 or B2 as described in Section~\ref{sec:simu}. The censoring time distribution follows scenario C2. }
\begin{tabular}{|rr|rr|rrr|rrr|rrr|rrr|}
  \hline
\multicolumn{2}{|c|}{Scenario} & \multirow{2}{*}{$t$} & \multirow{2}{*}{$N$}  & \multicolumn{3}{c|}{$\widehat \AUC_{\text{RC-IPW}}$} & \multicolumn{3}{c|}{$\widehat \AUC_{\text{RC-CIPW}}$} & \multicolumn{3}{c|}{$\widehat \AUC_{\text{KM}}$} &\multicolumn{3}{c|}{$\widehat \AUC_{\text{REG}}$} \\ 
\cline{1-2}\cline{5-16}
$T$ & $L$ & &  & Bias &$\sqrt{\text{MSE}}$ & Coverage & Bias &$\sqrt{\text{MSE}}$ & Coverage & Bias &$\sqrt{\text{MSE}}$ & Coverage & Bias &$\sqrt{\text{MSE}}$ & Coverage \\
  \hline
T1 & L1 & 1.0 & 300 & 0.037 & 0.065 & 84.7 & 0.037 & 0.064 & 84.7 & 0.006 & 0.063 & 93.2 & -0.018 & 0.036 & 90.9 \\ 
  T1 & L1 & 1.0 & 600 & 0.037 & 0.052 & 80.9 & 0.037 & 0.051 & 80.9 & 0.003 & 0.042 & 94.6 & -0.016 & 0.028 & 86.1 \\ 
  T1 & L1 & 1.6 & 300 & 0.014 & 0.036 & 93.1 & 0.014 & 0.036 & 93.5 & 0.009 & 0.041 & 95.8 & -0.025 & 0.045 & 87.3 \\ 
  T1 & L1 & 1.6 & 600 & 0.014 & 0.028 & 89.1 & 0.013 & 0.028 & 89.7 & 0.003 & 0.028 & 95.0 & -0.023 & 0.036 & 80.6 \\ 
  T1 & L1 & 2.3 & 300 & -0.015 & 0.034 & 92.5 & -0.016 & 0.034 & 92.3 & 0.010 & 0.040 & 95.5 & -0.036 & 0.058 & 79.2 \\ 
  T1 & L1 & 2.3 & 600 & -0.014 & 0.026 & 89.8 & -0.015 & 0.026 & 89.1 & 0.002 & 0.025 & 96.3 & -0.033 & 0.046 & 70.6 \\ 
  \hline
  T1 & L2 & 1.0 & 300 & -0.036 & 0.071 & 91.6 & -0.035 & 0.071 & 91.8 & -0.065 & 0.096 & 81.4 & -0.025 & 0.041 & 84.8 \\ 
  T1 & L2 & 1.0 & 600 & -0.036 & 0.057 & 86.2 & -0.036 & 0.057 & 86.4 & -0.069 & 0.084 & 66.1 & -0.022 & 0.033 & 80.2 \\ 
  T1 & L2 & 1.6 & 300 & -0.047 & 0.059 & 77.7 & -0.047 & 0.059 & 77.9 & -0.057 & 0.072 & 68.1 & -0.033 & 0.051 & 80.5 \\ 
  T1 & L2 & 1.6 & 600 & -0.047 & 0.055 & 56.0 & -0.047 & 0.055 & 56.9 & -0.063 & 0.070 & 40.7 & -0.030 & 0.042 & 72.4 \\ 
  T1 & L2 & 2.3 & 300 & -0.069 & 0.077 & 39.6 & -0.069 & 0.077 & 39.4 & -0.055 & 0.068 & 60.0 & -0.043 & 0.063 & 73.9 \\ 
  T1 & L2 & 2.3 & 600 & -0.070 & 0.073 & 10.7 & -0.070 & 0.073 & 10.9 & -0.064 & 0.070 & 21.0 & -0.040 & 0.053 & 63.4 \\ 
  \hline
  T1 & L3 & 1.0 & 300 & 0.005 & 0.049 & 92.4 & 0.004 & 0.049 & 92.6 & -0.040 & 0.069 & 86.8 & -0.027 & 0.044 & 82.3 \\ 
  T1 & L3 & 1.0 & 600 & 0.004 & 0.034 & 95.2 & 0.003 & 0.034 & 95.1 & -0.043 & 0.057 & 76.4 & -0.025 & 0.036 & 77.4 \\ 
  T1 & L3 & 1.6 & 300 & -0.005 & 0.033 & 95.0 & -0.007 & 0.034 & 95.2 & -0.028 & 0.048 & 85.6 & -0.035 & 0.055 & 79.4 \\ 
  T1 & L3 & 1.6 & 600 & -0.004 & 0.024 & 93.4 & -0.006 & 0.025 & 93.0 & -0.032 & 0.042 & 72.6 & -0.033 & 0.046 & 70.8 \\ 
  T1 & L3 & 2.3 & 300 & -0.016 & 0.035 & 91.1 & -0.019 & 0.036 & 90.0 & -0.007 & 0.038 & 93.2 & -0.046 & 0.068 & 75.8 \\ 
  T1 & L3 & 2.3 & 600 & -0.016 & 0.027 & 87.9 & -0.019 & 0.029 & 85.9 & -0.016 & 0.030 & 86.2 & -0.043 & 0.058 & 64.9 \\ 
  \hline
  T2 & L1 & 1.0 & 300 & 0.024 & 0.060 & 91.0 & 0.024 & 0.059 & 91.0 & 0.007 & 0.068 & 94.6 & -0.049 & 0.054 & 45.5 \\ 
  T2 & L1 & 1.0 & 600 & 0.024 & 0.045 & 88.9 & 0.024 & 0.045 & 88.9 & 0.004 & 0.044 & 94.9 & -0.047 & 0.051 & 19.4 \\ 
  T2 & L1 & 1.6 & 300 & 0.005 & 0.036 & 94.6 & 0.005 & 0.035 & 94.8 & 0.013 & 0.046 & 94.4 & -0.042 & 0.051 & 67.5 \\ 
  T2 & L1 & 1.6 & 600 & 0.005 & 0.025 & 94.8 & 0.005 & 0.025 & 95.0 & 0.006 & 0.029 & 94.6 & -0.040 & 0.045 & 44.9 \\ 
  T2 & L1 & 2.3 & 300 & -0.012 & 0.034 & 93.0 & -0.012 & 0.035 & 92.6 & 0.026 & 0.048 & 92.9 & -0.026 & 0.044 & 88.1 \\ 
  T2 & L1 & 2.3 & 600 & -0.012 & 0.025 & 91.6 & -0.012 & 0.026 & 91.0 & 0.013 & 0.029 & 93.5 & -0.023 & 0.034 & 84.2 \\ 
  \hline
  T2 & L2 & 1.0 & 300 & -0.065 & 0.089 & 82.0 & -0.065 & 0.089 & 82.3 & -0.086 & 0.114 & 72.2 & -0.058 & 0.063 & 30.8 \\ 
  T2 & L2 & 1.0 & 600 & -0.069 & 0.081 & 63.6 & -0.069 & 0.081 & 63.6 & -0.092 & 0.103 & 48.0 & -0.057 & 0.060 & 7.1 \\ 
  T2 & L2 & 1.6 & 300 & -0.066 & 0.075 & 57.4 & -0.066 & 0.075 & 58.2 & -0.070 & 0.083 & 56.1 & -0.053 & 0.061 & 51.2 \\ 
  T2 & L2 & 1.6 & 600 & -0.065 & 0.070 & 30.5 & -0.065 & 0.070 & 31.1 & -0.075 & 0.081 & 22.9 & -0.053 & 0.056 & 25.3 \\ 
  T2 & L2 & 2.3 & 300 & -0.069 & 0.077 & 44.1 & -0.069 & 0.076 & 44.3 & -0.049 & 0.065 & 67.4 & -0.040 & 0.054 & 75.3 \\ 
  T2 & L2 & 2.3 & 600 & -0.072 & 0.075 & 13.2 & -0.071 & 0.075 & 13.2 & -0.064 & 0.069 & 23.9 & -0.039 & 0.047 & 65.3 \\ 
  \hline
  T2 & L3 & 1.0 & 300 & -0.027 & 0.059 & 91.2 & -0.027 & 0.060 & 90.6 & -0.051 & 0.078 & 82.2 & -0.062 & 0.067 & 29.7 \\ 
  T2 & L3 & 1.0 & 600 & -0.027 & 0.045 & 87.1 & -0.028 & 0.045 & 86.9 & -0.055 & 0.067 & 67.1 & -0.061 & 0.064 & 7.0 \\ 
  T2 & L3 & 1.6 & 300 & -0.029 & 0.045 & 87.3 & -0.030 & 0.046 & 86.4 & -0.035 & 0.053 & 81.4 & -0.059 & 0.066 & 47.2 \\ 
  T2 & L3 & 1.6 & 600 & -0.028 & 0.038 & 78.2 & -0.029 & 0.039 & 77.2 & -0.040 & 0.048 & 62.8 & -0.058 & 0.062 & 21.6 \\ 
  T2 & L3 & 2.3 & 300 & -0.029 & 0.043 & 86.1 & -0.031 & 0.045 & 84.1 & -0.008 & 0.039 & 93.3 & -0.048 & 0.061 & 73.7 \\ 
  T2 & L3 & 2.3 & 600 & -0.030 & 0.039 & 74.0 & -0.033 & 0.041 & 71.0 & -0.022 & 0.034 & 81.1 & -0.046 & 0.054 & 56.3 \\ 
   \hline
\end{tabular}
\end{sidewaystable}

\newpage

\begin{sidewaystable}[!htbp]
    \centering   
    \caption{\label{tab:sim_common_b1b2_2}Bias, square root of MSE, and coverage rate of 95\% bootstrap intervals of $\widehat \AUC_{\text{IPW-1}}$, $\widehat \AUC_{\text{CIPW-1}}$, $\widehat \AUC_{\text{IPW-2}}$, and $\widehat \AUC_{\text{CIPW-2}}$ in simulation studies with 1,000 replications under different scenarios of event time and left truncation time distribution within Scenarios B1 or B2 as described in Section~\ref{sec:simu}. The censoring time distribution follows scenario C2. }
\begin{tabular}{|rr|rr|rrr|rrr|rrr|rrr|}
  \hline
\multicolumn{2}{|c|}{Scenario} & \multirow{2}{*}{$t$} & \multirow{2}{*}{$N$}  & \multicolumn{3}{c|}{$\widehat \AUC_{\text{IPW-1}}$} & \multicolumn{3}{c|}{$\widehat \AUC_{\text{CIPW-1}}$} & \multicolumn{3}{c|}{$\widehat \AUC_{\text{IPW-2}}$} &\multicolumn{3}{c|}{$\widehat \AUC_{\text{CIPW-2}}$} \\ 
\cline{1-2}\cline{5-16}
$T$ & $L$ & &  & Bias &$\sqrt{\text{MSE}}$ & Coverage & Bias &$\sqrt{\text{MSE}}$ & Coverage & Bias &$\sqrt{\text{MSE}}$ & Coverage & Bias &$\sqrt{\text{MSE}}$ & Coverage \\
  \hline
T1 & L1 & 1.0 & 300 & -0.003 & 0.062 & 93.1 & -0.004 & 0.064 & 94.2 & -0.003 & 0.063 & 94.1 & -0.003 & 0.065 & 93.7 \\ 
  T1 & L1 & 1.0 & 600 & -0.002 & 0.041 & 94.5 & -0.002 & 0.043 & 95.2 & -0.001 & 0.043 & 94.4 & -0.002 & 0.045 & 95.0 \\ 
  T1 & L1 & 1.6 & 300 & -0.004 & 0.039 & 95.1 & -0.004 & 0.042 & 95.8 & -0.003 & 0.041 & 94.6 & -0.003 & 0.044 & 95.5 \\ 
  T1 & L1 & 1.6 & 600 & -0.004 & 0.028 & 94.8 & -0.004 & 0.029 & 95.3 & -0.004 & 0.029 & 94.8 & -0.004 & 0.031 & 95.0 \\ 
  T1 & L1 & 2.3 & 300 & -0.011 & 0.036 & 94.2 & -0.011 & 0.039 & 94.7 & -0.010 & 0.036 & 94.6 & -0.010 & 0.040 & 94.9 \\ 
  T1 & L1 & 2.3 & 600 & -0.009 & 0.025 & 93.4 & -0.010 & 0.028 & 93.1 & -0.009 & 0.025 & 93.6 & -0.009 & 0.029 & 94.4 \\ 
  \hline
  T1 & L2 & 1.0 & 300 & -0.075 & 0.103 & 79.2 & -0.017 & 0.073 & 93.5 & -0.040 & 0.080 & 91.6 & -0.016 & 0.075 & 94.2 \\ 
  T1 & L2 & 1.0 & 600 & -0.074 & 0.088 & 64.3 & -0.009 & 0.051 & 93.1 & -0.040 & 0.063 & 86.7 & -0.008 & 0.052 & 94.0 \\ 
  T1 & L2 & 1.6 & 300 & -0.071 & 0.082 & 61.2 & -0.009 & 0.047 & 95.0 & -0.052 & 0.067 & 78.8 & -0.008 & 0.049 & 95.9 \\ 
  T1 & L2 & 1.6 & 600 & -0.070 & 0.077 & 33.3 & -0.006 & 0.034 & 94.4 & -0.052 & 0.060 & 59.7 & -0.005 & 0.035 & 94.7 \\ 
  T1 & L2 & 2.3 & 300 & -0.077 & 0.085 & 38.8 & -0.012 & 0.042 & 94.6 & -0.065 & 0.075 & 54.9 & -0.010 & 0.042 & 94.7 \\ 
  T1 & L2 & 2.3 & 600 & -0.076 & 0.080 & 11.5 & -0.011 & 0.030 & 93.4 & -0.066 & 0.070 & 22.2 & -0.010 & 0.031 & 94.0 \\ 
  \hline
  T1 & L3 & 1.0 & 300 & -0.048 & 0.074 & 84.7 & -0.053 & 0.080 & 87.3 & -0.061 & 0.088 & 82.3 & -0.059 & 0.090 & 86.8 \\ 
  T1 & L3 & 1.0 & 600 & -0.048 & 0.061 & 74.4 & -0.054 & 0.068 & 75.3 & -0.062 & 0.075 & 68.4 & -0.061 & 0.077 & 75.0 \\ 
  T1 & L3 & 1.6 & 300 & -0.040 & 0.055 & 81.7 & -0.049 & 0.064 & 79.3 & -0.056 & 0.069 & 71.6 & -0.059 & 0.073 & 75.2 \\ 
  T1 & L3 & 1.6 & 600 & -0.038 & 0.046 & 67.7 & -0.048 & 0.056 & 59.9 & -0.054 & 0.061 & 50.8 & -0.057 & 0.065 & 52.4 \\ 
  T1 & L3 & 2.3 & 300 & -0.028 & 0.043 & 86.5 & -0.042 & 0.055 & 80.6 & -0.042 & 0.055 & 78.9 & -0.050 & 0.063 & 76.6 \\ 
  T1 & L3 & 2.3 & 600 & -0.027 & 0.036 & 78.9 & -0.040 & 0.048 & 64.0 & -0.041 & 0.048 & 62.2 & -0.049 & 0.056 & 54.4 \\ 
  \hline
  T2 & L1 & 1.0 & 300 & -0.002 & 0.063 & 94.4 & -0.002 & 0.067 & 95.4 & -0.002 & 0.064 & 94.6 & -0.002 & 0.068 & 94.8 \\ 
  T2 & L1 & 1.0 & 600 & -0.001 & 0.043 & 95.3 & -0.001 & 0.046 & 94.4 & -0.001 & 0.045 & 94.1 & -0.001 & 0.048 & 94.6 \\ 
  T2 & L1 & 1.6 & 300 & -0.001 & 0.041 & 94.5 & -0.001 & 0.046 & 96.0 & -0.001 & 0.042 & 94.2 & -0.001 & 0.047 & 96.0 \\ 
  T2 & L1 & 1.6 & 600 & -0.001 & 0.028 & 95.3 & -0.001 & 0.031 & 95.9 & -0.001 & 0.029 & 94.7 & -0.001 & 0.033 & 94.9 \\ 
  T2 & L1 & 2.3 & 300 & -0.001 & 0.036 & 94.8 & 0.000 & 0.041 & 94.7 & 0.000 & 0.037 & 94.2 & 0.001 & 0.042 & 94.6 \\ 
  T2 & L1 & 2.3 & 600 & -0.001 & 0.025 & 94.4 & 0.000 & 0.028 & 95.3 & 0.000 & 0.025 & 95.6 & 0.000 & 0.029 & 94.6 \\ 
  \hline
  T2 & L2 & 1.0 & 300 & -0.095 & 0.117 & 67.9 & -0.017 & 0.079 & 92.6 & -0.061 & 0.091 & 85.0 & -0.016 & 0.080 & 92.8 \\ 
  T2 & L2 & 1.0 & 600 & -0.097 & 0.107 & 45.0 & -0.011 & 0.056 & 93.7 & -0.063 & 0.079 & 71.9 & -0.010 & 0.056 & 93.6 \\ 
  T2 & L2 & 1.6 & 300 & -0.084 & 0.093 & 48.2 & -0.007 & 0.050 & 94.7 & -0.064 & 0.076 & 67.4 & -0.006 & 0.052 & 95.2 \\ 
  T2 & L2 & 1.6 & 600 & -0.082 & 0.087 & 18.6 & -0.003 & 0.037 & 95.3 & -0.063 & 0.070 & 42.8 & -0.003 & 0.038 & 94.5 \\ 
  T2 & L2 & 2.3 & 300 & -0.075 & 0.084 & 45.1 & -0.002 & 0.044 & 95.2 & -0.063 & 0.074 & 60.0 & -0.002 & 0.045 & 96.0 \\ 
  T2 & L2 & 2.3 & 600 & -0.077 & 0.081 & 14.7 & -0.002 & 0.033 & 95.5 & -0.066 & 0.071 & 29.4 & -0.002 & 0.033 & 95.9 \\ 
  \hline
  T2 & L3 & 1.0 & 300 & -0.060 & 0.083 & 80.0 & -0.070 & 0.093 & 79.9 & -0.078 & 0.101 & 76.0 & -0.083 & 0.107 & 78.5 \\ 
  T2 & L3 & 1.0 & 600 & -0.059 & 0.070 & 66.1 & -0.072 & 0.083 & 59.6 & -0.078 & 0.089 & 56.4 & -0.085 & 0.097 & 56.1 \\ 
  T2 & L3 & 1.6 & 300 & -0.048 & 0.062 & 74.8 & -0.062 & 0.074 & 69.2 & -0.066 & 0.078 & 63.4 & -0.074 & 0.086 & 63.9 \\ 
  T2 & L3 & 1.6 & 600 & -0.046 & 0.053 & 57.8 & -0.061 & 0.068 & 41.9 & -0.064 & 0.071 & 37.8 & -0.074 & 0.080 & 33.1 \\ 
  T2 & L3 & 2.3 & 300 & -0.033 & 0.048 & 85.9 & -0.048 & 0.060 & 79.2 & -0.048 & 0.060 & 76.9 & -0.058 & 0.070 & 73.3 \\ 
  T2 & L3 & 2.3 & 600 & -0.033 & 0.042 & 72.6 & -0.050 & 0.057 & 54.5 & -0.049 & 0.056 & 55.2 & -0.060 & 0.067 & 44.8 \\  
   \hline
\end{tabular}
\end{sidewaystable}

\clearpage

\bmsection{Additional simulation studies}\label{append:addition-sim}

\bmsubsection{Scenarios with one covariate}\label{append:sim-onecov}

As the regression-based AUC estimator exhibited biases in all scenarios of Appendix~\ref{append:sim}, we evaluated $\widehat\AUC_{\text{REG}}$ in settings where the distribution of event times and left-truncation times depend on only one common covariate. In this case, the regression-based estimator reduces to the semiparametric estimator in Li 2016 $\widehat\AUC_{\text{LI-SP}}$. As in Section~\ref{sec:simu}, we consider two scenarios of event time distributions for correctly and incorrectly specified models of the event times, respectively:
\begin{enumerate}[label = {(T\arabic*)}]
    \item $T\mid Z\sim \text{Weibull}(2, \exp(-0.3Z_1))$;
    \item $T\mid Z\sim \text{Weibull}(2, 2\exp(-\{2(Z_1-0.33)^++(Z_1+0.33)^+\}/2)$;
\end{enumerate}
 We similarly consider the following scenarios for the distribution of left truncation times that depends on at most one covariate:
\begin{enumerate}[label = {(L\arabic*)}]
    \item $L\mid Z\sim \text{Unif}(0,5)$;
    \item $$S_{L\mid Z}(u\mid Z)= 1 - \left(\dfrac{5-u}{5}\right)^{\exp(0.5Z_1)}$$ if  $u\leq 5$ and $S_{L\mid Z}(u\mid Z)=0$ if $u>5$;
    \item $$S_{L\mid Z}(u\mid Z)= 1 - \left(\dfrac{5-u}{5}\right)^{\exp(0.4\times \text{sign}\{|Z_1|-0.33\})}$$
    if $u\leq 5$ and $S_{L\mid Z}(u\mid Z)=0$ if $u>5$, where $\text{sign}(a)$ is the sign of $a$. 
\end{enumerate}
Scenarios L1-L3 encode independent left truncation, covariate-induced dependent left truncation with correctly specified model (Cox proportional hazards model on the reverse time), and covariate-induced dependent left truncation with incorrectly specified model. 
Finally, we generate the censoring times as $C=L+D$ where $D\sim \text{Unif}(0.5,5)$. That is, we only consider the simulation settings under Scenario A.

Table~\ref{tab:sim_common_a1a2_regonly_v2_1} shows the results. As expected, the estimator $\widehat{\AUC}_{\text{LI-SP}}$ is unbiased with calibrated 95\% confidence intervals under Scenario T1 (correctly specified model for event times), regardless of the distribution of left truncation times, but is subject to bias under T2.

We also evaluated $\widehat{\AUC}_{\text{LI-SP}}$ under the scenario settings of Tables~\ref{tab:sim_common_a1a2_0}-\ref{tab:sim_common_a1a2_2} where the distribution of event times and left-truncation times depends on more than one  covariate. In those scenarios, the model for the event times in $\widehat{\AUC}_{\text{LI-SP}}$ is misspecified since it assumes the proportional hazard assumption relative to the marker alone. As expected, in Table~\ref{tab:sim_common_a1a2_regonly_v1}, we saw that $\widehat{\AUC}_{\text{LI-SP}}$ is biased in all scenarios.

\clearpage

\begin{table}[!htbp]
    \centering   
    \caption{\label{tab:sim_common_a1a2_regonly_v2_0} Summary of different simulation scenarios of event time and left truncation time distribution the scenarios in Appendix~\ref{append:sim-onecov} with only one covariate. In each scenario and each time point, we report the true value of the time-dependent AUC ($\AUC(t)$) the average number and IQR of total events ($\overline{\#\text{Total events}}$), number of subjects that have entered the study ($\overline{\#\text{Entered by $t$}}$), number at risk  ($\overline{\#\text{At risk at $t$}}$), and number of events cumulative by $t$ ($\overline{\#\text{Events by $t$}}$)}
\begin{tabular}{rr|rr|r|cccc}
  \hline
\multicolumn{2}{c|}{Scenario} & \multirow{2}{*}{$t$} & \multirow{2}{*}{$N$} & \multirow{2}{*}{$AUC(t)$}  & \multirow{2}{*}{$\#\text{ Entered by $t$}$} & \multirow{2}{*}{$\#\text{ At risk at }t$} & \multirow{2}{*}{$\#\text{ Events by $t$}$} & \multirow{2}{*}{$\#\text{ Total events}$} \\ 
\cline{1-2}
$T$ & $L$ & & & & & & \\
  \hline
T1 & L1 & 0.5 & 0.610 & 300 & 161 (155, 167) & 135 (129, 141) & 26 (23, 30) & 285 (282, 287) \\ 
  T1 & L1 & 0.5 & 0.610 & 600 & 322 (313, 331) & 270 (261, 279) & 52 (47, 57) & 569 (566, 573) \\ 
  T1 & L1 & 0.8 & 0.632 & 300 & 229 (224, 234) & 144 (138, 150) & 83 (78, 88) & 285 (282, 287) \\ 
  T1 & L1 & 0.8 & 0.632 & 600 & 458 (450, 465) & 288 (280, 296) & 166 (158, 173) & 569 (566, 573) \\ 
  T1 & L1 & 1.2 & 0.675 & 300 & 282 (279, 284) & 99 (93, 104) & 175 (169, 181) & 285 (282, 287) \\ 
  T1 & L1 & 1.2 & 0.675 & 600 & 562 (558, 566) & 198 (189, 205) & 350 (341, 358) & 569 (566, 573) \\ 
  \hline
  T1 & L2 & 0.5 & 0.610 & 300 & 166 (160, 172) & 142 (136, 148) & 24 (21, 27) & 283 (280, 286) \\ 
  T1 & L2 & 0.5 & 0.610 & 600 & 332 (324, 341) & 284 (275, 292) & 48 (44, 53) & 566 (562, 569) \\ 
  T1 & L2 & 0.8 & 0.632 & 300 & 232 (228, 237) & 151 (145, 157) & 79 (74, 84) & 283 (280, 286) \\ 
  T1 & L2 & 0.8 & 0.632 & 600 & 466 (458, 472) & 302 (294, 310) & 158 (151, 166) & 566 (562, 569) \\ 
  T1 & L2 & 1.2 & 0.675 & 300 & 283 (280, 286) & 102 (97, 107) & 172 (166, 178) & 283 (280, 286) \\ 
  T1 & L2 & 1.2 & 0.675 & 600 & 566 (563, 570) & 204 (197, 212) & 344 (335, 352) & 566 (562, 569) \\ 
  \hline
  T1 & L3 & 0.5 & 0.610 & 300 & 148 (142, 154) & 124 (118, 130) & 23 (21, 27) & 285 (283, 288) \\ 
  T1 & L3 & 0.5 & 0.610 & 600 & 297 (289, 305) & 249 (241, 258) & 48 (43, 52) & 571 (567, 574) \\ 
  T1 & L3 & 0.8 & 0.632 & 300 & 217 (212, 222) & 138 (132, 144) & 77 (72, 82) & 285 (283, 288) \\ 
  T1 & L3 & 0.8 & 0.632 & 600 & 434 (427, 441) & 277 (267, 285) & 154 (147, 161) & 571 (567, 574) \\ 
  T1 & L3 & 1.2 & 0.675 & 300 & 276 (273, 279) & 101 (95, 106) & 169 (163, 174) & 285 (283, 288) \\ 
  T1 & L3 & 1.2 & 0.675 & 600 & 553 (548, 557) & 201 (194, 209) & 337 (328, 345) & 571 (567, 574) \\ 
  \hline
  T2 & L1 & 0.5 & 0.659 & 300 & 165 (159, 171) & 135 (129, 140) & 30 (27, 34) & 285 (283, 288) \\ 
  T2 & L1 & 0.5 & 0.659 & 600 & 330 (322, 339) & 269 (261, 277) & 61 (56, 66) & 571 (567, 574) \\ 
  T2 & L1 & 0.8 & 0.674 & 300 & 232 (227, 237) & 139 (133, 145) & 91 (85, 96) & 285 (283, 288) \\ 
  T2 & L1 & 0.8 & 0.674 & 600 & 464 (457, 471) & 279 (270, 287) & 181 (174, 189) & 571 (567, 574) \\ 
  T2 & L1 & 1.2 & 0.702 & 300 & 282 (280, 285) & 94 (88, 100) & 180 (175, 186) & 285 (283, 288) \\ 
  T2 & L1 & 1.2 & 0.702 & 600 & 564 (561, 568) & 188 (180, 196) & 361 (353, 369) & 571 (567, 574) \\ 
  \hline
  T2 & L2 & 0.5 & 0.659 & 300 & 171 (165, 177) & 144 (138, 150) & 27 (24, 31) & 283 (281, 286) \\ 
  T2 & L2 & 0.5 & 0.659 & 600 & 342 (335, 351) & 288 (280, 296) & 55 (50, 59) & 567 (563, 570) \\ 
  T2 & L2 & 0.8 & 0.674 & 300 & 236 (232, 241) & 148 (142, 154) & 85 (80, 91) & 283 (281, 286) \\ 
  T2 & L2 & 0.8 & 0.674 & 600 & 473 (467, 480) & 297 (289, 305) & 171 (164, 179) & 567 (563, 570) \\ 
  T2 & L2 & 1.2 & 0.702 & 300 & 284 (281, 287) & 98 (93, 103) & 177 (172, 183) & 283 (281, 286) \\ 
  T2 & L2 & 1.2 & 0.702 & 600 & 569 (565, 572) & 196 (188, 203) & 355 (347, 363) & 567 (563, 570) \\ 
  \hline
  T2 & L3 & 0.5 & 0.659 & 300 & 153 (147, 159) & 126 (121, 132) & 26 (23, 30) & 285 (283, 288) \\ 
  T2 & L3 & 0.5 & 0.659 & 600 & 308 (299, 316) & 255 (246, 262) & 53 (49, 58) & 571 (567, 574) \\ 
  T2 & L3 & 0.8 & 0.674 & 300 & 221 (216, 226) & 136 (130, 142) & 83 (78, 88) & 285 (283, 288) \\ 
  T2 & L3 & 0.8 & 0.674 & 600 & 442 (435, 449) & 272 (264, 280) & 165 (158, 173) & 571 (567, 574) \\ 
  T2 & L3 & 1.2 & 0.702 & 300 & 277 (274, 281) & 97 (92, 102) & 173 (167, 179) & 285 (283, 288) \\ 
  T2 & L3 & 1.2 & 0.702 & 600 & 555 (551, 559) & 194 (187, 201) & 347 (339, 355) & 571 (567, 574) \\ 
   \hline
\end{tabular}
\end{table}

\clearpage 

\begin{table}[!htbp]
    \centering   
    \caption{\label{tab:sim_common_a1a2_regonly_v2_1}Bias, square root of MSE, and coverage rate of 95\% bootstrap intervals of $\widehat \AUC_{\text{LI-SP}}$ in simulation studies with 1,000 replications under the scenarios in Appendix~\ref{append:sim-onecov} with only one covariate. }
\begin{tabular}{|rr|rrr|rrr|}
  \hline
\multicolumn{2}{|c|}{Scenario} & \multirow{2}{*}{$t$} & \multirow{2}{*}{$\AUC(t)$} & \multirow{2}{*}{$N$}  & \multicolumn{3}{c|}{$\widehat \AUC_{\text{LI-SP}}$} \\ 
\cline{1-2}\cline{6-8}
$T$ & $L$ & &  & & Bias &$\sqrt{\text{MSE}}$ & Coverage \\
  \hline
T1 & L1 & 0.5 & 0.610 & 300 & 0.001 & 0.020 & 94.4 \\ 
  T1 & L1 & 0.5 & 0.610 & 600 & 0.000 & 0.014 & 94.8 \\ 
  T1 & L1 & 0.8 & 0.632 & 300 & 0.002 & 0.024 & 94.9 \\ 
  T1 & L1 & 0.8 & 0.632 & 600 & 0.000 & 0.017 & 95.1 \\ 
  T1 & L1 & 1.2 & 0.675 & 300 & 0.002 & 0.032 & 95.8 \\ 
  T1 & L1 & 1.2 & 0.675 & 600 & 0.000 & 0.023 & 95.5 \\ 
  \hline
  T1 & L2 & 0.5 & 0.610 & 300 & 0.001 & 0.021 & 94.3 \\ 
  T1 & L2 & 0.5 & 0.610 & 600 & 0.001 & 0.015 & 93.7 \\ 
  T1 & L2 & 0.8 & 0.632 & 300 & 0.001 & 0.026 & 94.9 \\ 
  T1 & L2 & 0.8 & 0.632 & 600 & 0.001 & 0.019 & 94.6 \\ 
  T1 & L2 & 1.2 & 0.675 & 300 & 0.002 & 0.034 & 95.6 \\ 
  T1 & L2 & 1.2 & 0.675 & 600 & 0.001 & 0.025 & 95.6 \\ 
  \hline
  T1 & L3 & 0.5 & 0.610 & 300 & 0.001 & 0.021 & 94.7 \\ 
  T1 & L3 & 0.5 & 0.610 & 600 & 0.001 & 0.016 & 92.9 \\ 
  T1 & L3 & 0.8 & 0.632 & 300 & 0.001 & 0.026 & 95.3 \\ 
  T1 & L3 & 0.8 & 0.632 & 600 & 0.001 & 0.020 & 93.5 \\ 
  T1 & L3 & 1.2 & 0.675 & 300 & 0.002 & 0.034 & 95.5 \\ 
  T1 & L3 & 1.2 & 0.675 & 600 & 0.002 & 0.026 & 94.6 \\ 
  \hline
  T2 & L1 & 0.5 & 0.659 & 300 & -0.036 & 0.041 & 51.7 \\ 
  T2 & L1 & 0.5 & 0.659 & 600 & -0.038 & 0.040 & 23.7 \\ 
  T2 & L1 & 0.8 & 0.674 & 300 & -0.026 & 0.035 & 79.4 \\ 
  T2 & L1 & 0.8 & 0.674 & 600 & -0.028 & 0.033 & 61.0 \\ 
  T2 & L1 & 1.2 & 0.702 & 300 & -0.005 & 0.032 & 94.8 \\ 
  T2 & L1 & 1.2 & 0.702 & 600 & -0.008 & 0.023 & 92.2 \\ 
  \hline
  T2 & L2 & 0.5 & 0.659 & 300 & -0.045 & 0.049 & 38.2 \\ 
  T2 & L2 & 0.5 & 0.659 & 600 & -0.046 & 0.048 & 13.5 \\ 
  T2 & L2 & 0.8 & 0.674 & 300 & -0.037 & 0.045 & 65.0 \\ 
  T2 & L2 & 0.8 & 0.674 & 600 & -0.038 & 0.042 & 41.3 \\ 
  T2 & L2 & 1.2 & 0.702 & 300 & -0.019 & 0.038 & 90.2 \\ 
  T2 & L2 & 1.2 & 0.702 & 600 & -0.020 & 0.031 & 83.9 \\ 
  \hline
  T2 & L3 & 0.5 & 0.659 & 300 & -0.045 & 0.050 & 41.6 \\ 
  T2 & L3 & 0.5 & 0.659 & 600 & -0.045 & 0.048 & 15.1 \\ 
  T2 & L3 & 0.8 & 0.674 & 300 & -0.037 & 0.046 & 67.7 \\ 
  T2 & L3 & 0.8 & 0.674 & 600 & -0.038 & 0.042 & 44.7 \\ 
  T2 & L3 & 1.2 & 0.702 & 300 & -0.020 & 0.039 & 89.6 \\ 
  T2 & L3 & 1.2 & 0.702 & 600 & -0.020 & 0.032 & 84.2 \\ 
   \hline
\end{tabular}
\end{table}

\clearpage

\begin{table}[!htbp]
    \centering   
    \caption{\label{tab:sim_common_a1a2_regonly_v1}Bias, square root of MSE, and coverage rate of 95\% bootstrap intervals of $\widehat \AUC_{\text{LI-SP}}$ in simulation studies with 1,000 replications under the same simulation scenarios in Tables~\ref{tab:sim_common_a1a2_0}-\ref{tab:sim_common_a1a2_2}. }
\begin{tabular}{|rr|rrr|rrr|}
  \hline
\multicolumn{2}{|c|}{Scenario} & \multirow{2}{*}{$t$} & \multirow{2}{*}{$\AUC(t)$} & \multirow{2}{*}{$N$}  & \multicolumn{3}{c|}{$\widehat \AUC_{\text{LI-SP}}$} \\ 
\cline{1-2}\cline{6-8}
$T$ & $L$ & &  & & Bias &$\sqrt{\text{MSE}}$ & Coverage \\
  \hline
T1 & L1 & 1.0 & 0.679 & 300 & -0.019 & 0.031 & 85.7 \\ 
  T1 & L1 & 1.0 & 0.679 & 600 & -0.019 & 0.025 & 79.8 \\ 
  T1 & L1 & 1.6 & 0.707 & 300 & -0.021 & 0.036 & 87.3 \\ 
  T1 & L1 & 1.6 & 0.707 & 600 & -0.020 & 0.029 & 81.7 \\ 
  T1 & L1 & 2.3 & 0.745 & 300 & -0.021 & 0.041 & 87.9 \\ 
  T1 & L1 & 2.3 & 0.745 & 600 & -0.021 & 0.032 & 83.6 \\ 
  \hline
  T1 & L2 & 1.0 & 0.679 & 300 & -0.024 & 0.035 & 81.6 \\ 
  T1 & L2 & 1.0 & 0.679 & 600 & -0.023 & 0.030 & 72.1 \\ 
  T1 & L2 & 1.6 & 0.707 & 300 & -0.027 & 0.041 & 82.3 \\ 
  T1 & L2 & 1.6 & 0.707 & 600 & -0.027 & 0.035 & 74.9 \\ 
  T1 & L2 & 2.3 & 0.745 & 300 & -0.030 & 0.047 & 84.4 \\ 
  T1 & L2 & 2.3 & 0.745 & 600 & -0.030 & 0.040 & 78.0 \\ 
  \hline
  T1 & L3 & 1.0 & 0.679 & 300 & -0.026 & 0.038 & 81.1 \\ 
  T1 & L3 & 1.0 & 0.679 & 600 & -0.025 & 0.032 & 73.3 \\ 
  T1 & L3 & 1.6 & 0.707 & 300 & -0.029 & 0.045 & 81.6 \\ 
  T1 & L3 & 1.6 & 0.707 & 600 & -0.029 & 0.038 & 74.2 \\ 
  T1 & L3 & 2.3 & 0.745 & 300 & -0.033 & 0.051 & 83.5 \\ 
  T1 & L3 & 2.3 & 0.745 & 600 & -0.033 & 0.043 & 76.4 \\ 
  \hline
  T2 & L1 & 1.0 & 0.658 & 300 & -0.044 & 0.049 & 44.3 \\ 
  T2 & L1 & 1.0 & 0.658 & 600 & -0.043 & 0.046 & 16.8 \\ 
  T2 & L1 & 1.6 & 0.674 & 300 & -0.037 & 0.045 & 68.0 \\ 
  T2 & L1 & 1.6 & 0.674 & 600 & -0.036 & 0.040 & 46.8 \\ 
  T2 & L1 & 2.3 & 0.697 & 300 & -0.021 & 0.038 & 90.0 \\ 
  T2 & L1 & 2.3 & 0.697 & 600 & -0.020 & 0.029 & 85.7 \\
  \hline
  T2 & L2 & 1.0 & 0.658 & 300 & -0.054 & 0.059 & 29.8 \\ 
  T2 & L2 & 1.0 & 0.658 & 600 & -0.054 & 0.056 & 6.6 \\ 
  T2 & L2 & 1.6 & 0.674 & 300 & -0.050 & 0.056 & 50.8 \\ 
  T2 & L2 & 1.6 & 0.674 & 600 & -0.049 & 0.052 & 25.8 \\ 
  T2 & L2 & 2.3 & 0.697 & 300 & -0.038 & 0.051 & 76.8 \\ 
  T2 & L2 & 2.3 & 0.697 & 600 & -0.037 & 0.043 & 67.2 \\ 
  \hline
  T2 & L3 & 1.0 & 0.658 & 300 & -0.057 & 0.062 & 29.6 \\ 
  T2 & L3 & 1.0 & 0.658 & 600 & -0.057 & 0.059 & 6.8 \\ 
  T2 & L3 & 1.6 & 0.674 & 300 & -0.054 & 0.061 & 49.8 \\ 
  T2 & L3 & 1.6 & 0.674 & 600 & -0.054 & 0.057 & 22.6 \\ 
  T2 & L3 & 2.3 & 0.697 & 300 & -0.044 & 0.057 & 74.7 \\ 
  T2 & L3 & 2.3 & 0.697 & 600 & -0.043 & 0.050 & 57.9 \\ 
   \hline
\end{tabular}
\end{table}

\clearpage

\bmsubsection{Scenarios with less frequent outcomes}\label{append:sim-rare}

The simulation settings of Tables~\ref{tab:sim_common_a1a2_0}-\ref{tab:sim_common_b1b2_2} consider primarily frequent outcomes: by the end on the follow-up, from 77\% to 88\% of the subjects in the study sample experienced the event of interest. In this appendix, we evaluated the estimators with less frequent outcomes. We generate the covariates $Z_1$ and $Z_2$, left-truncation times $L$, and right-censoring times $C$ using the same distributions as in Tables~\ref{tab:sim_common_a1a2_0}-\ref{tab:sim_common_b1b2_2}. We simulate the event times $T$ under a mixture cure model~\citep{boag1949maximum} as follows:

\begin{align*}
    S_0\mid Z &\sim \text{Bernoulli}(\exp\{-0.85 + 1.2Z_1 + 0.2Z_2\}/\left[1 + \exp\left\{-0.85 + 1.2Z_1 + 0.2Z_2\right\}\right]),\\
    T_0\mid Z &\sim \text{Weibull}(2, 2\exp(-\{Z_1+Z_2/2\}/2),\\
    T &= S_0T_0 + (1-S_0)\times(+\infty).
\end{align*}

Under the above model, the event time of a subject follows a Weibull distribution if $S_0=1$ and never experiences an event if $S_0=0$. Note that the distribution of event times does not satisfy a Cox proportional hazards model relative to $Z_1$ and $Z_2$.

Tables~\ref{tab:sim_rare_a1a2_0} and \ref{tab:sim_rare_b1b2_0} display the details of the simulation settings. By introducing a cure proportion, less than 15\% subjects in the study sample experience the event. Tables \ref{tab:sim_rare_a1a2_1}, \ref{tab:sim_rare_a1a2_2}, \ref{tab:sim_rare_b1b2_1}, and \ref{tab:sim_rare_b1b2_2} further compared different AUC estimators under the various settings.

\clearpage

\begin{table}[!htbp]
    \centering   
    \caption{\label{tab:sim_rare_a1a2_0} Summary of different simulation scenarios of event time and left truncation time distribution within Scenarios A1 or A2 with less frequent outcomes. In each scenario and each time point, we report the true value of the time-dependent AUC ($\AUC(t)$) the average number and IQR of total events ($\overline{\#\text{Total events}}$), number of subjects that have entered the study ($\overline{\#\text{Entered by $t$}}$), number at risk  ($\overline{\#\text{At risk at $t$}}$), and number of events cumulative by $t$ ($\overline{\#\text{Events by $t$}}$)}
\begin{tabular}{r|rr|r|rrrr}
  \hline
Scenario & $t$ & $N$ & $AUC(t)$ & $\#\text{ Total events}$ & $\#\text{ Entered by $t$}$ & $\#\text{ At risk at }t$ & $\#\text{ Events by $t$}$ \\ 
  \hline
  L1 & 1.5 & 500 & 0.756 & 55 (51, 60) & 179 (171, 187) & 148 (140, 155) & 18 (16, 22) \\ 
  L1 & 1.5 & 1000 & 0.756 & 111 (104, 117) & 356 (346, 366) & 294 (284, 304) & 37 (33, 41) \\ 
  L1 & 2.7 & 500 & 0.730 & 55 (51, 60) & 296 (288, 304) & 199 (192, 207) & 42 (38, 47) \\ 
  L1 & 2.7 & 1000 & 0.730 & 111 (104, 117) & 591 (580, 602) & 399 (389, 409) & 85 (79, 91) \\ 
  \hline
  L2 & 1.5 & 500 & 0.756 & 49 (45, 54) & 184 (177, 191) & 155 (148, 162) & 14 (12, 17) \\ 
  L2 & 1.5 & 1000 & 0.756 & 99 (93, 105) & 368 (358, 378) & 311 (301, 320) & 29 (26, 33) \\ 
  L2 & 2.7 & 500 & 0.730 & 49 (45, 54) & 299 (292, 307) & 204 (197, 211) & 37 (33, 41) \\ 
  L2 & 2.7 & 1000 & 0.730 & 99 (93, 105) & 598 (589, 609) & 407 (398, 419) & 74 (68, 79) \\ 
  \hline
  L3 & 1.5 & 500 & 0.756 & 49 (45, 54) & 157 (150, 164) & 130 (124, 137) & 15 (12, 18) \\ 
  L3 & 1.5 & 1000 & 0.756 & 98 (91, 104) & 315 (305, 326) & 261 (252, 271) & 30 (26, 33) \\ 
  L3 & 2.7 & 500 & 0.730 & 49 (45, 54) & 269 (262, 277) & 184 (176, 191) & 37 (33, 41) \\ 
  L3 & 2.7 & 1000 & 0.730 & 98 (91, 104) & 539 (529, 550) & 368 (358, 378) & 73 (67, 78) \\ 
   \hline
\end{tabular}
\end{table}

\begin{sidewaystable}[!htbp]
    \centering   
    \caption{\label{tab:sim_rare_a1a2_1}Bias, square root of MSE, and coverage rate of 95\% bootstrap intervals of $\widehat \AUC_{\text{RC-IPW}}$, $\widehat \AUC_{\text{RC-CIPW}}$, $\widehat \AUC_{\text{KM}}$, and $\widehat \AUC_{\text{REG}}$ in simulation studies with 1,000 replications under different scenarios of event time and left truncation time distribution within Scenarios A1 or A2 as described in Section~\ref{sec:simu}. The censoring time distribution follows scenario C1. }
\begin{tabular}{|crr|rrr|rrr|rrr|rrr|}
  \hline
\multirow{2}{*}{Scenario} & \multirow{2}{*}{$t$} & \multirow{2}{*}{$N$}  & \multicolumn{3}{c|}{$\widehat \AUC_{\text{RC-IPW}}$} & \multicolumn{3}{c|}{$\widehat \AUC_{\text{RC-CIPW}}$} & \multicolumn{3}{c|}{$\widehat \AUC_{\text{KM}}$} &\multicolumn{3}{c|}{$\widehat \AUC_{\text{REG}}$} \\ 
\cline{4-15}
& &  & Bias &$\sqrt{\text{MSE}}$ & Coverage & Bias &$\sqrt{\text{MSE}}$ & Coverage & Bias &$\sqrt{\text{MSE}}$ & Coverage & Bias &$\sqrt{\text{MSE}}$ & Coverage \\
  \hline
L1 & 1.5 & 500 & 0.001 & 0.056 & 91.0 & 0.001 & 0.056 & 91.0 & -0.632 & 0.632 & 0.0 & -0.121 & 0.127 & 16.0 \\ 
  L1 & 1.5 & 1000 & 0.000 & 0.038 & 93.9 & 0.000 & 0.038 & 94.0 & -0.632 & 0.632 & 0.0 & -0.118 & 0.122 & 1.6 \\ 
  L1 & 2.7 & 500 & -0.029 & 0.049 & 89.9 & -0.029 & 0.049 & 89.6 & -0.683 & 0.683 & 0.0 & -0.085 & 0.096 & 54.7 \\ 
  L1 & 2.7 & 1000 & -0.028 & 0.040 & 83.1 & -0.028 & 0.040 & 82.8 & -0.684 & 0.684 & 0.0 & -0.083 & 0.089 & 23.6 \\ 
  \hline
  L2 & 1.5 & 500 & -0.077 & 0.103 & 80.5 & -0.077 & 0.103 & 80.7 & -0.649 & 0.650 & 0.0 & -0.122 & 0.129 & 22.8 \\ 
  L2 & 1.5 & 1000 & -0.078 & 0.092 & 63.5 & -0.078 & 0.092 & 64.0 & -0.651 & 0.651 & 0.0 & -0.121 & 0.126 & 3.1 \\ 
  L2 & 2.7 & 500 & -0.099 & 0.109 & 39.1 & -0.096 & 0.106 & 42.2 & -0.688 & 0.688 & 0.0 & -0.086 & 0.098 & 58.3 \\ 
  L2 & 2.7 & 1000 & -0.098 & 0.103 & 10.6 & -0.095 & 0.101 & 12.8 & -0.688 & 0.688 & 0.0 & -0.086 & 0.093 & 28.2 \\ 
  \hline
  L3 & 1.5 & 500 & -0.060 & 0.081 & 79.9 & -0.060 & 0.081 & 79.8 & -0.642 & 0.643 & 0.0 & -0.150 & 0.159 & 16.1 \\ 
  L3 & 1.5 & 1000 & -0.059 & 0.070 & 65.7 & -0.059 & 0.071 & 65.2 & -0.642 & 0.643 & 0.0 & -0.152 & 0.157 & 1.0 \\ 
  L3 & 2.7 & 500 & -0.064 & 0.074 & 61.3 & -0.064 & 0.075 & 60.4 & -0.685 & 0.685 & 0.0 & -0.117 & 0.129 & 45.5 \\ 
  L3 & 2.7 & 1000 & -0.063 & 0.069 & 33.6 & -0.064 & 0.069 & 31.7 & -0.685 & 0.685 & 0.0 & -0.119 & 0.125 & 12.1 \\ 
   \hline
\end{tabular}
\end{sidewaystable}

\begin{sidewaystable}[!htbp]
    \centering   
    \caption{\label{tab:sim_rare_a1a2_2}Bias, square root of MSE, and coverage rate of 95\% bootstrap intervals of $\widehat \AUC_{\text{IPW-1}}$, $\widehat \AUC_{\text{CIPW-1}}$, $\widehat \AUC_{\text{IPW-2}}$, and $\widehat \AUC_{\text{CIPW-2}}$ in simulation studies with 1,000 replications under different scenarios of event time and left truncation time distribution within Scenarios A1 or A2 with less frequent outcomes. The censoring time distribution follows scenario C1. }
\begin{tabular}{|crr|rrr|rrr|rrr|rrr|}
  \hline
\multirow{2}{*}{Scenario} & \multirow{2}{*}{$t$} & \multirow{2}{*}{$N$}  & \multicolumn{3}{c|}{$\widehat \AUC_{\text{IPW-1}}$} & \multicolumn{3}{c|}{$\widehat \AUC_{\text{CIPW-1}}$} & \multicolumn{3}{c|}{$\widehat \AUC_{\text{IPW-2}}$} &\multicolumn{3}{c|}{$\widehat \AUC_{\text{CIPW-2}}$} \\ 
\cline{4-15}
 & &  & Bias &$\sqrt{\text{MSE}}$ & Coverage & Bias &$\sqrt{\text{MSE}}$ & Coverage & Bias &$\sqrt{\text{MSE}}$ & Coverage & Bias &$\sqrt{\text{MSE}}$ & Coverage \\
  \hline
L1 & 1.5 & 500 & -0.060 & 0.105 & 88.0 & -0.067 & 0.127 & 92.5 & -0.002 & 0.064 & 92.2 & -0.002 & 0.081 & 95.1 \\ 
  L1 & 1.5 & 1000 & -0.064 & 0.087 & 80.5 & -0.068 & 0.098 & 84.1 & -0.003 & 0.045 & 93.9 & -0.004 & 0.052 & 95.8 \\ 
  L1 & 2.7 & 500 & 0.076 & 0.106 & 75.0 & 0.075 & 0.119 & 85.4 & -0.001 & 0.046 & 94.1 & -0.004 & 0.066 & 97.0 \\ 
  L1 & 2.7 & 1000 & 0.075 & 0.091 & 66.3 & 0.075 & 0.097 & 77.7 & -0.001 & 0.033 & 94.0 & -0.003 & 0.044 & 96.1 \\ 
  \hline
  L2 & 1.5 & 500 & -0.119 & 0.154 & 77.3 & -0.089 & 0.148 & 90.8 & -0.024 & 0.078 & 93.8 & -0.019 & 0.087 & 96.6 \\ 
  L2 & 1.5 & 1000 & -0.122 & 0.139 & 54.4 & -0.081 & 0.113 & 83.2 & -0.027 & 0.059 & 93.7 & -0.015 & 0.062 & 95.7 \\ 
  L2 & 2.7 & 500 & 0.010 & 0.087 & 91.3 & 0.067 & 0.116 & 89.8 & -0.048 & 0.068 & 85.4 & -0.008 & 0.068 & 98.1 \\ 
  L2 & 2.7 & 1000 & 0.009 & 0.059 & 93.2 & 0.067 & 0.093 & 84.6 & -0.049 & 0.061 & 73.7 & -0.008 & 0.050 & 96.1 \\ 
  \hline
  L3 & 1.5 & 500 & -0.130 & 0.159 & 68.5 & -0.123 & 0.164 & 86.1 & -0.068 & 0.098 & 84.5 & -0.067 & 0.106 & 94.1 \\ 
  L3 & 1.5 & 1000 & -0.126 & 0.141 & 46.1 & -0.121 & 0.142 & 63.8 & -0.067 & 0.085 & 74.4 & -0.069 & 0.090 & 83.2 \\ 
  L3 & 2.7 & 500 & 0.040 & 0.097 & 86.5 & 0.060 & 0.114 & 91.1 & -0.051 & 0.067 & 80.2 & -0.043 & 0.076 & 94.5 \\ 
  L3 & 2.7 & 1000 & 0.042 & 0.073 & 85.8 & 0.056 & 0.086 & 87.4 & -0.050 & 0.060 & 64.8 & -0.045 & 0.062 & 83.4 \\   
   \hline
\end{tabular}
\end{sidewaystable}

\newpage

\begin{table}[!htbp]
    \centering   
    \caption{\label{tab:sim_rare_b1b2_0} Summary of different simulation scenarios of event time and left truncation time distribution within Scenarios B1 or B2 with less frequent outcomes. In each scenario and each time point, we report the true value of the time-dependent AUC ($\AUC(t)$) the average number and IQR of total events ($\overline{\#\text{Total events}}$), number of subjects that have entered the study ($\overline{\#\text{Entered by $t$}}$), number at risk  ($\overline{\#\text{At risk at $t$}}$), and number of events cumulative by $t$ ($\overline{\#\text{Events by $t$}}$)}
\begin{tabular}{r|rr|r|rrrr}
  \hline
Scenario & $t$ & $N$ & $AUC(t)$ & $\#\text{ Total events}$ & $\#\text{ Entered by $t$}$ & $\#\text{ At risk at }t$ & $\#\text{ Events by $t$}$ \\ 
  \hline
 L1 & 1.5 & 500 & 0.756 & 66 (62, 71) & 214 (206, 222) & 177 (169, 184) & 22 (19, 26) \\ 
  L1 & 1.5 & 1000 & 0.756 & 132 (125, 140) & 427 (417, 439) & 352 (343, 364) & 45 (41, 50) \\ 
  L1 & 2.7 & 500 & 0.730 & 66 (62, 71) & 335 (329, 343) & 237 (230, 245) & 51 (46, 55) \\ 
  L1 & 2.7 & 1000 & 0.730 & 132 (125, 140) & 671 (660, 681) & 475 (464, 485) & 102 (96, 109) \\ 
  \hline
  L2 & 1.5 & 500 & 0.756 & 60 (55, 64) & 220 (213, 227) & 186 (180, 193) & 17 (14, 20) \\ 
  L2 & 1.5 & 1000 & 0.756 & 119 (112, 127) & 441 (430, 452) & 373 (363, 384) & 35 (31, 39) \\ 
  L2 & 2.7 & 500 & 0.730 & 60 (55, 64) & 339 (332, 346) & 246 (239, 253) & 44 (40, 48) \\ 
  L2 & 2.7 & 1000 & 0.730 & 119 (112, 127) & 678 (669, 688) & 492 (481, 503) & 88 (82, 94) \\ 
  \hline
  L3 & 1.5 & 500 & 0.756 & 60 (55, 65) & 193 (186, 200) & 161 (154, 168) & 18 (15, 21) \\ 
  L3 & 1.5 & 1000 & 0.756 & 120 (113, 127) & 387 (377, 398) & 323 (313, 334) & 37 (33, 41) \\ 
  L3 & 2.7 & 500 & 0.730 & 60 (55, 65) & 311 (304, 319) & 223 (215, 230) & 45 (41, 49) \\ 
  L3 & 2.7 & 1000 & 0.730 & 120 (113, 127) & 624 (614, 634) & 448 (437, 458) & 90 (84, 96) \\ 
   \hline
\end{tabular}
\end{table}

\begin{sidewaystable}[!htbp]
    \centering   
    \caption{\label{tab:sim_rare_b1b2_1}Bias, square root of MSE, and coverage rate of 95\% bootstrap intervals of $\widehat \AUC_{\text{RC-IPW}}$, $\widehat \AUC_{\text{RC-CIPW}}$, $\widehat \AUC_{\text{KM}}$, and $\widehat \AUC_{\text{REG}}$ in simulation studies with 1,000 replications under different scenarios of event time and left truncation time distribution within Scenarios B1 or B2 with less frequent outcomes. The censoring time distribution follows scenario C2. }
\begin{tabular}{|crr|rrr|rrr|rrr|rrr|}
  \hline
\multirow{2}{*}{Scenario} & \multirow{2}{*}{$t$} & \multirow{2}{*}{$N$}  & \multicolumn{3}{c|}{$\widehat \AUC_{\text{RC-IPW}}$} & \multicolumn{3}{c|}{$\widehat \AUC_{\text{RC-CIPW}}$} & \multicolumn{3}{c|}{$\widehat \AUC_{\text{KM}}$} &\multicolumn{3}{c|}{$\widehat \AUC_{\text{REG}}$} \\ 
\cline{4-15}
& &  & Bias &$\sqrt{\text{MSE}}$ & Coverage & Bias &$\sqrt{\text{MSE}}$ & Coverage & Bias &$\sqrt{\text{MSE}}$ & Coverage & Bias &$\sqrt{\text{MSE}}$ & Coverage \\
  \hline
L1 & 1.5 & 500 & 0.000 & 0.050 & 93.2 & 0.000 & 0.050 & 93.1 & -0.632 & 0.632 & 0.0 & -0.120 & 0.125 & 11.6 \\ 
  L1 & 1.5 & 1000 & -0.001 & 0.035 & 92.4 & -0.001 & 0.035 & 92.4 & -0.631 & 0.631 & 0.0 & -0.119 & 0.122 & 0.8 \\ 
  L1 & 2.7 & 500 & -0.027 & 0.046 & 89.1 & -0.027 & 0.046 & 89.5 & -0.683 & 0.683 & 0.0 & -0.084 & 0.093 & 47.4 \\ 
  L1 & 2.7 & 1000 & -0.027 & 0.038 & 81.4 & -0.028 & 0.038 & 81.2 & -0.684 & 0.684 & 0.0 & -0.084 & 0.089 & 18.2 \\ 
  \hline
  L2 & 1.5 & 500 & -0.070 & 0.094 & 81.6 & -0.070 & 0.094 & 81.7 & -0.648 & 0.648 & 0.0 & -0.122 & 0.128 & 14.1 \\ 
  L2 & 1.5 & 1000 & -0.070 & 0.082 & 65.0 & -0.070 & 0.082 & 65.2 & -0.649 & 0.649 & 0.0 & -0.122 & 0.125 & 0.8 \\ 
  L2 & 2.7 & 500 & -0.089 & 0.098 & 42.3 & -0.088 & 0.097 & 42.4 & -0.686 & 0.687 & 0.0 & -0.087 & 0.097 & 50.5 \\ 
  L2 & 2.7 & 1000 & -0.088 & 0.093 & 12.3 & -0.087 & 0.092 & 12.9 & -0.687 & 0.687 & 0.0 & -0.087 & 0.092 & 18.8 \\ 
  \hline
  L3 & 1.5 & 500 & -0.061 & 0.078 & 75.6 & -0.061 & 0.078 & 75.4 & -0.642 & 0.642 & 0.0 & -0.145 & 0.153 & 12.9 \\ 
  L3 & 1.5 & 1000 & -0.059 & 0.070 & 60.1 & -0.060 & 0.070 & 60.0 & -0.642 & 0.642 & 0.0 & -0.145 & 0.149 & 0.9 \\ 
  L3 & 2.7 & 500 & -0.065 & 0.074 & 54.9 & -0.065 & 0.074 & 55.0 & -0.685 & 0.685 & 0.0 & -0.111 & 0.122 & 43.9 \\ 
  L3 & 2.7 & 1000 & -0.064 & 0.069 & 27.3 & -0.064 & 0.069 & 27.0 & -0.685 & 0.685 & 0.0 & -0.112 & 0.117 & 12.3 \\  
   \hline
\end{tabular}
\end{sidewaystable}

\newpage

\begin{sidewaystable}[!htbp]
    \centering   
    \caption{\label{tab:sim_rare_b1b2_2}Bias, square root of MSE, and coverage rate of 95\% bootstrap intervals of $\widehat \AUC_{\text{IPW-1}}$, $\widehat \AUC_{\text{CIPW-1}}$, $\widehat \AUC_{\text{IPW-2}}$, and $\widehat \AUC_{\text{CIPW-2}}$ in simulation studies with 1,000 replications under different scenarios of event time and left truncation time distribution within Scenarios B1 or B2 with less frequent outcomes. The censoring time distribution follows scenario C2. }
\begin{tabular}{|crr|rrr|rrr|rrr|rrr|}
  \hline
\multirow{2}{*}{Scenario} & \multirow{2}{*}{$t$} & \multirow{2}{*}{$N$}  & \multicolumn{3}{c|}{$\widehat \AUC_{\text{IPW-1}}$} & \multicolumn{3}{c|}{$\widehat \AUC_{\text{CIPW-1}}$} & \multicolumn{3}{c|}{$\widehat \AUC_{\text{IPW-2}}$} &\multicolumn{3}{c|}{$\widehat \AUC_{\text{CIPW-2}}$} \\ 
\cline{4-15}
 & &  & Bias &$\sqrt{\text{MSE}}$ & Coverage & Bias &$\sqrt{\text{MSE}}$ & Coverage & Bias &$\sqrt{\text{MSE}}$ & Coverage & Bias &$\sqrt{\text{MSE}}$ & Coverage \\
  \hline
L1 & 1.5 & 500 & -0.060 & 0.098 & 86.9 & -0.060 & 0.099 & 87.4 & -0.002 & 0.059 & 92.8 & -0.002 & 0.060 & 92.8 \\ 
  L1 & 1.5 & 1000 & -0.062 & 0.082 & 78.3 & -0.062 & 0.083 & 79.5 & -0.003 & 0.041 & 92.9 & -0.003 & 0.042 & 92.4 \\ 
  L1 & 2.7 & 500 & 0.077 & 0.101 & 74.6 & 0.077 & 0.102 & 75.5 & -0.001 & 0.042 & 94.5 & -0.001 & 0.044 & 95.0 \\ 
  L1 & 2.7 & 1000 & 0.076 & 0.090 & 60.1 & 0.076 & 0.090 & 61.3 & -0.001 & 0.030 & 94.4 & -0.001 & 0.031 & 94.7 \\ 
  \hline
  L2 & 1.5 & 500 & -0.111 & 0.142 & 75.7 & -0.070 & 0.115 & 87.5 & -0.021 & 0.071 & 93.8 & -0.009 & 0.067 & 94.3 \\ 
  L2 & 1.5 & 1000 & -0.113 & 0.128 & 53.4 & -0.065 & 0.090 & 79.9 & -0.023 & 0.053 & 92.8 & -0.007 & 0.046 & 93.8 \\ 
  L2 & 2.7 & 500 & 0.018 & 0.076 & 91.7 & 0.075 & 0.102 & 76.2 & -0.041 & 0.061 & 86.1 & -0.003 & 0.048 & 94.5 \\ 
  L2 & 2.7 & 1000 & 0.017 & 0.054 & 92.4 & 0.075 & 0.089 & 64.5 & -0.041 & 0.053 & 75.5 & -0.003 & 0.035 & 93.4 \\ 
  \hline
  L3 & 1.5 & 500 & -0.128 & 0.152 & 63.4 & -0.133 & 0.157 & 63.3 & -0.067 & 0.093 & 83.3 & -0.068 & 0.095 & 83.6 \\ 
  L3 & 1.5 & 1000 & -0.125 & 0.137 & 38.5 & -0.131 & 0.143 & 37.6 & -0.066 & 0.081 & 69.7 & -0.067 & 0.083 & 70.0 \\ 
  L3 & 2.7 & 500 & 0.033 & 0.084 & 89.2 & 0.026 & 0.083 & 90.3 & -0.052 & 0.066 & 75.9 & -0.056 & 0.070 & 75.3 \\ 
  L3 & 2.7 & 1000 & 0.033 & 0.062 & 88.4 & 0.025 & 0.060 & 91.7 & -0.052 & 0.060 & 57.9 & -0.056 & 0.064 & 55.0 \\  
   \hline
\end{tabular}
\end{sidewaystable}

\clearpage

\bmsection{Results for the OPTN data analysis}\label{append:optn_roc}

\begin{table}[!htbp]
\centering
\caption{Characteristics of the OPTN data study sample. Categorical variables were summarized as Frequency (Percentage), and numeric variables were summarized as Median (IQR).}
\label{tab:optn_summary}
\begin{tabular}{lll}
  \hline
Characteristic & Training data
(n=5,487) & Test data (n=2,351) \\ 
  \hline
Gender &  &  \\ 
  $\qquad$F & 1,981 (36\%) & 828 (35\%) \\ 
  $\qquad$M & 3,506 (64\%) & 1,523 (65\%) \\ 
  Race/ethnicity &  &  \\ 
  $\qquad$Non-Hispanic White & 2,984 (54\%) & 1,279 (54\%) \\ 
  $\qquad$Non-Hispanic Black & 1,266 (23\%) & 532 (23\%) \\ 
  $\qquad$Hispanic/Latino & 743 (14\%) & 326 (14\%) \\ 
  $\qquad$Others or Unknown & 494 (9.0\%) & 214 (9.1\%) \\ 
  Age at dialysis initiation & 64.6 (62.2, 67.7) & 64.4 (62.2, 67.3) \\ 
  Time to waitlist registration & 0.79 (0.44, 1.24) & 0.80 (0.44, 1.26) \\ 
  Death by 12/1/2019 & 2,918 (53\%) & 1,226 (52\%) \\ 
  Time to death & 5.01 (3.21, 6.75) & 5.03 (3.19, 6.83) \\ 
  Unknown & 2,569 & 1,125 \\ 
  BMI &  &  \\ 
  $\qquad$Healthy weight & 1,256 (23\%) & 562 (24\%) \\ 
  $\qquad$Underweight & 55 (1.0\%) & 26 (1.1\%) \\ 
  $\qquad$Overweight & 2,057 (37\%) & 869 (37\%) \\ 
  $\qquad$Obese & 2,119 (39\%) & 894 (38\%) \\ 
  Serum Albumin Level &  &  \\ 
  $\qquad$Normal albumin & 4,122 (75\%) & 1,747 (74\%) \\ 
  $\qquad$Low albumin & 1,355 (25\%) & 596 (25\%) \\ 
  $\qquad$High albumin & 10 (0.2\%) & 8 (0.3\%) \\ 
  Diabetes & 3,213 (59\%) & 1,424 (61\%) \\ 
   \hline
\end{tabular}
\end{table}

\clearpage

\begin{table}[!htbp]
\centering
\caption{Results of Cox proportional hazards regression for death using the training data of the OPTN study sample.}
\label{tab:optn_reg}

\begin{tabular}{lrrr}
  \hline
Characteristic & log(HR) & 95\% CI & p-value \\ 
  \hline
Age at dialysis initiation &  0.038 &  0.029,  0.048 & $<$0.001 \\ 
  Gender &      &      &  \\ 
  $\qquad$F &     - &    -  &  \\ 
  $\qquad$M &  0.106 &  0.028,  0.183 & 0.008 \\ 
  Race/ethnicity &      &      &  \\ 
  $\qquad$Non-Hispanic White &     - &     - &  \\ 
  $\qquad$Non-Hispanic Black & -0.250 & -0.343, -0.157 & $<$0.001 \\ 
  $\qquad$Hispanic/Latino & -0.293 & -0.408, -0.178 & $<$0.001 \\ 
  $\qquad$Others or Unknown & -0.405 & -0.546, -0.264 & $<$0.001 \\ 
  Serum Albumin Level &      &      &  \\ 
  $\qquad$Normal albumin &     - &     - &  \\ 
  $\qquad$Low albumin &  0.299 &  0.217,  0.380 & $<$0.001 \\ 
  $\qquad$High albumin &  0.834 &  0.178,  1.490 & 0.013 \\ 
  BMI &     &      &  \\ 
  $\qquad$Healthy weight &     - &    - &  \\ 
  $\qquad$Underweight &  0.257 & -0.100,  0.615 & 0.158 \\ 
  $\qquad$Overweight & -0.177 & -0.274, -0.079 & $<$0.001 \\ 
  $\qquad$Obese & -0.173 & -0.272, -0.074 & $<$0.001 \\ 
  Diabetes &  0.424 &  0.344,  0.504 & $<$0.001 \\ 
   \hline
\end{tabular}
\end{table}

\clearpage
\begin{figure}[!htbp]
    \centering
    \includegraphics[width = 0.6\textwidth]{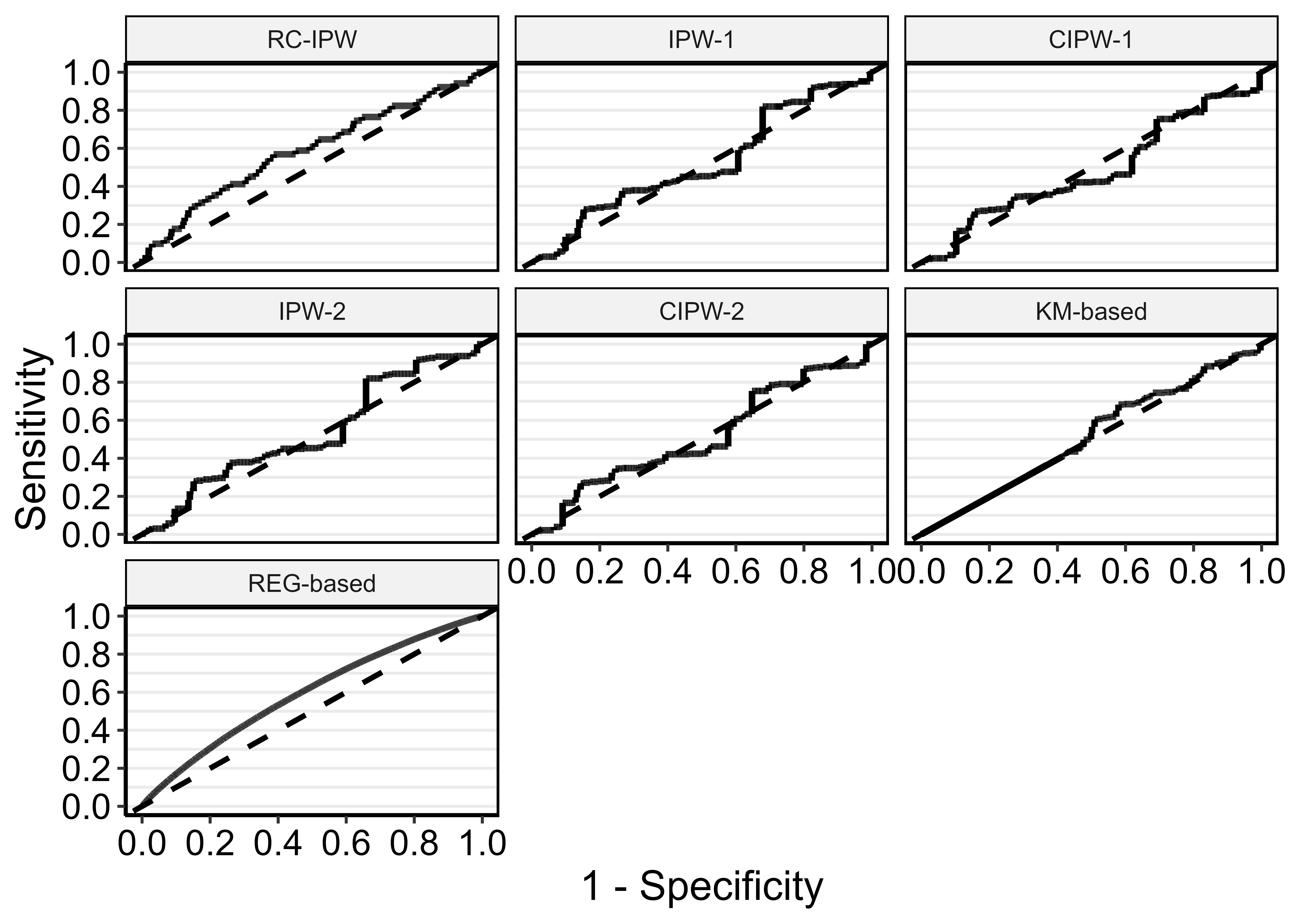}
    \caption{Time-dependent ROC curves  for the descriminatory performances of the mortality prediction model at 1 year after dialysis initiation in elderly ESRD patients using the OPTN data. RC-IPW: the IPW estimators adjusting for only right censoring by Uno et al.~\citep{uno2007evaluating}; IPW-1: the IPW estimators in Equations \eqref{eq:se-ipw-1} and \eqref{eq:sp-ipw-1}; CIPW-1: the conditional IPW estimator in Equations \eqref{eq:se-cipw} and \eqref{eq:sp-cipw-1}; IPW-2: the IPW estimator in Equations \eqref{eq:se-ipw-1} and \eqref{eq:sp-ipw-2}; CIPW-2: the conditional IPW estimator in Equations \eqref{eq:se-cipw} and \eqref{eq:sp-cipw-2}; KM-based: the KM-based estimator in Equations~\eqref{eq:se-km} and \eqref{eq:sp-km}; REG-based: the regression-based estimator in Equations~\eqref{eq:se-reg} and \eqref{eq:sp-reg}.}
    \label{fig:roc-optn-1}
\end{figure}

\clearpage

\begin{figure}[!htbp]
    \centering
    \includegraphics[width = 0.6\textwidth]{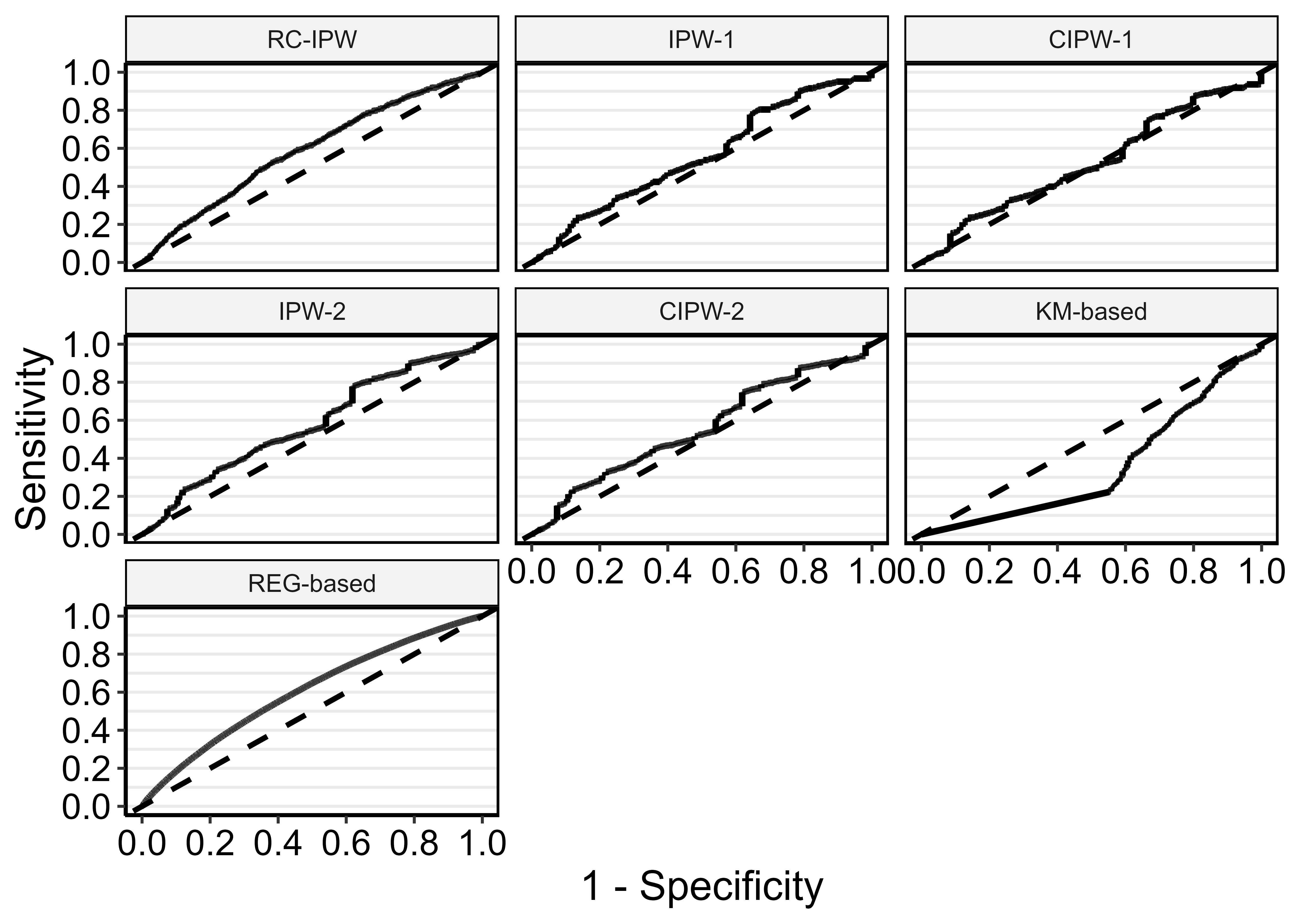}
    \caption{Time-dependent ROC curves  for the descriminatory performances of the mortality prediction model at 5 years after dialysis initiation in elderly ESRD patients using the OPTN data. RC-IPW: the IPW estimators adjusting for only right censoring by Uno et al.~\citep{uno2007evaluating}; IPW-1: the IPW estimators in Equations \eqref{eq:se-ipw-1} and \eqref{eq:sp-ipw-1}; CIPW-1: the conditional IPW estimator in Equations \eqref{eq:se-cipw} and \eqref{eq:sp-cipw-1}; IPW-2: the IPW estimator in Equations \eqref{eq:se-ipw-1} and \eqref{eq:sp-ipw-2}; CIPW-2: the conditional IPW estimator in Equations \eqref{eq:se-cipw} and \eqref{eq:sp-cipw-2}; KM-based: the KM-based estimator in Equations~\eqref{eq:se-km} and \eqref{eq:sp-km}; REG-based: the regression-based estimator in Equations~\eqref{eq:se-reg} and \eqref{eq:sp-reg}.}
    \label{fig:roc-optn-5}
\end{figure}

\clearpage

\begin{figure}[!htbp]
    \centering
    \includegraphics[width = 0.6\textwidth]{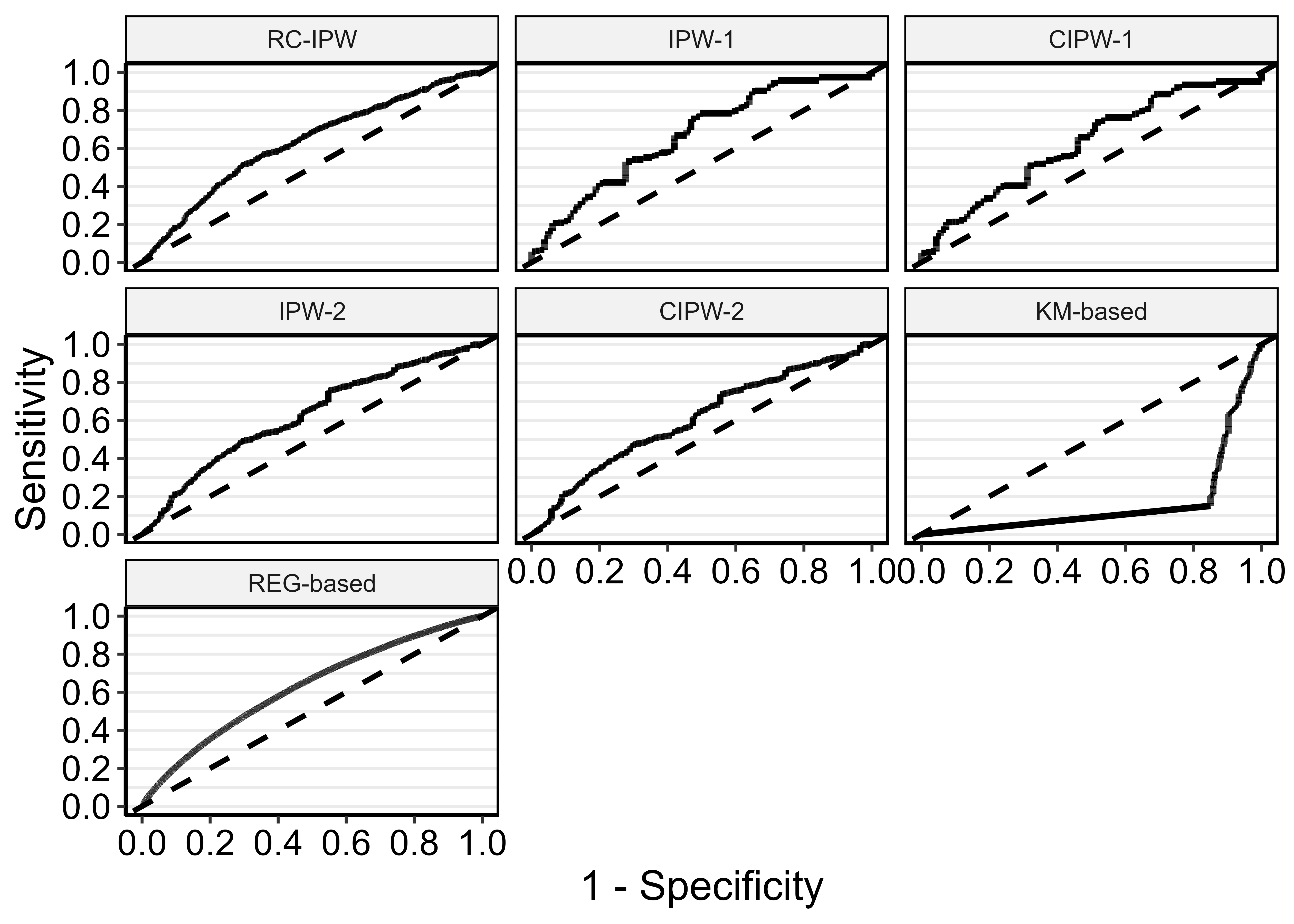}
    \caption{Time-dependent ROC curves  for the descriminatory performances of the mortality prediction model at 9 years after dialysis initiation in elderly ESRD patients using the OPTN data. RC-IPW: the IPW estimators adjusting for only right censoring by Uno et al.~\citep{uno2007evaluating}; IPW-1: the IPW estimators in Equations \eqref{eq:se-ipw-1} and \eqref{eq:sp-ipw-1}; CIPW-1: the conditional IPW estimator in Equations \eqref{eq:se-cipw} and \eqref{eq:sp-cipw-1}; IPW-2: the IPW estimator in Equations \eqref{eq:se-ipw-1} and \eqref{eq:sp-ipw-2}; CIPW-2: the conditional IPW estimator in Equations \eqref{eq:se-cipw} and \eqref{eq:sp-cipw-2}; KM-based: the KM-based estimator in Equations~\eqref{eq:se-km} and \eqref{eq:sp-km}; REG-based: the regression-based estimator in Equations~\eqref{eq:se-reg} and \eqref{eq:sp-reg}.}
    \label{fig:roc-optn-9}
\end{figure}

\end{document}